\documentclass[structabstract]{aa}  
\usepackage{graphicx}
\usepackage{lscape}
\usepackage{txfonts}
\usepackage{natbib}
\usepackage{url}
\newcommand{\Funits}{10$^{-16}$ erg~s$^{-1}$~cm$^{-2}$}

\newcommand{\nodata}{ ~$\cdots$~ }

\newcommand{\upstar}{$^{\star}$}

\begin{document}
   \title{Integral Field Spectroscopy of a sample of nearby galaxies.}

   \subtitle{I. Sample, Observations and Data Reduction\thanks{Based on observations collected at the Centro
   Astron\'omico Hispano Alem\'an (CAHA) at Calar Alto, operated jointly by the Max-Planck Institut f\"ur Astronomie
   and the Instituto de Astrof\'{\i}sica de Andaluc\'{\i}a (CSIC).} }

   \author{E. M\'{a}rmol-Queralt\'{o}\inst{1,2}
     \and
     S.F. S\'anchez\inst{1}
     \and
     R.A. Marino\inst{1,2}
     \and
     D. Mast\inst{1,3}
     \and
     K. Viironen\inst{1,4}
     \and
     A. Gil de Paz\inst{2}
     \and
     J. Iglesias-P\'{a}ramo\inst{3,1}
     \and
     F.F. Rosales-Ortega\inst{5,1}
     \and
     J.M. Vilchez\inst{3}
          }

   \institute{
        Centro Astron\'omico Hispano Alem\'an, Calar Alto, (CSIC-MPG),
        C/Jes\'{u}s Durb\'{a}n Rem\'{o}n 2-2, E-04004 Almer\'{\i}a, Spain 
        \email{emq@astrax.fis.ucm.es}
       \and
        Departamento de Astrof\'{\i}sica y CC. de la Atm\'{o}sfera, 
        Universidad Complutense de Madrid, E28040-Madrid, Spain
       \and
        Instituto de Astrof\'{\i}sica de Andaluc\'{\i}a (CSIC), Camino Bajo 
        de Huetor s/n, Aptdo. 3004, E18080-Granada, Spain
       \and
        Centro de Estudios de F\'isica del Cosmos de Arag\'on (CEFCA), C/
        Pizarro 1, 3$^a$, E-41001 Teruel, Spain
       \and
        Departamento de F\'{\i}sica Te\'{o}rica, 
        Universidad Aut\'{on}oma de Madrid, E28049-Cantoblanco (Madrid), Spain
              }

   \date{Received ----- ; accepted ---- }

 
  \abstract
   {}
   {Integral Field Spectroscopy (IFS) is a powerful approach for the study of
   nearby galaxies since it enables a detailed analysis of their resolved
   physical properties. Here we present the sample of nearby galaxies selected
   to exploit the two dimensional information provided by the IFS.}
   {We observed a sample of 48 galaxies from the Local Universe with the PPAK
   Integral Field Spectroscopy unit (IFU), of the PMAS spectrograph, mounted at
   the 3.5~m telescope at Calar Alto Observatory (Almeria, Spain). Two
   different setups were used during these studies (low -V300- and medium
   -V600- resolution mode) covering a spectral range of around
   $3700-7000~\AA\AA$. We developed a full automatic pipeline for the data
   reduction, that includes an analysis of the quality of the final data
   products.  We applied a decoupling method to obtain the ionised gas and
   stellar content of these galaxies, and to derive the main physical
   properties of the galaxies. To asses the accuracy in the measurements of the
   different parameters, we performed a set of simulations to derive the
   expected relative errors obtained with these data. In addition, we extracted
   two aperture, central and integrated spectra, from the datacubes. The main
   properties of the stellar populations and ionised gas of these galaxies and
   an estimate of their relative errors are derived from those spectra, as well
   as from the whole datacubes.}
   {The comparison of the central spectrum extracted from the datacubes and the
   SDSS spectrum for those galaxies in common shows a good agreement between
   the derived values from both samples. We find differences in the properties
   of galaxies when comparing a central and an integrated spectra, showing the
   effects of the extracted aperture in the interpretation of the data.
   Finally, we present two dimensional maps of some of the main properties
   derived with the decoupling procedure.}
   {}

   \keywords{surveys -- techniques: spectroscopic -- galaxies:
     abundances -- stars: 
     formation -- galaxies: 
     ISM -- galaxies: stellar content}
   \maketitle
%

\section{Introduction}

Galaxies in the Local Universe are the final product of their cosmological
evolution, and their actual properties are a direct consequence of their
assembling, star formation, enrichment and environmental effects
(interactions/mergers). Since all these processes are imprinted in the spectral
features along their whole extension, a two dimensional analysis can help us to
understand the evolution that galaxies have undergone.  The use of Integral
Field Spectroscopy (IFS) enables us to build two dimensional maps with a
detailed characterisation of their spectroscopic properties, and therefore IFS
is a powerful tool for the study of nearby galaxies.

In this paper, we present an exploratory study of 48 nearby galaxies observed
using the PMAS spectrograph \citep[Postdam Multi Aperture
Spectrograph,][]{PMAS_1} in the PPAK mode \citep{Verheijen2004,PMAS_1}, mounted
at the 3.5~m telescope of the Centro Astron\'{o}mico Hispano Alem\'{a}n (CAHA,
Spain). The wide field-of-view (FOV) of PPAK ($74\arcsec\times 65\arcsec$)
allows us to sample galaxies in the Local Universe covering most of their
optical extent. We have explored two different setups (low and medium spectral
resolution) to finally obtain $\sim 40000$ spectra, covering an area of
$\sim 50~{\rm arcmin}^2$. To exploit at maximum the capabilities of this
instrument, we have observed galaxies in a redshift range of $0.005 < z <
0.025$, in a compromise between the covered area and their sampling (spatial
resolution). This redshift range corresponds to a luminosity distances of
$D_{\rm L} = 19-106$~Mpc, and one arcsec would correspond to a linear scale of
$\sim 92-527$~pc, assuming a standard $\Lambda$CDM cosmology \citep[
$H_0=70.5$, $\Omega=0.27$, $\Lambda=0.73$,][]{Hinshaw2009}.
Thus, this study enables to distinguish structures in the galaxies (arms, bars,
bulge/disk), although smaller interesting regions, as individual
H\,{\footnotesize II} regions, will not be resolved with these observations. 

Previous works have explored the use of different integral field
spectrophotometers for a detailed study of nearby galaxies. In particular, the
SAURON project \citep{SAURON_1, SAURON_2}, and its extension Atlas3D
\citep{Atlas3D} are focused on the analysis of early-type galaxies and bulges
of spirals at $z<0.01$, to study their kinematics and stellar populations. Due
to the distance of their objects ($D_{\rm L} < 42$~Mpc) and the FOV covered by 
the SAURON instrument ($33\arcsec\times41\arcsec$), this
study is mainly restricted to the central part of galaxies. The study of nearby
spiral galaxies has been addressed by the VENGA survey \citep[32 nearby spiral
galaxies,][]{VENGA} using the VIRUS-P spectrograph \citep{VIRUS-P}, the 
DiskMass Survey \citep[146 nearly face-on galaxies]{Bershady2010a} combining PPAK and the 
SparsePak spectrograph \citep{SparsePak2004, SparsePak2005}, and the
PINGS survey \citep[17 nearby disky galaxies,][]{PINGS}, using PPAK, as in this
work. In particular, the mosaicking designed for the PINGS survey allowed to
map the H\,{\footnotesize II} regions along the whole extension their galaxies
and to explore the two-dimensional metallicity structure of disks. Our sample
comprises galaxies of different morphological types in the considered redshift
range, with predominance of spiral galaxies. In addition, they were selected in
size to fit in the FOV of the PPAK instrument, and therefore, to map their
physical properties in their whole extension in just one pointing. The
results obtained from the PING survey can be compared with the work presented in
this article to analyse the distance effects in determination and
interpretation of their physical properties.

The sample selection, observational strategy, data reduction and analysis of
the accuracy of the final data set presented here formed part of the
exploratory studies for the final selection for the optimal instrumental setup
and sample of the CALIFA survey \citep[Calar Alto Legacy Integral Field
spectroscopy Area
survey\footnote{\url{http://www.caha.es/CALIFA/}},][]{CALIFA_SEA,CALIFA_survey}. CALIFA is an
ongoing survey for observing $\sim 600$ nearby galaxies to explore the
mechanisms who drives their evolution along the colour-magnitude diagram by
studying their kinematics, ionised gas and stellar content. As part of this
work, we have performed a wide set of simulations to check the quality of the
data and the recovering of the physical properties of the observed galaxies.
Althoug this analysis has produced a bench mark for the observational strategy and data
reduction scheme for future samples, e.g. in the CALIFA survey,
the data described in this article comprise a well separated data
set, which contains interesting physical information of the considered
galaxies, and therefore we present them as a separated study.

In this paper we focus on the sample (\S~\ref{sec:sample}), observations
(\S~\ref{sec:observations}), and data reduction (\S~\ref{sec:reduction}). The
first analysis and results from these data, including a detailed study of the
accuracy of the derived physical parameters from the simulations, are presented
in \S~\ref{sec:analysis}. Further analysis for different aspects of galaxies,
as the properties of the ionised gas, and particular interesting objects will
be presented in forthcoming papers.


\section{Sample} \label{sec:sample}

\begin{figure}
\centering
\includegraphics[width=9cm]{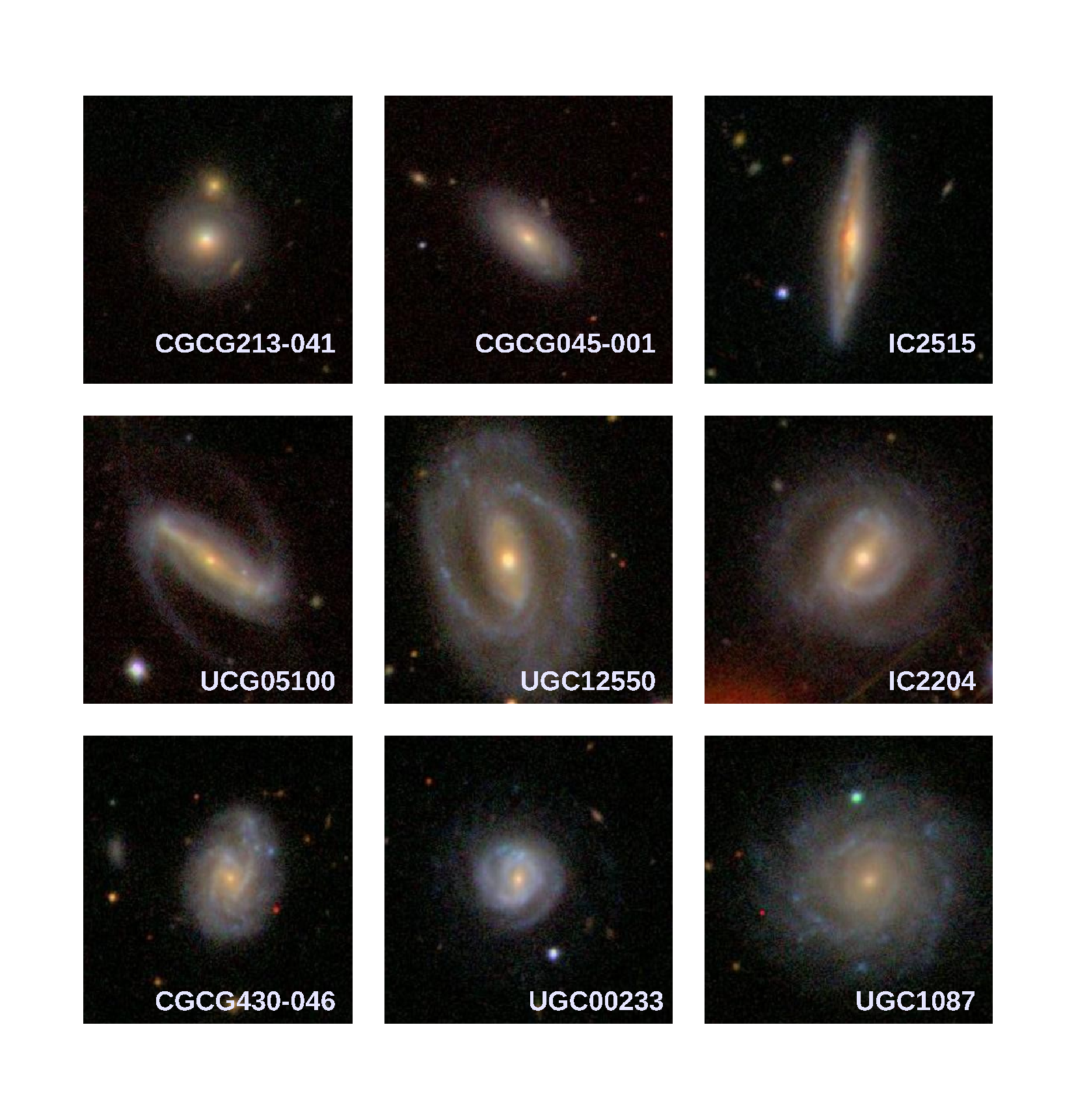}
\caption{Subset of 9 of the observed galaxies, where the difference in
projected sizes can be seen. All images are $1.5\arcmin \times 1.5\arcmin$
(north is up, east to left). Images from SDSS survey.}
\label{example_galaxies}
\end{figure}

Two different sample selections have been explored in this work: (1) galaxies
extracted from the SDSS DR4 imaging sample brighter than $r<15.75$ mag, at a
redshift slice between $0.005 < z < 0.025$ (selection in volume and limiting
magnitude) and (2) face-on disk galaxies included in the DiskMass Survey
\citep[DMS,][Verheijen, private communications]{Bershady2010a} with appropriate sizes to fill the field-of-view
(FOV) of the instrument (angular isophotal diameter selection). The first
criterion is the standard method to select galaxies at a redshift range. In
this case, the angular size of each galaxy is different and, therefore the two
dimensional information available for the sample is limited. On the other hand,
the face-on galaxy sub-sample was selected to solve this problem, but in this
case the sample is not complete and representative of the Local Universe, and
particular care has to be taken when interpreting the results in terms of
global evolution.

In total, a subset of 48 galaxies were finally observed, randomly selected from
those two samples depending on their visibility at the considered nights. Four
galaxies were partially observed and they are not included in this work.
Table~\ref{data_sample} lists the observed galaxies, including some basic
information as the name, coordinates, a basic morphological classification
(when available), redshift, and $V$-band absolute magnitudes. Although there
is a bias towards gas-rich late-type galaxies due to the selection of the
sample, it is an heterogeneous sample that includes bulge-dominated and almost
pure-disk galaxies, face-on and edge-on objects, and galaxies with and without
bars. This is shown in Fig.~\ref{example_galaxies}, where some galaxies
included in this study (images from the SDSS catalogue) are presented.

\begin{figure}[h]
\centering
\includegraphics[width=9.0cm]{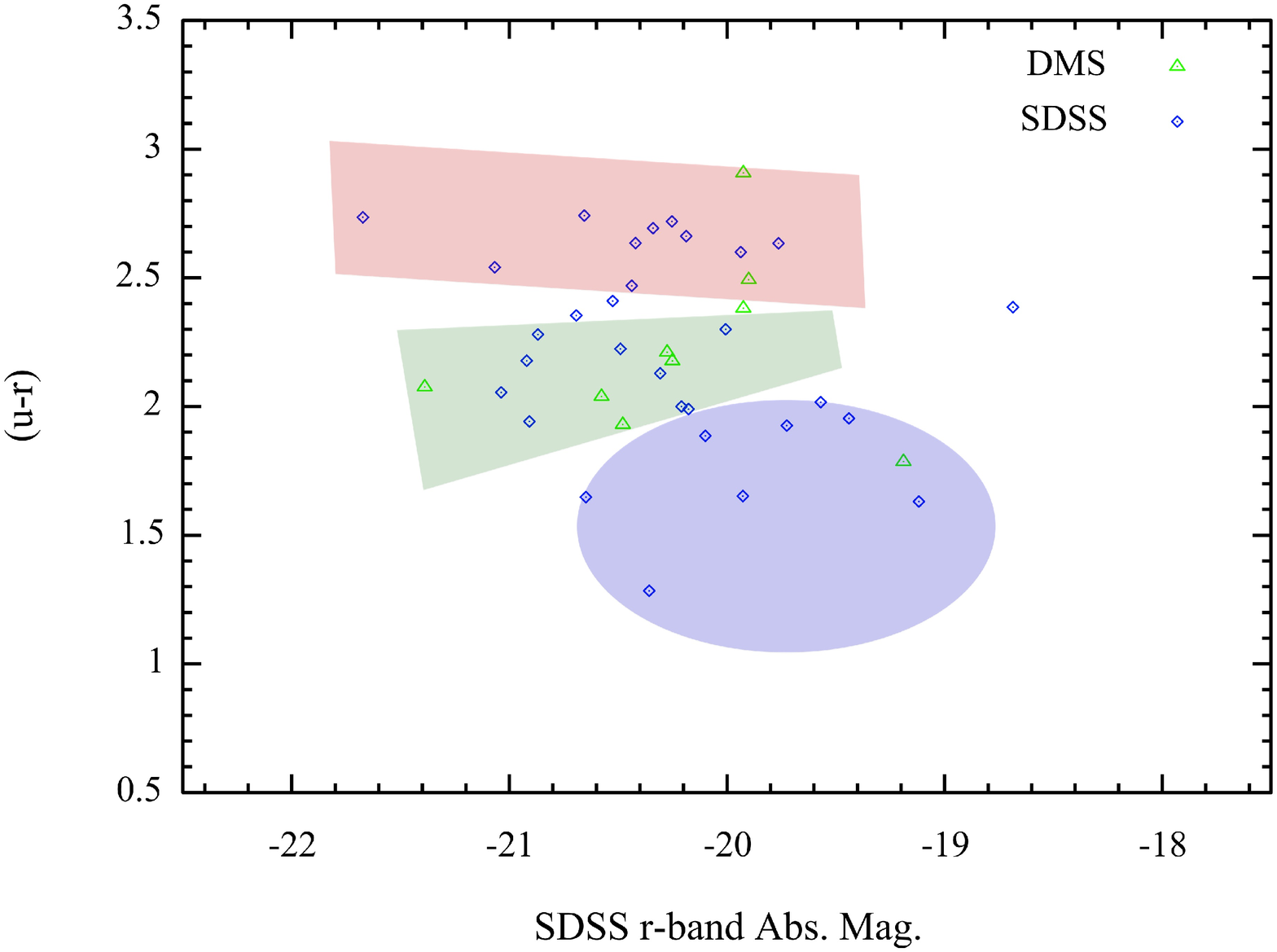}
\caption{u-r vs. M$_r$ colour-magnitude diagram of the observed sample. In blue
circles we have indicated the DR4 selected sample, and in green triangles those
galaxies from the Disk Mass Survey (see text for more detail). The red
sequence, green valley and blue cloud are noted with shaded colour areas. It is
clear the observed sample still covers each region with a significant number of
objects.}
\label{C_M_diag}
\end{figure}

Finally, we present in Figure~\ref{C_M_diag} the colour-magnitude diagram of
the observed objects, indicating those ones coming from each sub-sample.
Regions corresponding to the red sequence, the green valley and the blue cloud
are also indicated \citep[e.g.,][]{SDSS_bimodality, Bell2004, Bell2006,
Faber2007, Chen2010}. Although the sample presented here is small, there is a
significant number observed galaxies in both the red sequence and the blue
cloud, as well as the green valley. Therefore, this sample will allow a
preliminary study of the global evolution of galaxies in the Local Universe.

\begin{table*}
\centering
{\scriptsize
\begin{tabular}{lccccccccccc}
\hline\hline
Galaxy      		&  RA (J2000)                     & DEC (J2000)                      & Type          & $z$    & R$_{25}$    & M$_{\rm V}$  & (B-V) &  $\rm{Observing Date}$ & Grid     & Seeing    & Airmass \\
           		&                                 &                                  &               &        & (mag)       & (mag)        & (mag) &   (yy-mm-dd)           &          & ($^{''}$) & (mag)   \\
  (1)       		&    (2)                          &     (3)                          &  (4)          & (5)    & (6)         & (7)	   & (8)   &  (9)      	       & (10)     & (11)      & (12)    \\
\hline                                                                                                                          
KUG~1033+403                 & $10^{\rm h}36^{\rm m}12.4^{s}$ & $+40^{\rm o}03^{\rm '}12.6^{''}$ & S?            & 0.0229 & 0.56         &      16.5    & 0.87  &  $2009-03-23$ & V300       & \nodata  & 1.12    \\
2MASX~J13062093+5318232~\dag & $13^{\rm h}06^{\rm m}21.0^{s}$ & $+53^{\rm o}18^{\rm '}23.2^{''}$ & AGN           & 0.0237 & \nodata      &      15.9    & 1.25  &  $2009-03-22$ & V300       & 2.0      & 1.11    \\
2MASX~J13193595+5330102~\dag & $13^{\rm h}19^{\rm m}35.9^{s}$ & $+53^{\rm o}30^{\rm '}09.8^{''}$ & IrS           & 0.0248 & \nodata      &      12.4    & 0.72  &  $2009-03-24$ & V300       & 2.0      & 1.13    \\
CGCG~045-001 	             & $13^{\rm h}30^{\rm m}43.5^{s}$ & $+07^{\rm o}31^{\rm '}48.8^{''}$ & \nodata       & 0.0220 & \nodata      &      15.4    & 1.07  &  $2009-03-20$ & V300       & 1.3      & 1.19    \\
CGCG~071-096~\dag            & $13^{\rm h}00^{\rm m}33.2^{s}$ & $+10^{\rm o}07^{\rm '}47.8^{''}$ & \nodata       & 0.0239 & \nodata      &      15.1    & 0.79  &  $2009-03-20$ & V300       & 1.1      & 1.25    \\
CGCG~148-006~\dag            & $07^{\rm h}44^{\rm m}57.4^{s}$ & $+28^{\rm o}55^{\rm '}39.0^{''}$ & \nodata       & 0.0234 & \nodata      &      15.3    & 0.75  &  $2009-10-23$ & V600       & 1.1      & 1.30    \\
CGCG~181-068~\dag 	     & $09^{\rm h}37^{\rm m}19.2^{s}$ & $+33^{\rm o}49^{\rm '}25.8^{''}$ & \nodata       & 0.0227 & \nodata      &      15.7    & 1.17  &  $2009-03-20$ & V300       & 1.2      & 1.06    \\
CGCG~213-041~\dag            & $11^{\rm h}21^{\rm m}16.8^{s}$ & $+40^{\rm o}20^{\rm '}43.4^{''}$ & \nodata       & 0.0208 & \nodata      &      15.5    & 1.16  &  $2009-03-21$ & V300       & 1.7      & 1.04    \\
CGCG~293-023~\dag            & $02^{\rm h}30^{\rm m}21.5^{s}$ & $+56^{\rm o}47^{\rm '}29.5^{''}$ & \nodata       & 0.0156 & \nodata      &      15.6    & 0.99  &  $2009-03-23$ & V300       & \nodata  & 1.18    \\
CGCG~428-059~\dag 	     & $22^{\rm h}14^{\rm m}51.4^{s}$ & $+13^{\rm o}42^{\rm '}54.9^{''}$ & \nodata       & 0.0242 & \nodata      &      15.7    & 0.90  &  $2009-06-27$ & V300       & 1.0      & 1.28    \\
CGCG~428-060 		     & $22^{\rm h}14^{\rm m}57.9^{s}$ & $+13^{\rm o}45^{\rm '}42.9^{''}$ & \nodata       & 0.0242 & \nodata      &      13.7    & 0.92  &  $2009-06-28$ & V300       & 1.2      & 1.21    \\
CGCG~430-046~\dag 	     & $23^{\rm h}00^{\rm m}46.2^{s}$ & $+13^{\rm o}37^{\rm '}07.9^{''}$ & \nodata       & 0.0243 & \nodata      &      15.3    & 0.86  &  $2009-10-23$ & V600       & 1.1      & 1.10    \\
IC~2204~\dag      	     & $07^{\rm h}41^{\rm m}18.1^{s}$ & $+34^{\rm o}13^{\rm '}55.9^{''}$ & (R)SB(r)ab    & 0.0155 & 0.00         &      14.6    & 0.93  &  $2009-10-22$ & V600       & 1.5      & 1.12    \\
IC~2500~\dag     	     & $09^{\rm h}42^{\rm m}23.4^{s}$ & $+36^{\rm o}20^{\rm '}59.1^{''}$ & \nodata       & 0.0221 & \nodata      &      15.2    & 1.17  &  $2009-03-21$ & V300       & 1.7      & 1.14    \\
IC~2515~\dag      	     & $09^{\rm h}54^{\rm m}39.4^{s}$ & $+37^{\rm o}24^{\rm '}30.8^{''}$ & Sb            & 0.0193 & 0.72         &      15.1    & 1.11  &  $2009-03-22$ & V300       & 1.6      & 1.05    \\
MRK~1477~\dag 		     & $13^{\rm h}16^{\rm m}14.7^{s}$ & $+41^{\rm o}29^{\rm '}40.1^{''}$ & \nodata       & 0.0207 & \nodata      &      15.4    & 0.47  &  $2009-03-24$ & V300       & 1.8      & 1.04    \\
NGC~99~\dag,$\star$          & $00^{\rm h}23^{\rm m}59.5^{s}$ & $+15^{\rm o}46^{\rm '}13.7^{''}$ & Scd           & 0.0177 & 0.04         &      14.0    & 0.62  &  $2009-10-22$ & V600       & 2.3      & 1.11    \\
NGC~3820~\dag                & $11^{\rm h}42^{\rm m}04.9^{s}$ & $+10^{\rm o}23^{\rm '}03.3^{''}$ & S?            & 0.0203 & 0.17         &      14.9    & 0.95  &  $2009-03-22$ & V300       & 1.2      & 1.40    \\
NGC~4109~\dag,$\star$        & $12^{\rm h}06^{\rm m}51.1^{s}$ & $+42^{\rm o}59^{\rm '}44.3^{''}$ & Sa?           & 0.0235 & 0.05         &      14.8    & 0.98  &  $2009-03-19$ & V300       & \nodata  & 1.05    \\
NGC~7570~\dag,$\star$        & $23^{\rm h}16^{\rm m}44.7^{s}$ & $+13^{\rm o}28^{\rm '}58.8^{''}$ & SBa           & 0.0157 & 0.23         &      13.9    & 0.81  &  $2009-10-22$ & V600       & 2.0      & 1.13    \\
UGC~74 			     & $00^{\rm h}08^{\rm m}44.7^{s}$ & $+04^{\rm o}36^{\rm '}45.1^{''}$ & SAB(rs)c?     & 0.0131 & 0.05         &      14.5    & 0.85  &  $2009-10-17$ & V600       & 1.3      & 1.23    \\
UGC~233    		     & $00^{\rm h}24^{\rm m}42.7^{s}$ & $+14^{\rm o}49^{\rm '}28.8^{''}$ & SBbc D        & 0.0176 & 0.00         &      14.6    & 0.62  &  $2009-10-23$ & V600       & 1.1      & 1.20    \\
UGC~448 		     & $00^{\rm h}42^{\rm m}22.0^{s}$ & $+29^{\rm o}38^{\rm '}29.9^{''}$ & SABc          & 0.0162 & 0.04         &      14.4    & 1.16  &  $2009-10-19$ & V600       & 2.0      & 1.08    \\ 
UGC~463~\dag 		     & $00^{\rm h}43^{\rm m}32.4^{s}$ & $+14^{\rm o}20^{\rm '}33.2^{''}$ & SAB(rs)c      & 0.0148 & 0.00         &      13.5    & 0.85  &  $2009-10-18$ & V600       & 1.8      & 1.29    \\
UGC~1081 	             & $01^{\rm h}30^{\rm m}46.6^{s}$ & $+21^{\rm o}26^{\rm '}25.5^{''}$ & SB(rs)c       & 0.0104 & 0.03         &      13.8    & 0.83  &  $2009-10-22$ & V600       & 2.4      & 1.15    \\ 
UGC~1087~\dag 		     & $01^{\rm h}31^{\rm m}26.6^{s}$ & $+14^{\rm o}16^{\rm '}39.0^{''}$ & SA(rs)c       & 0.0149 & 0.03         &      15.1    & 0.69  &  $2009-10-17$ & V600       & 1.0      & 1.13    \\
UGC~1529 		     & $02^{\rm h}02^{\rm m}31.0^{s}$ & $+11^{\rm o}05^{\rm '}35.1^{''}$ & SA(rs)c       & 0.0155 & 0.07         &      14.7    & 1.01  &  $2009-10-18$ & V600       & 1.5      & 1.17    \\
UGC~1635 		     & $02^{\rm h}08^{\rm m}27.7^{s}$ & $+06^{\rm o}23^{\rm '}41.7^{''}$ & SAbc          & 0.0115 & 0.01         &      14.8    & 0.96  &  $2009-10-23$ & V600       & 1.2      & 1.10    \\
UGC~1862 		     & $02^{\rm h}24^{\rm m}24.8^{s}$ & $-02^{\rm o}09^{\rm '}44.5^{''}$ & SAB(rs)d pec? & 0.0046 & 0.11         &      14.0    & 0.83  &  $2009-10-23$ & V600       & 1.3      & 1.20    \\
UGC~3091 		     & $04^{\rm h}33^{\rm m}56.1^{s}$ & $+01^{\rm o}06^{\rm '}49.5^{''}$ & SAB(s)d?      & 0.0184 & \nodata      &      15.4    & 0.88  &  $2009-10-17$ & V600       & 0.9      & 1.40    \\
UGC~3140 		     & $04^{\rm h}42^{\rm m}54.9^{s}$ & $+00^{\rm o}37^{\rm '}06.9^{''}$ & SA(rs)c?      & 0.0154 & 0.06         &      13.6    & 0.97  &  $2009-10-19$ & V600       & 1.8      & 1.36    \\
UGC~3701 		     & $07^{\rm h}11^{\rm m}42.7^{s}$ & $+72^{\rm o}10^{\rm '}09.5^{''}$ & SA(rs)cd?     & 0.0097 & 0.00         &      15.2    & 0.87  &  $2009-10-17$ & V600       & 1.5      & 1.30    \\
UGC~3997~$\star$  	     & $07^{\rm h}44^{\rm m}38.7^{s}$ & $+40^{\rm o}21^{\rm '}58.5^{''}$ & Im?           & 0.0197 & 0.06         &      16.0    & 1.07  &  $2009-10-18$ & V600       & 1.5      & 1.13    \\
UGC~4036 		     & $07^{\rm h}51^{\rm m}54.7^{s}$ & $+73^{\rm o}00^{\rm '}56.5^{''}$ & SAB(r)b?      & 0.0116 & 0.06         &      12.7    & 0.98  &  $2009-10-19$ & V600       & 1.6      & 1.38    \\
UGC~4107 		     & $07^{\rm h}57^{\rm m}01.9^{s}$ & $+49^{\rm o}34^{\rm '}02.5^{''}$ & SA(rs)c       & 0.0117 & 0.01         &      13.9    & 1.06  &  $2009-10-22$ & V600       & 1.2      & 1.13    \\
UGC~5100~\dag    	     & $09^{\rm h}34^{\rm m}38.6^{s}$ & $+05^{\rm o}50^{\rm '}29.9^{''}$ & SB(s)b        & 0.0184 & 0.22         &      14.9    & 1.02  &  $2009-03-19$ & V300       & 1.1      & 1.23    \\
UGC~6156~\dag    	     & $11^{\rm h}06^{\rm m}31.3^{s}$ & $+48^{\rm o}39^{\rm '}05.7^{''}$ & S?            & 0.0245 & 0.56         &      15.1    & 0.54  &  $2009-03-24$ & V300       & 1.5      & 1.11    \\
UGC~6410~\dag                & $11^{\rm h}24^{\rm m}05.9^{s}$ & $+45^{\rm o}48^{\rm '}39.9^{''}$ & SABc          & 0.0187 & 0.11         &      14.8    & 0.78  &  $2009-03-20$ & V300       & 1.5      & 1.08    \\
UGC~7993~\dag                & $12^{\rm h}50^{\rm m}31.6^{s}$ & $+52^{\rm o}07^{\rm '}22.5^{''}$ & Scd?          & 0.0161 & 0.90         &      15.3    & 0.79  &  $2009-03-19$ & V300       & 1.3      & 1.25    \\
UGC~9837~\dag 		     & $15^{\rm h}23^{\rm m}51.7^{s}$ & $+58^{\rm o}03^{\rm '}10.6^{''}$ & SAB(s)c       & 0.0089 & 0.02         &      14.6    & 0.57  &  $2007-06-23$ & V300       & \nodata  & 1.10    \\
UGC~9965 		     & $15^{\rm h}40^{\rm m}06.7^{s}$ & $+20^{\rm o}40^{\rm '}50.2^{''}$ & SA(rs)c       & 0.0151 & 0.06         &      14.4    & 0.72  &  $2007-06-22$ & V300       & \nodata  & 1.39    \\
UGC~11318 	             & $18^{\rm h}39^{\rm m}12.2^{s}$ & $+55^{\rm o}38^{\rm '}30.5^{''}$ & SB(rs)bc      & 0.0196 & 0.03         &      14.1    & 0.72  &  $2007-06-23$ & V300       & \nodata  & 1.10    \\
UGC~12250~\dag 		     & $22^{\rm h}55^{\rm m}35.9^{s}$ & $+12^{\rm o}47^{\rm '}25.1^{''}$ & SBb           & 0.0242 & 0.22         &      14.1    & 1.04  &  $2009-10-18$ & V600       & 1.1      & 1.33    \\
UGC~12391 		     & $23^{\rm h}08^{\rm m}57.2^{s}$ & $+12^{\rm o}02^{\rm '}52.9^{''}$ & SAB(s)c       & 0.0163 & 0.04         &      14.7    & 0.77  &  $2009-10-17$ & V600       & 1.5      & 1.20    \\
\hline\hline
\end{tabular}
} 
\begin{minipage}{18.0cm}
\caption[General information about the observed galaxies]{\footnotesize General
information about the observed galaxies. 
All the galaxies are included in 2MASS\footnote{Two Micron All Sky Survey \citep{2MASS}: 
\url{http://www.ipac.caltech.edu/2mass/}}. Galaxies marked with \dag have available SDSS spectroscopy. Galaxies marked with $\star$
have been observed with GALEX\footnote{The GALEX Ultraviolet Atlas of Nearby Galaxies \citep{GALEX}}.

(1) Galaxy identification; 
Coordinates: (2) Right ascension and (3) Declination in J2000 Equinox from NED\footnote{NASA/IPAC Extragalactic Database: \url{http://nedwww.ipac.caltech.edu/}}; 
(4) Morphological type from RC3 catalogue\footnote{Third Reference Catalogue of Bright Galaxies: \url{http://vizier.u-strasbg.fr/viz-bin/VizieR?-source=VII/155}};
(5) Redshift from NED;
(6) $R_{25}$: Radius where the surface brightness reaches $\mu_B =
25.0~{\rm mag}/{\rm arcsec}^2$, obtained from RC3 catalogue; 
(7) M$_{\rm V}$: Absolute B-band magnitude from NED;
(8) (B-V);
(9) Observing date;
(10) PPAK mode utilised for each observation (see Table~\ref{setups}); 
(11) Average seeing value during the corresponding night observation. Data from the Calar Alto Observatory archive.
(12) Airmass for the each observation.}
\label{data_sample}
\end{minipage}
\end{table*}

\section{Observations} \label{sec:observations}
\begin{table}
{\scriptsize
\caption{Atmospheric conditions during the observations. For each
night, we present average values of the seeing, humidity, transparency 
of the atmosphere, and the extinction in the $V-$band.}
\label{weather}
\begin{tabular}{llcccc}
\hline\hline
Observing       & &  Seeing  & Humidity & Tranparency & Extinction  \\
date            & & ($^{''}$) &  (\%) & (\%) & in $V-$band (mag)  \\
\hline
2007-06-22\upstar      & Clear  &    --   &  56  &  95  & 0.22  \\ 
2007-06-23\upstar      & Clear  &    --   &  48  &  95  & 0.22  \\ 
2009-03-19      & Clear  &    1.2  &  43  &  95  & 0.15  \\ 
2009-03-20      & Clear  &    1.3  &  62  &  85  & 0.25  \\ 
2009-03-21      & Clouds &    1.7  &  71  &  75  & 0.28  \\ 
2009-03-22      & Clear  &    2.0  &  70  &  95  & 0.15  \\ 
2009-03-23\upstar      & Clouds &    --   &  47  &  75  & 0.25  \\ 
2009-03-24      & Clouds &    1.8  &  62  &  65  & 0.25  \\ 
2009-06-27      & Clear  &    1.0  &  44  &  95  & 0.20  \\ 
2009-06-28      & Clear  &    1.2  &  70  &  90  & 0.20  \\ 
2009-10-17      & Clouds &    1.2  &  79  &  85  & 0.25  \\ 
2009-10-18      & Clouds &    1.5  &  58  &  70  & 0.30  \\ 
2009-10-19      & Clouds &    1.8  &  75  &  75  & 0.20  \\ 
2009-10-22      & Clear  &    1.6  &  86  &  90  & 0.20  \\ 
2009-10-23      & Clear  &    1.2  &  50  &  95  & 0.15  \\ 
\hline
\end{tabular}
\begin{minipage}{9cm}
{\scriptsize\upstar Due to technical problems, there are not data from the seeing monitor for these nights.}
\end{minipage}
}
\end{table}

Observations were carried out during 15 nights in several observing runs. 
Information related to the weather condition during the observations are 
presented in Table~\ref{weather}. The full sample was observed
at the 3.5~m telescope of the Calar Alto
observatory with PMAS spectrograph \citep[Postdam Multi Aperture
Spectrograph,][]{PMAS_1} in the PPAK mode \citep{Verheijen2004,PMAS_1}. The
PPAK fibre bundle consists of 382 fibres of $2.7\arcsec$ diameter each
\citep[see Fig.~5 in][]{PMAS_1}. 
The science fibres (331 fibres) are concentrated in a single hexagonal bundle, with a filling factor of
$\sim 65$~\%. The sky background is sampled by 36 additional fibres,
distributed in 6 bundles of 6 fibres each, distributed along a circle
$\sim 90\arcsec$ from the centre of the instrument FOV. The sky-fibres are
distributed among the science fibres within the pseudo-slit in order to have a
good characterisation of the sky; the remaining 15 fibres are used for
calibration purposes. 

\begin{table}
\caption{PMAS/PPAK instrumental setups.}
\label{setups}
\begin{tabular}{l|cc}
\hline\hline
Grid type                         & V300                  & V600 \\
\hline
Spectral coverage                 & 3620--7056 \AA\AA     & 3845--7014 \AA\AA \\
Dispersion                        & $3.2$ \AA /px         & $1.5$ \AA /px \\
FWHM \tablefootmark{a}            & 10.7 \AA              & 5.4 \AA \\
Exposure time \tablefootmark{b} & $600~{\rm s}\times 3$ & $600~{\rm s}\times 3$ \\
\hline
\end{tabular}
\tablefoot{ 
\tablefoottext{a} Spectral resolution (FWHM) for PPAK IFU with $150~\mu{\rm m}$ fibers. 
\tablefoottext{b} Exposure time for each dithering pointing.}
\end{table}

A dithering scheme with three pointings was adopted during the different
observing runs, in order to cover the complete FOV of the bundle and to
increase the spatial resolution of the data. This scheme has been already
adopted in previous studies using PPAK, with a considerable increase in the
quality of the data \citep[e.g.,][]{Sanchez2007c, Castillo-Morales2010,
Perez-Gallego2010,PINGS}. The offsets in RA-dec of the different pointings,
with respect to the nominal coordinates of the targets, were (in arcsec): (1)
$(0,0)$,(2) $(+1.56,+0.78)$ and (3) $+(1.56,-0.78)$, which allows to cover the
holes between fibers in the central bundle. The spatial re-composition of the
three pointings is included in the standard data reduction scheme adopted for
these data (described in \S~\ref{sec:reduction}). 

\begin{figure}
\centering
\includegraphics[angle=-90,width=0.90\columnwidth]{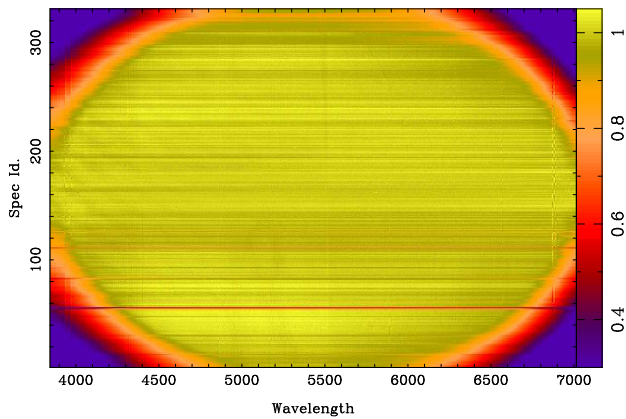}
\caption{Intensity map of the differential transmission fiber-to-fiber for the
V600 grating. The drop of the transmission at the edges of the image is due to
the vignetting. The $\sim 15\%$ of the CCD is affected this effect.}
\label{vigneting}
\end{figure}

Two different setups were used during these observations, listed in
Table~\ref{setups}. For the runs before October 2010, it was used the V300
grating, with a nominal spectral resolution of $\sim 10.7~\AA$ (Full Width at
Half Maximun --FWHM) and a wavelength range of $\lambda\lambda3620-7056~\AA\AA$.
All the targets observed during this run correspond to the first sub-sample
described in the previous section. After that date, the V600 grating was used,
with a nominal resolution of $\sim 5.4~\AA$ (FWHM) and a wavelength range of
$\lambda\lambda3845-7014~\AA\AA$. Most of the targets observed during this run
correspond to the DMS sub-sample. The main reason for that change was the
upgrade of the PMAS CCD, from a 2k~$\times$~4k chip to a 4k~$\times$~4k chip,
that nominally increases the wavelength range by a factor two at the same
spectral resolution. This increase was hampered by a vignetting effect that
affects to a $\sim 30$~\% of the fibers/spectra, at the edges of the CCD (see
Fig.~\ref{vigneting}). Fibers positioning in the entrance slit are in such a way that the central
and outer regions of the hexagon are located in the central regions of the
slit, while the outer regions of the slit corresponds to an annular ring.
Due to this positioning of the fibers over the detector for
this instrument, the vignetted regions correspond spatially to an annulus 
ring at about $\sim 15 \arcsec$ from the centre of the IFU. 
Despite this effect, all the fibers present unvignetting more than $\sim 50$~\% of their spectral
range. 


\section{Data reduction} \label{sec:reduction}

The large amount of data generated in IFS observations made that the
establishment of a precise and quick procedure for the data reduction was one
of the key points to test during this study. The reduction was performed
using a fully automatic pipeline, that operates without human intervention,
producing both the scientific useful frames and a set of quality control
measurements (on a web-based interface). The pipeline was based on the routines
included in the {\sc R3D} \citep{R3D} and {\sc E3D} \citep{E3D} packages. The
reduction consists of the standard steps for fibre-based integral-field
spectroscopy. For the V300 data, a master bias frame was created by averaging
all the bias frames observed during the night and subtracted from the science
frames. In the case of the V600, observed with the new PMAS CCD, which has
basically no structure in the bias, a single value bias-level (derived on the
basis of the over-scan values), is subtracted to the frames. The different
exposures taken at the same position on the sky were then combined, clipping
the cosmic rays using IRAF\footnote{IRAF is distributed by the National Optical
Astronomy Observatories, which are operated by the Association of Universities
for Research in Astronomy, Inc., under cooperative agreement with the National
Science Foundation.} routines. The next step was to determine the locations of
the spectra on the CCD, using in this case a continuum illuminated exposure
taken before the science exposures. Each spectrum was then extracted from the
science frames. 

In order to reduce the effects of the cross-talk, we adopted a modified
version of the Gaussian-suppression proposed by \citet{R3D}, and fully
described in \citet{Sanchez2011}. This technique assumes a Gaussian profile for
the projection of each fibre spectrum along the cross-dispersion axis, and it
performs a Gaussian fitting to each of the fibres after subtracting the
contribution of the adjacent fibres in an iterative process. The cross-talk is
reduced to less than a 1~\% when adopting this method.

Each spectrum extracted along the dispersion direction is stored in a
row-stacked-spectrum file \citep[RSS,][]{E3D}. Wavelength calibration was
performed using HeHgCd lamp exposures obtained before and after each pointing,
yielding an accuracy of $\sim 10 \%$ of the nominal pixel scale (i.e.,
$\sim 0.3~\AA$ for the V300 grating and $\sim 0.15~\AA$ for the V600 one).
Differences in the relative fibre-to-fibre transmission throughput were
corrected by comparing the wavelength-calibrated RSS science frames with the
corresponding frames derived from sky exposures taken during the twilight. 

As mentioned before, PPAK is equipped with 36 fibres to sample the sky,
distributed around the science fiber-bundle, in six small bundles of six fibers
each one, at a distance of $\sim 75~\arcsec$ from the centre of the FOV
\citep[seen Fig.~5 in][]{PMAS_2}. The objects selected to observe in these
studies cover a substantial fraction of the FOV of the central PPAK bundle, but
most of the sky fibers are free from emission by the targets. The procedure
adopted to derive the night-sky spectrum was to combine the spectra
corresponding to these fibers, performing a 2$\sigma$ clipping rejection to
remove any possible contamination\footnote{We adopted this criterion since
a 3$\sigma$ clipping rejection does not remove the contamination by stars in
the field, external regions of the galaxies, or a companion galaxy in IFU
observations, due to the positions of the fibers.}. Finally, the sky-spectrum
was subtracted to all the spectra the corresponding frame. 

Spectrophotometric standard stars from \citet{Oke1990} were observed during the
nights to perform the flux calibration. The extracted spectra of these stars
were compared with the flux calibrated spectra available in the
webpage of the
observatory\footnote{\url{http://www.caha.es/pedraz/SSS/Oke/oke.html}}, 
obtaining the transformation function from 
observed counts to intensity. In this step, the routines in R3D package 
take into account the airmass and extinction of the observation of both the
spectrophotometric standard stars and the science targets. This procedure
ensures a good relative flux calibration from the blue to the red part of the
spectra, if the weather conditions along the night were stable. However, an
absolute offset between the derived and real fluxes are expected due to the
incomplete spatial coverage of the FOV in the observations of the calibration
star despite the large fiber size of PPAK ($2.7\arcsec$), and second
order effects like small inaccuracies in the centering of the star and the
effects of the variation of the seeing.

After reducing the frames corresponding to each particular pointing, the
science spectra corresponding to the three dithered exposures were combined in
a single frame of 993 spectra. In order to take into account possible
variations in the atmosphere transmission during the exposures, the procedure
rescales the spectra to a common intensity by comparing the integrated spectra
within an aperture of $20 \arcsec$/diameter, chosen in a compromise
between the signal-to-noise and the depth of the images, and to avoid seeing
effects. The final position tables for the individual fibers are 
generated taking into account their relative positions to the PPAK central 
bundle and the offsets provided for the dither scheme. The information included 
in the combined spectra and position tables are used to create the final 
datacube.

Next step in the reduction process is to perform the 
differential atmospheric refraction (DAR) correction. 
To perform this correction, the pipeline creates an 2D image by co-adding the
flux within $4500-5500\AA\AA$ (approximately the V-band), and looks for
the peak emission within a region between [15:62,15:62] pixels. This centroid 
is used as initial reference, that will look for any shift in this centroid 
along the wavelength within a box of $6\times6$ pixels. Once the shift is 
determined, a polynomial order fitting of order 3 is applied to smooth the 
correction, and the shift is applied wavelength to wavelength, re-centering 
the cube and correcting for the DAR.

In order to get the best possible absolute flux calibration, we re-calibrated
our data using SDSS photometry, whenever is available (37 targets, 33 of them
with quality enough for the analysis). Of the five SDSS filters ($ugriz$), our
spectrum covers the passbands of two, namely the $g$ ($\lambda_{\rm
eff}=4770~\AA$) and $r$ ($\lambda_{\rm eff} = 6231~\AA$) filters. We measured
the counts of each galaxy in these SDSS images inside a 30$\arcsec$ diameter
aperture. These counts were converted to flux following the counts-to-magnitude
prescription in SDSS documentation\footnote{
\url{http://www.sdss.org/dr7/algorithms/fluxcal.html\#counts2mag}}. We
crosschecked this method measuring magnitudes of stars on each field and
comparing them with SDSS DR6 photometric catalogue, obtaining less than 0.05
magnitude dispersion for $g < 17.5$~mag.

We then extracted the spectrophotometry from our reduced cubes, summing the
flux of individual spectra inside a $30 \arcsec$ diameter aperture  around the
peak intensity of the object, and convolving this spectrum with the SDSS $g$
and $r$ filter passbands \citep{calibracion_SDSS}. Using these two data pairs,
a scaling solution of the form: 
\begin{equation} 
F_{\rm corr}(\lambda) =F_0(\lambda) \times a 
\end{equation}
was derived for each band, where $F_{\rm corr}(\lambda)$ and $F_0(\lambda)$ are
the re-calibrated and original fluxes, respectively, and $a$ is the calibration
constant. The {\it root-mean-square} ($rms$) for these values  
is $\sim 4\%$ in the $g$ band and $\sim 5\%$. Finally, the 
average of the two scaling constants derived for the $g$ and $r$ filters is 
adopted. In this way we were able to recalibrate the spectrum to
match the SDSS photometry. 

\begin{figure}
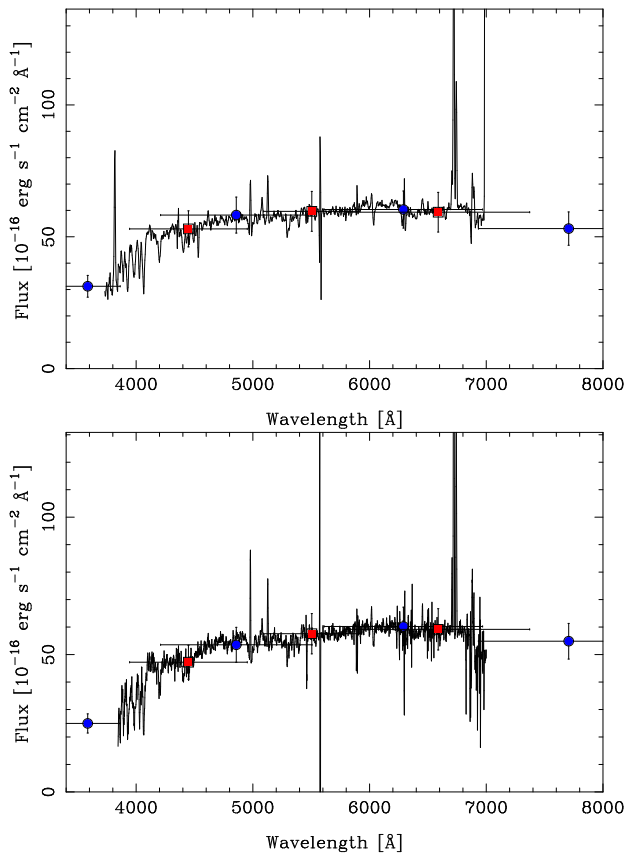

\centering
 \includegraphics[angle=-90,width=0.90\columnwidth]{new_plots/comp_phot_CGC071-096.ps}
\includegraphics[angle=-90,width=0.90\columnwidth]{new_plots/comp_phot_CGCG430-046.ps}
\caption{Examples of the flux re-calibration based on SDSS photometry for
CGC071-96 (observed with the V300 grating, {\it top panel}) and CGCG430-046
(V600 grating, {\it bottom panel}). The values for the SDSS $g$ and $r$ filters
used in the re-calibration are presented with blue solid circles, as well the
derived values for SDSS $u$ and $i$ filters. Johnson $B,V$ and Cousin $R_{\rm
C}$ magnitudes are also shown (red solid squares) to test the flux
calibration.}

\label{phot_comparison_SDSS}
\end{figure}

Figure~\ref{phot_comparison_SDSS} shows two
examples for the resulting re-calibrated spectra observed with the V300 and
V600 grating. Together with the values for the $g$ and $r$ filters used in
this re-calibration, we present the derived values for SDSS $u$ and $i$ bands.
In addition to these measurements, the figure includes the fluxes
corresponding to the Johnson $B, V$ and Cousin $R_C$ magnitudes determined on
the basis of the SDSS photometry and the transformation equations presented by
\citet{calibracion_SDSS} \footnote{Summary of the transformations in
\url{http://www.sdss.org/dr7/algorithms/sdssUBVRITransform.html}.}. There is a
good agreement in all cases suggesting that this re-calibration method is
reliable.

\begin{figure}
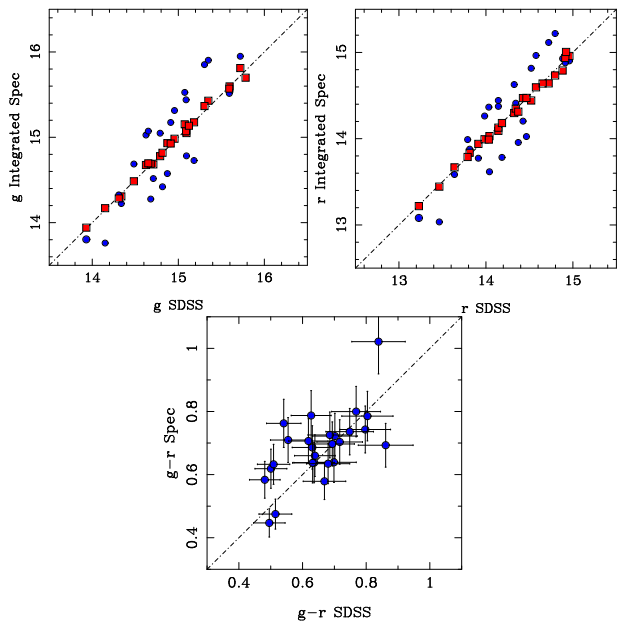

\centering
\includegraphics[width=0.45\columnwidth, angle = -90]{comp_g_SDSS_spec.ps}
\includegraphics[width=0.45\columnwidth, angle = -90]{comp_r_SDSS_spec.ps}
\includegraphics[width=0.45\columnwidth, angle = -90]{color_effect.ps}
\caption{Top: Comparison of $g$ (right panel) and $r$-band (left panel)
magnitudes derived from the integrated spectrum within an aperture of
$30\arcsec$/diameter extracted from the V600 datacubes and the corresponding
magnitudes derived from the SDSS images, before (blue solid circles) and after
(red solid squares) the flux calibration described in the text. Bottom:
Comparison of $g-r$ colour. Considering the errors in the absolute flux
calibration ($\sim 8$~\%) and between $g$ and $r$ band (around $4-5$~\%), the
expected errors in the colours are $\sim 0.1$~mag, consistent with the errors in
SDSS photometric data.}
\label{color_effects}
\end{figure}

Figure~\ref{color_effects} (top panels) shows the $g$ and $r$-band magnitudes
derived from the datacubes and the SDSS images before (blue solid circles) and
after (red solid squares) the recalibration process. Both photometric data
match for the 85\% of the galaxies within a range of $\pm 0.15$~mag before the
recalibration (just 4 objects present a higher dispersion below $ \pm
0.25$~mag). The possible differences in colours is explored in
Fig.~\ref{color_effects}, bottom panel. Any possible error in the relative
flux calibration from blue to red would produce a change in colour. The
dispersion (rms) around the one-to-one relation is $\sim 0.1$~mag, consistent with the expected errors from propagating
both the SDSS photometric errors ($\sim 4-5$~\%) and the absolute flux
calibration accuracy of our data ($\sim 8$~\%).

In the few cases (11 targets) when the SDSS photometry is not available for a
galaxy, we obtained broad-band images using CAFOS at the 2.2~m telescope of the
Calar Alto Observatory, Spain. The observations were carried out for 6 nights
from January 28th to March 30th, 2010, when images with the Johnson $B$, $V$,
$I$, and Cousin $R_{\rm C}$ filters were obtained. The CCD camera was equipped
with a SITe pixel chip. The pixel size was $24~\mu$m, corresponding to
$0.53\arcsec$. Sky flats were exposed in every dawn and dusk phase. Bias
subtraction, flat-field division, and trimming were performed in the usual
manner, using the IRAF package. 

We photometrically calibrated these images using GSC2.3 \citep{GSC2.3} and
USNO-B \citep{USNO-B} catalogues as it follows. We obtained $B$ and $V$
photometry for our galaxies using the {\tt qphot} routine of the IRAF and
adjusted the zero-point (ZP) magnitude to the GSC2.3 and USNO-B photometry. For
these ZP calibration we measured the magnitudes of several stars (7-8) in the
same field of our galaxies and then we determined the difference between our
magnitudes and the catalogued ones. In general, the dispersion obtained for the
ZP determined using the USNO-B catalogue was nearly 0.4~mag on average, so we
decided to use the GSC2.3 ZP with a dispersion of 0.2~mag on average. One ZP
was calculated for each image with this method. Following the same procedure
described before, we then convolved a similar aperture spectra extracted from
the datacubes with the $B$ and $V$ filter responses, obtaining another pair of
data sets for recalibrating these cubes. Considering that the stars from
the GSC2.3 catalogue used in this work presented errors of $0.1-0.4$~mag, we
expect that our recalibration procedure ensures an absolute spectrophotometric
calibration better than a $\sim 15$~\% in all the spectral range.

Finally, the pipeline performs a set of quality control measurements, that are
accessible to the user on a web-based interface, as indicated above. Among these
measurements the most relevant are (i) cross-checks of the focus accuracy of
the instrument, (ii) independent estimations of the accuracy of the wavelength
calibration (by comparing the nominal wavelength of the most prominent 
emission lines included in the 
sampled wavelength range with the measured ones on the data, for each 
individual spectra within the reduced RSS), 
(iii) estimations of accuracy of the
sky-subtraction (from the residual flux of the strongest and more variable night-sky emission 
line in the considered wavelength range -OI~$5577~\AA$),
(iv) estimations of the night-sky brightness at the moment of
the observation, (v) estimations of the depth of the final datacubes (by computing the 
signal-to-noise ratio in the spectral range $5670-5730~\AA\AA$). For the
observations presented in this paper, the average $5\sigma$ detection limit is
$\sim 22.5-23.0$~mag/arcsec$^2$ ($V-$band) and the accuracy of the wavelength
calibration is $\sim 0.3$~\AA. In Table~\ref{data_pipeline} we present these values, together
the $V-$band magnitude of the sky, for each observing night. On the basis of these quality control
procedures, it is possible to discard seven datacubes due to their poor
quality. Most of these objects were observed under non photometric 
conditions,
and the control has clearly identified them to be discarded, as expected.

\begin{table}
\caption{Authomatic quality control measurements along the pipeline. 
The average $5\sigma$ limit $V-$band magnitude, the $V-$band magnitude of the sky and the accuracy
of the wavelength calibration are presented for each night.}
\label{data_pipeline}
\begin{tabular}{lccc}
\hline\hline
Observing       & $5\sigma$ limit & Sky  &  Wavelength       \\
date            & $V-$band        & $V-$band & calibration  \\
                & (mag)           & magnitude & ($\AA$)      \\
\hline
2007-06-22 &  22.73  &  20.50  &  0.30 \\ 
2007-06-23 &  22.50  &  20.50  &  0.30 \\ 
2009-03-19      &  23.02  &  20.25  &  0.43 \\ 
2009-03-20      &  23.02  &  20.28  &  0.34 \\ 
2009-03-21      &  22.92  &  20.10  &  0.26 \\ 
2009-03-22      &  22.84  &  20.12  &  0.39 \\ 
2009-03-23      &  21.54  &  19.07  &  0.22 \\ 
2009-03-24      &  23.19  &  19.01  &  0.32 \\ 
2009-06-27      &  23.05  &  20.13  &  0.35 \\ 
2009-06-28      &  23.75  &  18.35  &  0.30 \\ 
2009-10-17      &  22.55  &  20.05  &  0.15 \\ 
2009-10-18      &  21.58  &  19.42  &  0.20 \\ 
2009-10-19      &  22.00  &  19.83  &  0.15 \\ 
2009-10-22      &  22.42  &  20.13  &  0.15 \\ 
2009-10-23      &  22.32  &  19.73  &  0.23 \\ 
\hline
\end{tabular}
\end{table}


\section{Analysis and results} \label{sec:analysis}

In order to extract physical properties of the galaxy from the data set, it is
necessary to identify the emission lines produced by ionised gas in the
galaxies, and decouple this emission from the underlying stellar population.
Particular care has to be taken in this decoupling technique, since some of the
emission lines (e.g. H$\beta$) may be strongly affected by underlying
absorption features. Several authors has developed different fitting codes to
obtain information about the stellar content and star formation history of
galaxies \citep[e.g.,][]{Starlight, Stecmap, Sarzi2006, Ulyss, MacArthur2009}.
In this paper we adopted the decoupling method proposed by \citet{Sanchez2011},
based in a full spectrum modelling. This method involves a linear combination of
multiple stellar populations and the non-linear effects of dust attenuation to
obtain information about the ionised gas and the stellar content of the
galaxies, and can be applied to each of the individual spectra of the datacubes
to obtain the two dimensional distribution of the parameters derived with this
procedure.

In this section we present the method (\S~\ref{sec:method}) and analysis of the
accuracy of the physical properties derived with this method
(\S~\ref{mock-galaxies}). We present in \S~\ref{sec:comparison_SDSS} a
comparison with the SDSS data, and we explore differences in the properties
derived from two different aperture obtained from the data in
\S~\ref{sec:apertures}. Finally, a first analysis from the two dimensional
spectroscopic data is presented (\S~\ref{sec:resolved_properties}).

\subsection{Decoupling the stellar population and gas content}
\label{sec:method}

The method for the decoupling of the stellar populations from the emission
lines can be summarised as follows \citep[for a more detailed
description, see][]{Sanchez2011}: (i) A set of detected emission lines was
identified in the spectrum. (ii) The underlying stellar population was fitted
by a linear combination of a simple grid of six single-stellar populations
(SSPs) templates by \citet{Vazdekis2010}, with three ages (0.089, 1.000 and
17.7828~Gyrs) and two metallicites ($Z = 0.0004$ and 0.03)\footnote{This grid
of templates has been also adopted by \citet{Viironen2011} in their analysis of
UGC9837 galaxy from this data set.}. These templates were first corrected for
the appropriate systemic velocity and velocity dispersion (including the
instrumental dispersion), taking into account the dust attenuation\footnote{The
dust attenuation is also a free parameter in this decoupling method. The
extinction law of \citet{extintion_law} is adopted, with a ratio of total to
selective attenuation $R_V=3.1$ \citep{Jenkins87}.}. In addition, a spectral
region of 30\,\AA\ width around each detected emission line was masked prior to
the linear fitting, including also the regions around bright sky-lines
\citep{Sanchez2007a}. (iii) Once we derived a first approximation of the
spectrum of the underlying stellar population, this was subtracted from the
original spectrum to obtain a pure emission-line spectrum. (iv) To derive the
intensity of each detected emission line, this emission-lines spectrum was
fitted to a single Gaussian function per emission line plus a low order
polynomial function for the local continuum. (v) A pure gas-emission spectrum
was created, based on the results of the last fitting procedure, using only the
combination of Gaussian functions. The pure gas-emission model was then
subtracted from the original spectrum to produce a spectrum of gas-free
spectra. (vi) This spectrum is fitted again by a combination of SSPs, as
described before (but without masking the spectral range around the emission
lines, in this case), deriving the luminosity-weighted age, metallicity and
dust content of the composite stellar population. (vii) Finally, individual
emission-line fluxes are measured in each spectrum by considering spectral
window regions around the most prominent emission lines (fitting of Gaussian
funtions to [O\,{\sc ii}]$\lambda$3727, H$\beta$, [O\,{\sc
iii}]$\lambda\lambda$4959,5007, H$\alpha$, [N\,{\sc ii}]$\lambda$6548,83,
[S\,{\sc ii}]$\lambda\lambda$6717,6731). With this method, we obtain
information about the ionised gas and the stellar content in the analysed
spectrum.  

\begin{figure}
\centering
\includegraphics[angle=-90,width=0.85\columnwidth]{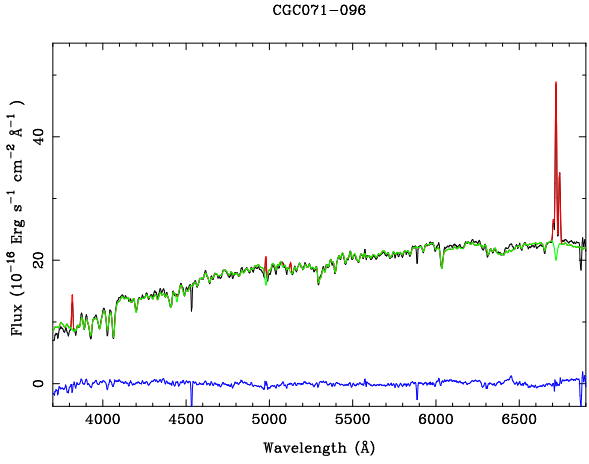}
\caption{Example of the decoupling of emission lines and underlying stellar
population for CGC071-096. We show the central spectrum in black line, the
best-fit model for the emission lines in red, and the best fitted stellar
population, modelled as a linear combination of different SSPs, plus the
computed dust extintion, in green. Finally, the differences between the real
spectrum and the final fitting is shown in blue.}
\label{fitting}
\end{figure}

This decoupling method can be applied to any galaxy spectrum, either integrated
spectrum or individual spectra in the datacubes, to analyse the spatially
resolved spectroscopic properties. As an example, we show in Fig.~\ref{fitting}
the best fitted gas-emission (red line) and stellar population (green line)
spectra obtained when applying this decoupling method on the central spectrum
of CGC071-096 galaxy (black line). The differences between the real spectrum
and the final fitting are shown with a blue line. These residuals represent an
error of $\sim 7$\% over the signal of the global spectrum, and it shows the
good agreement of the global fitting.

\subsubsection{Star formation history}\label{sec:SFH}

To explore a more realistic interpretation of the computed stellar population,
we decided to compare the galaxy data not only with SSPs, but also with spectra
obtained from a given star formation history (SFH). With this aim, we used the
population synthesis models by \citet{Vazdekis2010}. These models provide the
time ($T$) evolution of the synthesised spectra of galaxies characterised by an
initial metallicity ($Z$), a star formation history (SFH) and an initial mass
function (IMF) ($F_\lambda(T) [Z,SFH, IMF]$). We considered an
pure exponential-declining SFH \citep{Sandage1986}: 

\begin{equation} 
{\rm SFR}~(t) =  \exp(-(T_0-t)/\tau) 
\end{equation} 
where the steepness of the decay are regulated by a single parameter $\tau$.
To obtain the final spectra we considered a simple (not delayed) exponential,
being $T_0 = 15.5$~Gyr the age of the galaxy today, and a Salpeter IMF (a 
unimodal IMF with $\mu =1.3$). 
With these input data and using the tools provided in the
MILES (Medium-resolution Isaac Newton Telescope library of empirical 
spectra)
website\footnote{\url{http://www.iac.es/proyecto/miles/pages/webtools/get-spectra-for-a-sfh.php}.},
we created a grid of spectra with solar and sub-solar metallicities ($Z =
0.019, 0.0004$) and $\tau$ from 1 to 19 in steps of 2 Gyr, with a S/N~$\sim 50$
per pixel in the continuum, and spectral resolutions corresponding to 
the V300 and V600 grism.

\begin{figure}
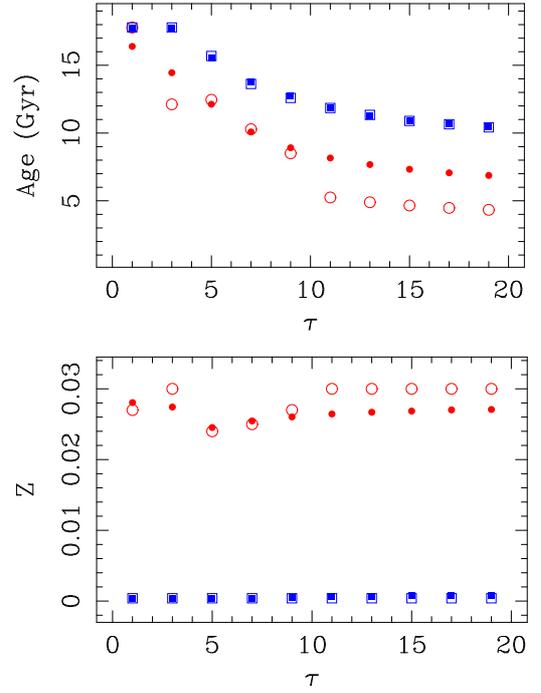

\centering
\includegraphics[angle=-90,width=0.75\columnwidth]{V600_equivalence_age.ps}
\vspace{0.3cm}

\includegraphics[angle=-90,width=0.75\columnwidth]{V600_equivalence_metallicity.ps}
\caption{Equivalence between the stellar population parameters, i.e the
luminosity-weighted age (top panel) and the luminosity-weighted metallicity
(bottom panel), and the characteristic time scale parameter $\tau$ for the two
SFH considered (red circle for solar metallicity, blue squares for sub-solar).
We present the results of the fitting procedure applied to mock galaxies with
the V600 resolution, when the dust attenuation is fixed (filled symbols) and
computed (open symbols).}

\label{equivalence_tau}
\end{figure}

In order to obtain the equivalence between the characteristic time scale
parameter $\tau$ and the stellar population parameters, i.e.,
luminosity-weighted age and metallicity, we applied the fitting procedure 
explained in the previous section, using
as input the spectra generated from these SFHs. In Fig.~\ref{equivalence_tau}
we represent the luminosity-weighted age and metallicity derived for the grid
of spectra from the different SFH parametrised by $\tau$ as a function of this
parameter. As it can be seen, we obtain a clear dichotomy for the metallicity.
This is expected since we are considering two different metallicities for the
spectra created from a given SFH. In this sense, the metallicity is well
recovered for all the values of $\tau$, although the metallicity is
over-estimated in the case of the solar metallicity due mainly to the limited
available values in the grid of models using in this experiment.

In the case of the age, we obtain old stellar ages for low values of
$\tau$ and younger ages until $\tau > 10$, where the age becomes almost
constant. This agrees with \citet{Gavazzi2002}, who found that dwarf and normal
elliptical galaxies, i.e. galaxies with older stellar populations, are well
characterised by $\tau \le 3$. They also found that galaxies with
younger stellar populations present $\tau \ge 4$, as spiral galaxies (Sa-Sb)
with values of $4 < \tau < 6$, and blue compact dwarfs with $\tau \ge 7$.
The equivalences obtained here (Fig.~\ref{equivalence_tau}) will be
used for a characterisation of the stellar populations in \S~\ref{sub_sec:pop}.

\subsection{Accuracy of the derived properties} \label{mock-galaxies}


Following the idea of recent works based on IFU observations, 
\citep[e.g.,][]{Bershady2010b, Sanchez2011}, we explore in this section the 
accuracy in the measurements of different parameters of our interest
(e.g., emission line fluxes, redshift, velocity dispersion, dust extinction,
average stellar populations). With this aim, we created a set of mock spectra of galaxies with
different kinematic galactic parameters (velocity dispersion, redshift),
underlying stellar populations (old or young SSP), ionised gas and
signal-to-noise (S/N) in the continuum. These mock spectra were generated as
follows. First, we consider an underlying stellar population using the stellar
population models provided by
\citet{Vazdekis2010}\footnote{\url{http://www.iac.es/proyecto/miles/}}. We
consider four possibilities for the underlying stellar population: a single
stellar population (young
--0.7~Gyr and solar metallicity ($Z = 0.019$) -- or old --12.6~Gyr and
metallicity of $Z = 0.004$), or two stellar populations mixed in different
ratios (10\% of young plus 90\% of old stellar population -2SSP-10-, and 1\% of
young plus 99\% of old stellar population -2SSP-01). Next, we add an
emission-line spectrum to simulate the ionised gas in the mock galaxy. To
obtain more realistic data, we measured the emission lines on several regions
of NGC628 from the 2D spectroscopic data presented by \citet{Sanchez2011}, and
we scaled these values to obtain similar values than the ones measured in a
real galaxy of our sample \citep[see e.g.][]{Viironen2011}. The
nebular continuum is considered negligible. Then we apply different values of
dust attenuation, same for the stellar continuum and the emission lines. Next,
we consider the kinematics of the mock galaxy, i.e., we broad the spectrum to a
given velocity dispersion $\sigma$ and shift it to a given redshift $z$.
Finally, we add noise to the spectrum to simulate a real galaxy observation. As
result of this process, we obtain a spectrum of a mock galaxy with well-known
parameters that we can change to check the recovering of the information for
these studies.

In next sections we present the different studies that we have carried out to
analyse the accuracy in measurements of the dust attenuation
(\S~\ref{sub_sec:dust}), kinematics (\S~\ref{sub_sec:kinematics}), ionised gas
(\S~\ref{sub_sec:gas}), and underlying stellar population
(\S~\ref{sub_sec:pop}), of the galaxies observed with the two setups (V300 and
V600) used in this work. 

\subsubsection{Dust attenuation}\label{sub_sec:dust}

\begin{figure}
\centering
\includegraphics[width=1.08\columnwidth]{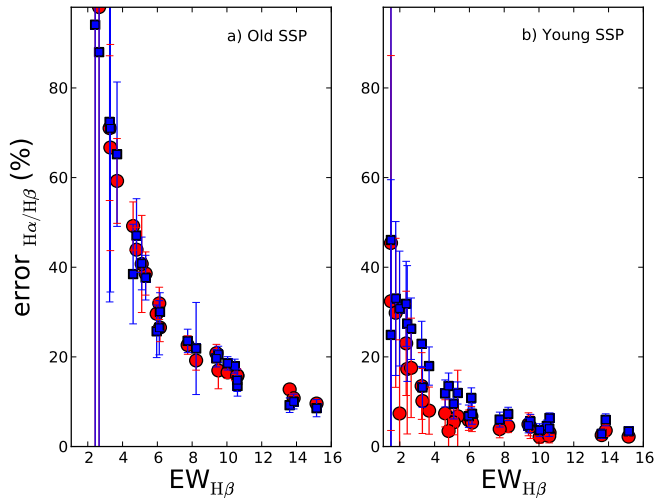}
\caption{Relative errors in the H$\alpha$/H$\beta$ ratio when the dust
attenuation is fitted (blue squares) or fixed (red circle) for different
equivalent width of the H$\beta$ emission line (EW$_{\rm H\beta}$), for a
spectrum of a mock galaxy with a pure old (left panel) or young underlying
stellar population (as explained in the text).}
\label{comparison_emission}
\end{figure}

The treatment of the dust attenuation may affect the resulting derived
parameters of the stellar population \citep[i.e., the luminosity-weighted age
and metallicity,][]{Calzetti2001, Munoz-Mateos2007, Munoz-Mateos2009a}. As part
of the decoupling method explained above, we adopted the \citet{extintion_law}
law, which may be not the optimal solution to study the dust attenuation in
star-forming galaxies \citep[e.g.,][]{Calzetti2001}. Recently,
\citet{MacArthur2009} considered the attenuation law based on the two component
dust models by \citet{Charlot_Fall2000}, especially developed for
star-formation galaxies. Despite the conceptual differences between these
attenuation laws, both of them present a very similar shape in the wavelength
range of our data, and it is not wide enough to distinguish between these
different laws. 

To explore the impact of fitting the dust attenuation on the derived
parameters, we generated a set of mock spectra of galaxies following the
procedure explained above. Since values of equivalent widths of the H$\beta$
emission line EW$_{\rm H\beta} > 3.5~\AA$ are suitable to perform an abundance
analysis in galaxies despite the dependence on other physical properties of the
ionising cluster (e.g., its age), we decided to explore a range of equivalent
widths of EW$_{\rm H\beta}\sim 1.5-15~\AA$ for our mock galaxies. Then we
applied the fitting technique (i) fixing the value for the dust attenuation, or
(ii) computing $A_V$ with the fitting program, with values of $A_V$ varying
between $A_V =0.1$ and $1.2$, in steps of 0.1. 

We present the results of this experiment in Fig.~\ref{comparison_emission},
where we show the relative random errors obtained for the H$\alpha$/H$\beta$ ratio (in
absolute values) when the dust attenuation fixed (red circles) and fitted (blue
squares), as a function of EW$_{\rm H\beta}$. In all cases, the
H$\alpha$/H$\beta$ ratio is recovered with a relative error $< 20$~\% for
EW$_{\rm H\beta} > 8~\AA$ for an old underlying stellar population (left
panel), while the relative errors for a young stellar population is $< 20$~\%
for EW$_{\rm H\beta} > 3.5~\AA$ (right panel). Although there are not
differences in the determination of H$\alpha$/H$\beta$ for an old population
depending on the method (fitting or not the dust attenuation), slightly higher
errors are found when the dust attenuation is fitted in the case of a young
underlying stellar population (right panel in Fig.~\ref{comparison_emission}).
This is due to the effect of the dust attenuation, more important for bluer
wavelengths where stronger Balmer absorption lines are found, especially for
young stellar populations. In this case, slightly variations in the dust
attenuation determination can produce the differences in the H$\alpha$/H$\beta$
ratio, more evident for small equivalent widths of H$\beta$. We see in
Fig.~\ref{comparison_emission} that, in general, the relative errors in the
determination of the H$\alpha$/H$\beta$ ratio are similar regardless the dust
attenuation fitting, and then, the accuracy in the determination of the gas
content is not depending on the procedure applied.

\subsubsection{Kinematics}\label{sub_sec:kinematics}

\begin{figure}
\centering
\includegraphics[width=4.2cm]{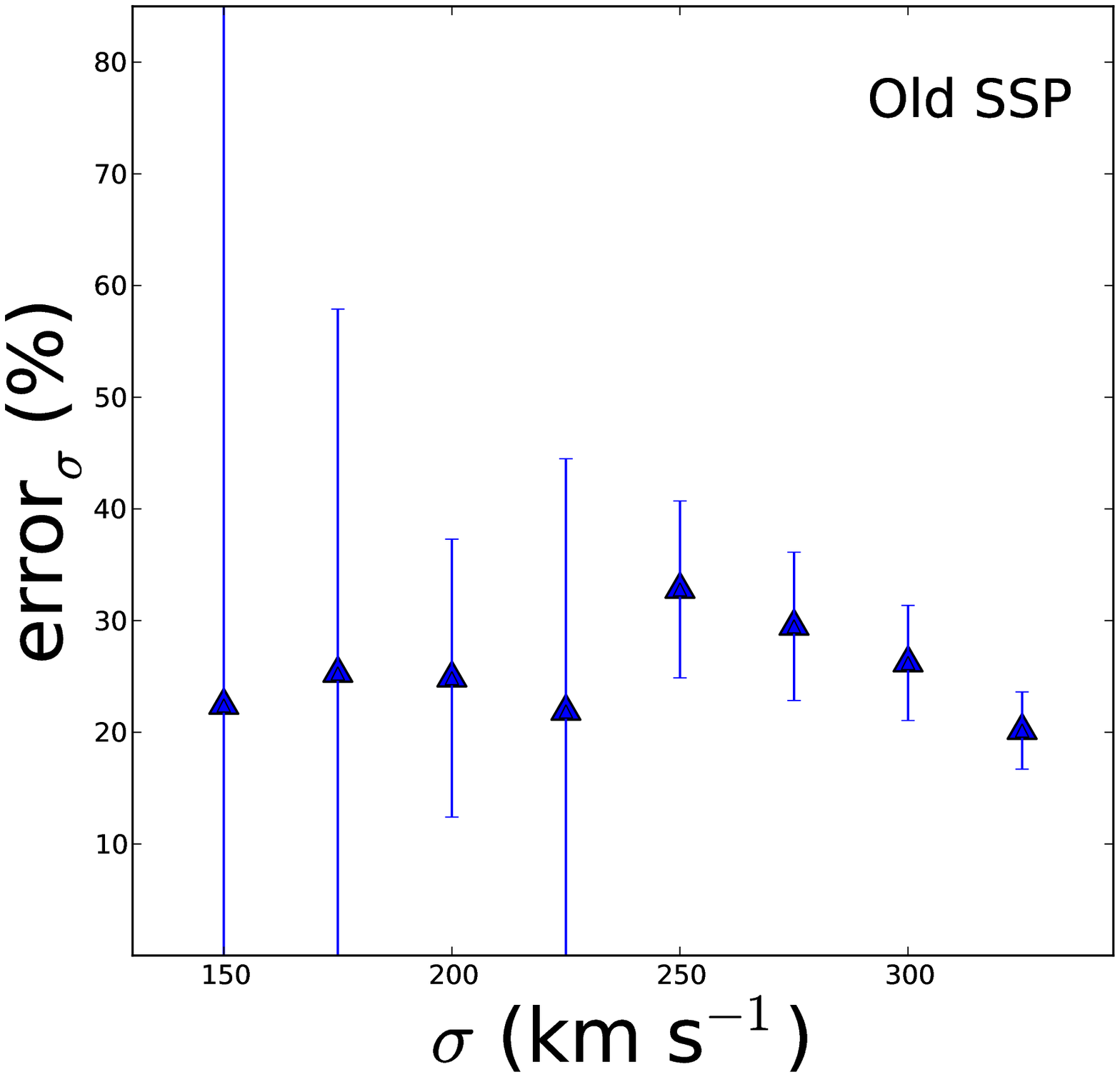}
\includegraphics[width=4.2cm]{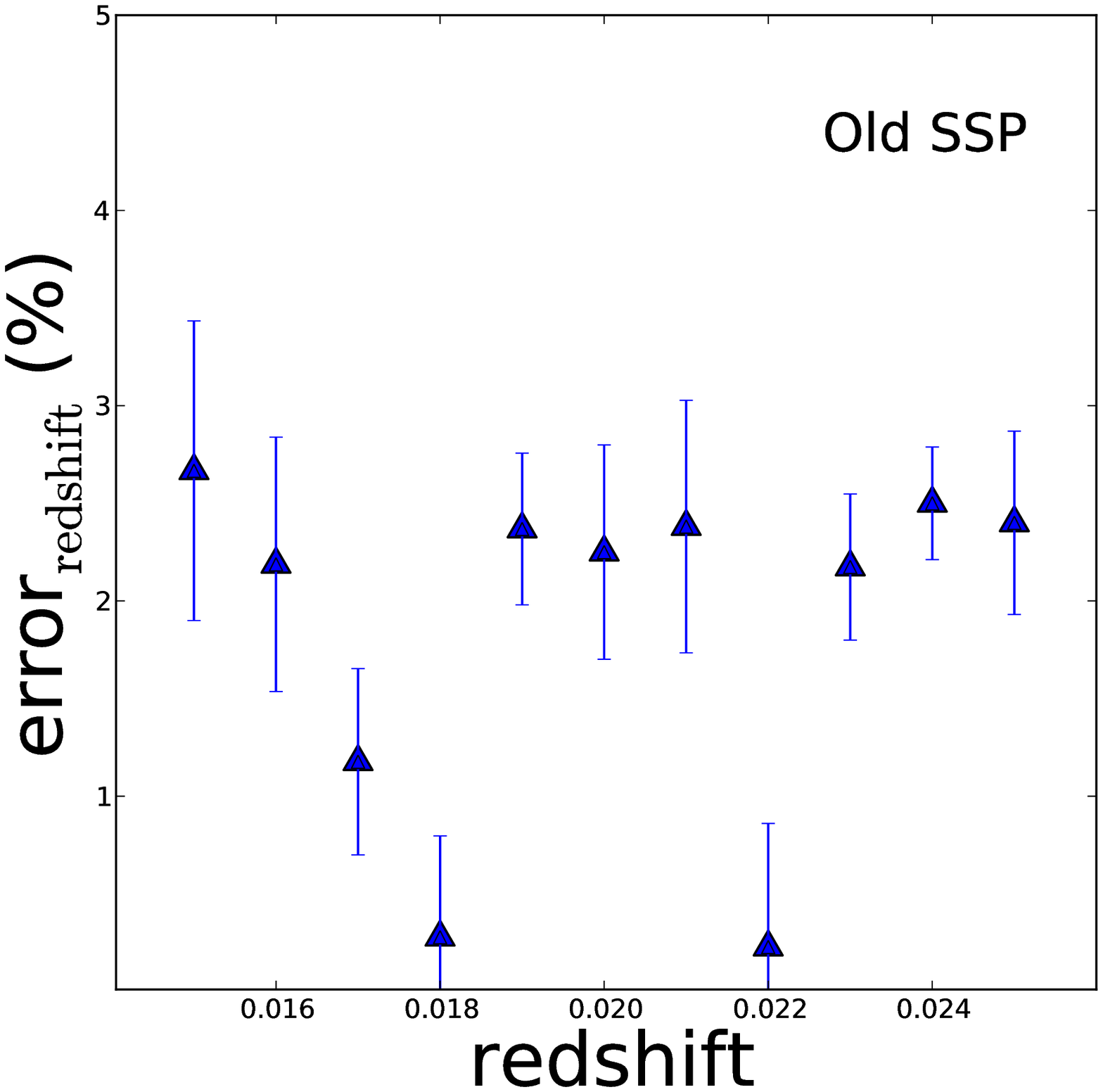}
\includegraphics[width=4.2cm]{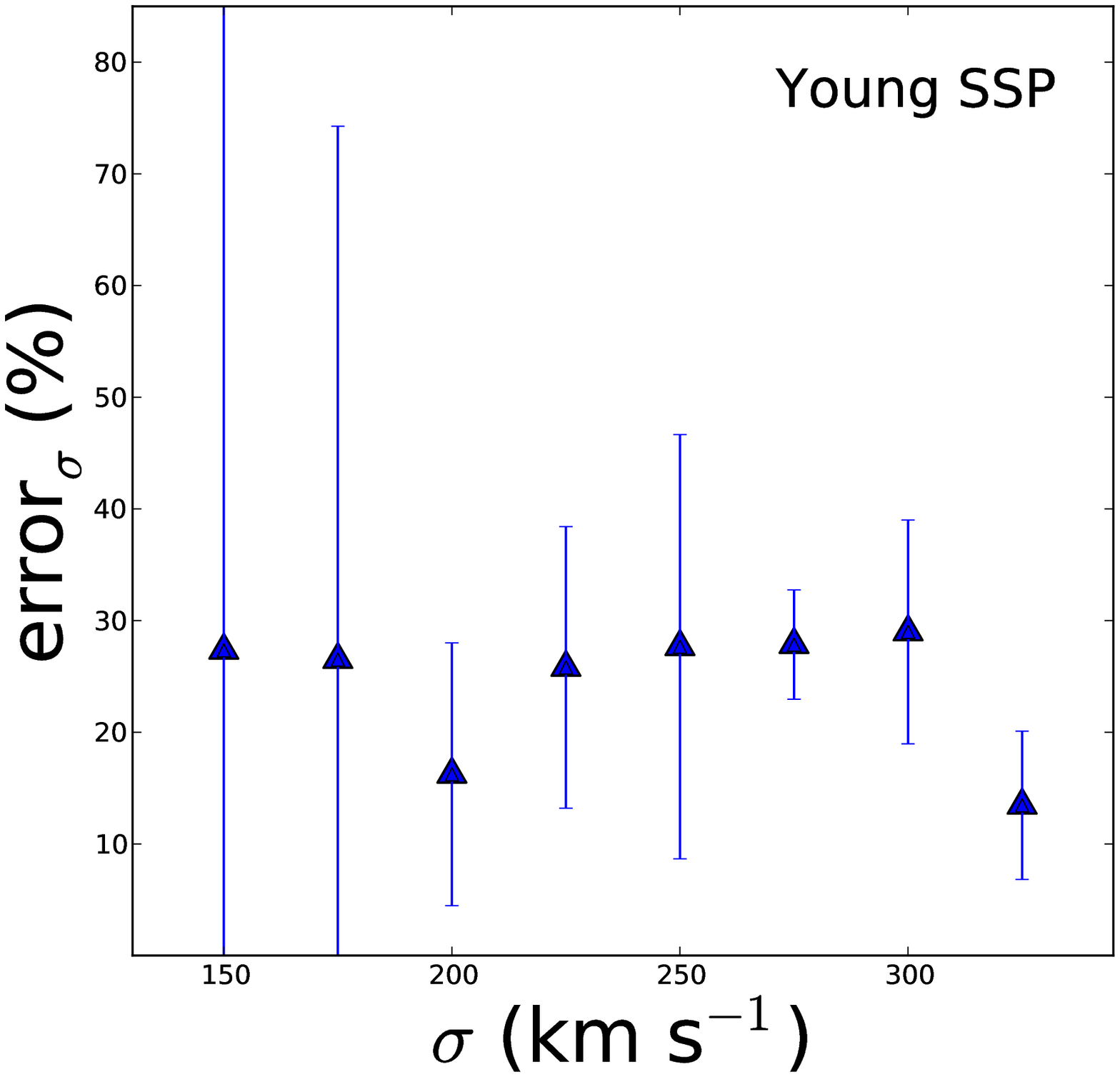}
\includegraphics[width=4.2cm]{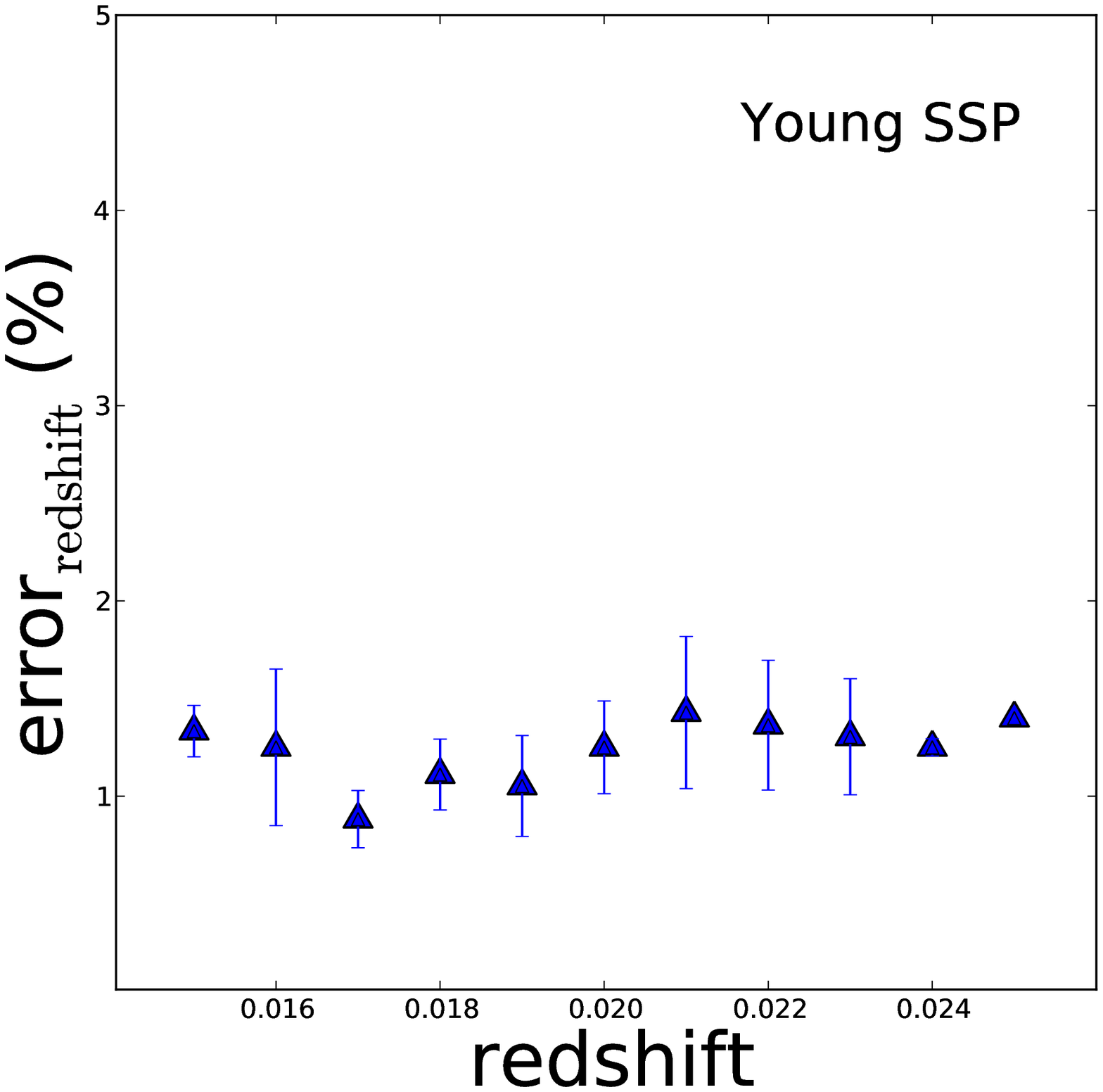}
\caption{Estimated relative random errors in the measurements of velocity dispersion
(left panels) and redshift (right panels) for mock galaxies with an old (upper
panels) and young (lower panels) underlying single stellar populations, as
explained in the text, for observations with the medium resolution grism
(V600).} 
\label{kinematics_plot}
\end{figure}

One of the advantages of using IFU data is the possibility to analyse spatially
resolved properties of the galaxies. In particular, kinematic studies can
reveal substructures in galaxies such as kinematically decoupled cores in the
stellar velocity field and twists in the gas velocity field, and also more
hidden substructures as disks and bars. The instrumental setups used in these
observations, although limited in spectral resolution, allow some studies of
this kind. Data obtained with V300 and V600 grism allow the measurement of
redshifts, while the resolution of the V600 grism (FWHM$ = 5.4~\AA$) will allow
measurements of velocity dispersion above $\sim 140$~km~s$^{-1}$. In order to
asses the accuracy in the measurements of redshift and velocity dispersion for
these observing setups, we applied the fitting method explained above to a set
of mock spectra from single stellar populations (an old and a young SSP), and
S/N = 50 (per pix) in the continuum. In this case, we considered our mock
spectra of galaxies at different redshifts (from 0.015 to 0.025, in steps of
0.001) and they were broadened to different velocity dispersions (from 150 to
350~km~s$^{-1}$, in steps of 25~km~s$^{-1}$). We applied the decoupling method
in the spectral range $\lambda\lambda3950-4600~\AA\AA$, which comprises clear
absorption features suited for this study. In addition, due to the short
wavelength considered in this experiment, a fixed value for the dust
attenuation was considered, and then, it is not fitted in this case. Finally,
for an estimation of the random error in these measurements, we simulated a grid of
spectra of each mock galaxy as follows. An average $rms$ of the mock galaxy in
areas free of sky lines, and galactic emission, or absorption lines was
calculated and added by bootstrapping method to the SSP+gas model spectrum.
This was repeated 100 times, and the fitting was done then on the resulting
spectra. The results are presented in Fig.~\ref{kinematics_plot}, where we show
the median relative errors obtained for the velocity dispersion and the
redshift for each mock galaxy. The error bars represent the standard deviation
computed from the simulations. We find that for both underlying stellar
populations, we recover the redshifts with errors $<5\%$. As expected, the
relative errors in determination of the velocity dispersion are high for values
close to the instrumental setups. It is possible to measure velocity dispersion
with a relative error of $\sim 30\%$ for $\sigma > 200$~km~s$^{-1}$ (i.e.,
intrinsic values of the velocity dispersion of galaxies above 140~km~s$^{-1}$).

\subsubsection{Ionised Gas}\label{sub_sec:gas}

\begin{figure}
\centering
\resizebox{0.49\hsize}{!}{\includegraphics{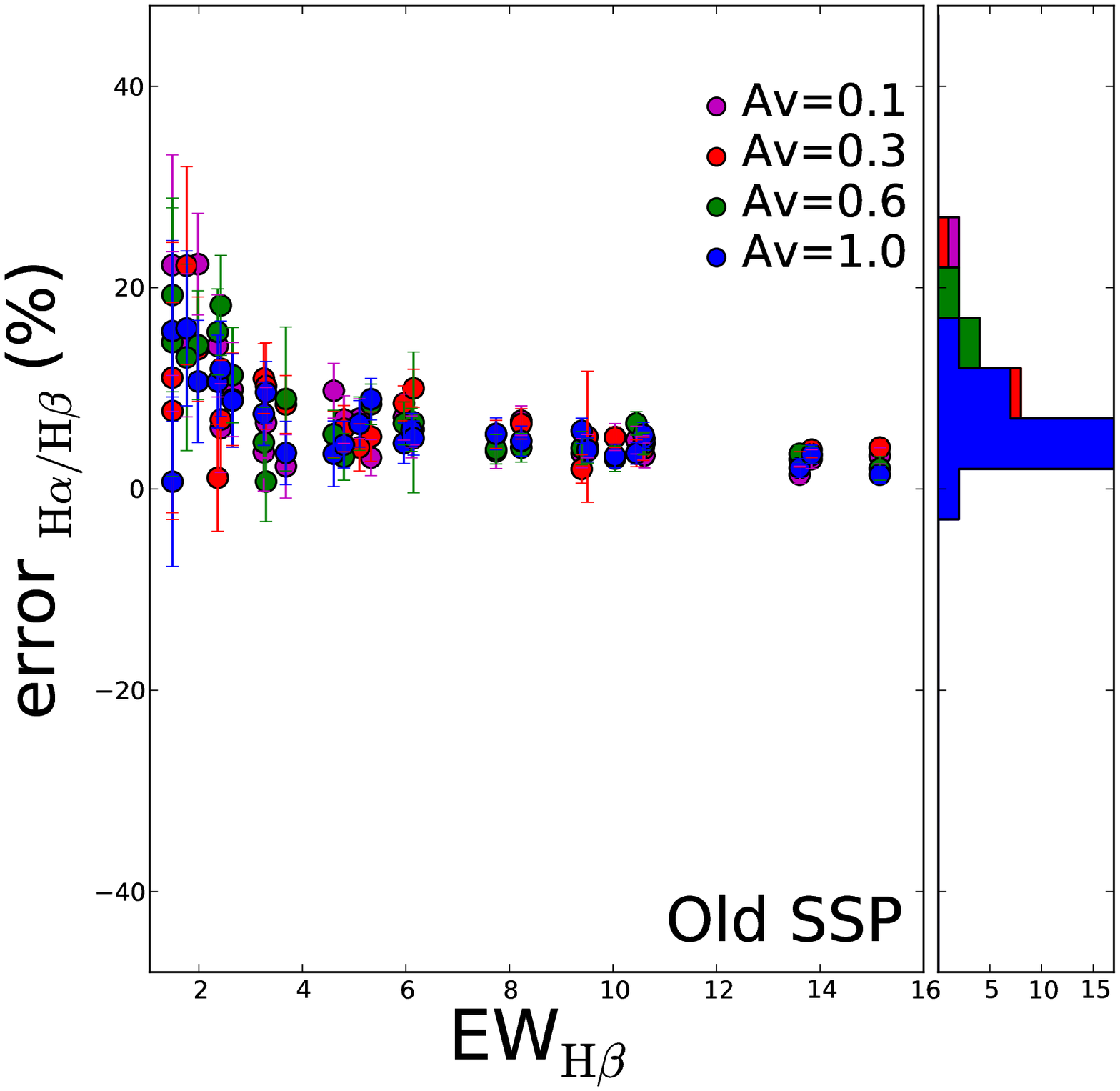}}
\resizebox{0.49\hsize}{!}{\includegraphics{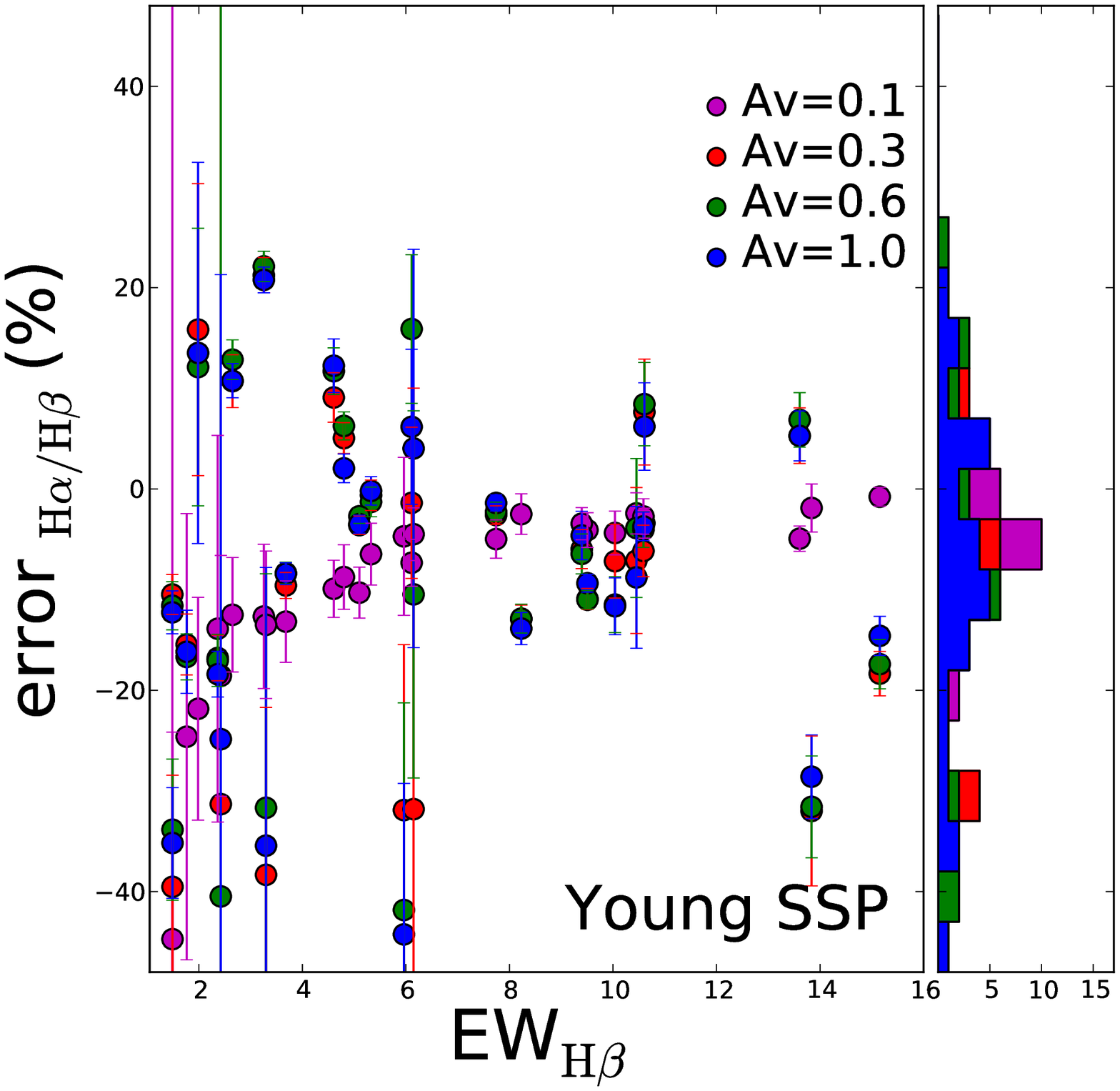}}
\resizebox{0.49\hsize}{!}{\includegraphics{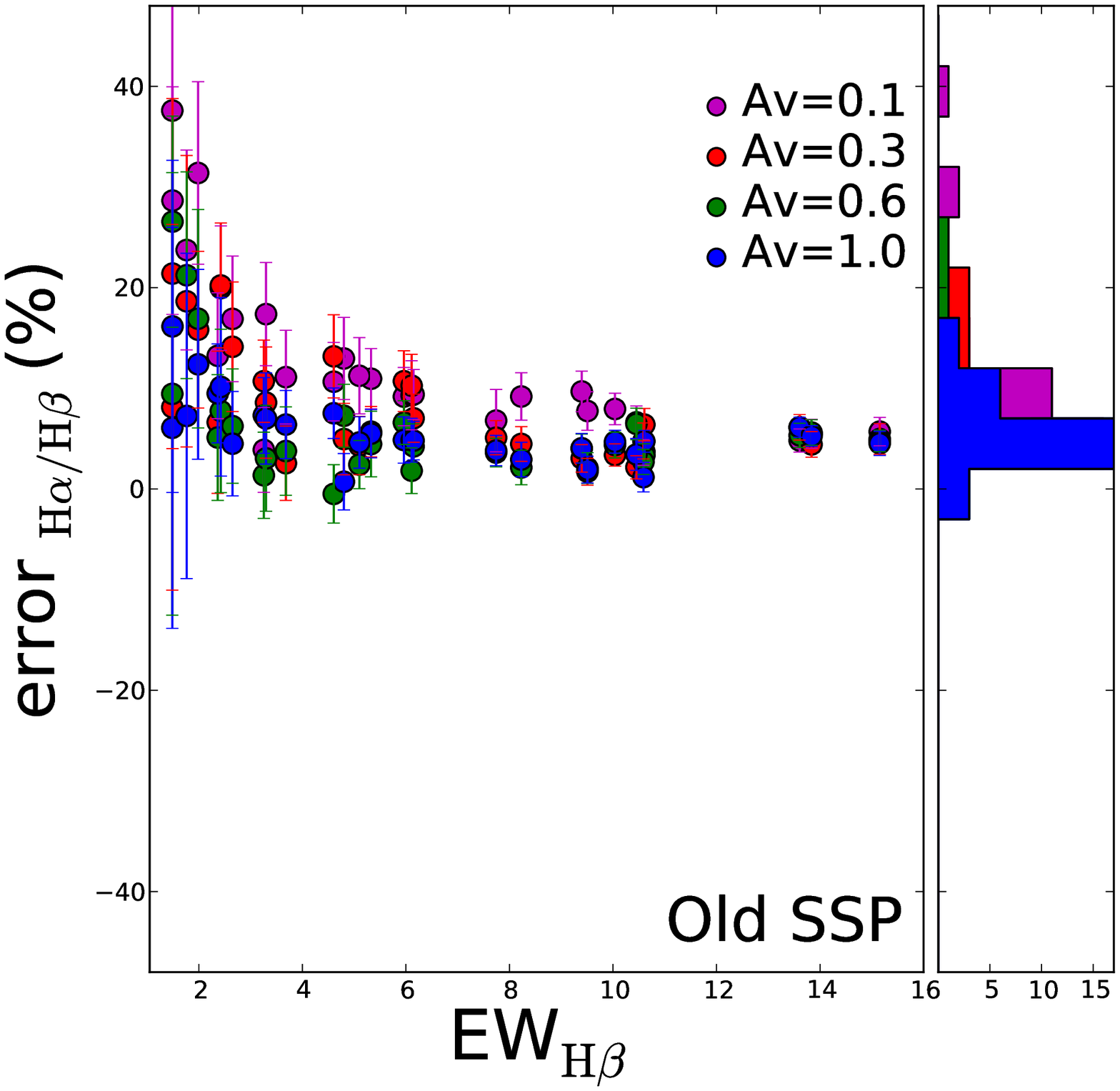}}
\resizebox{0.49\hsize}{!}{\includegraphics{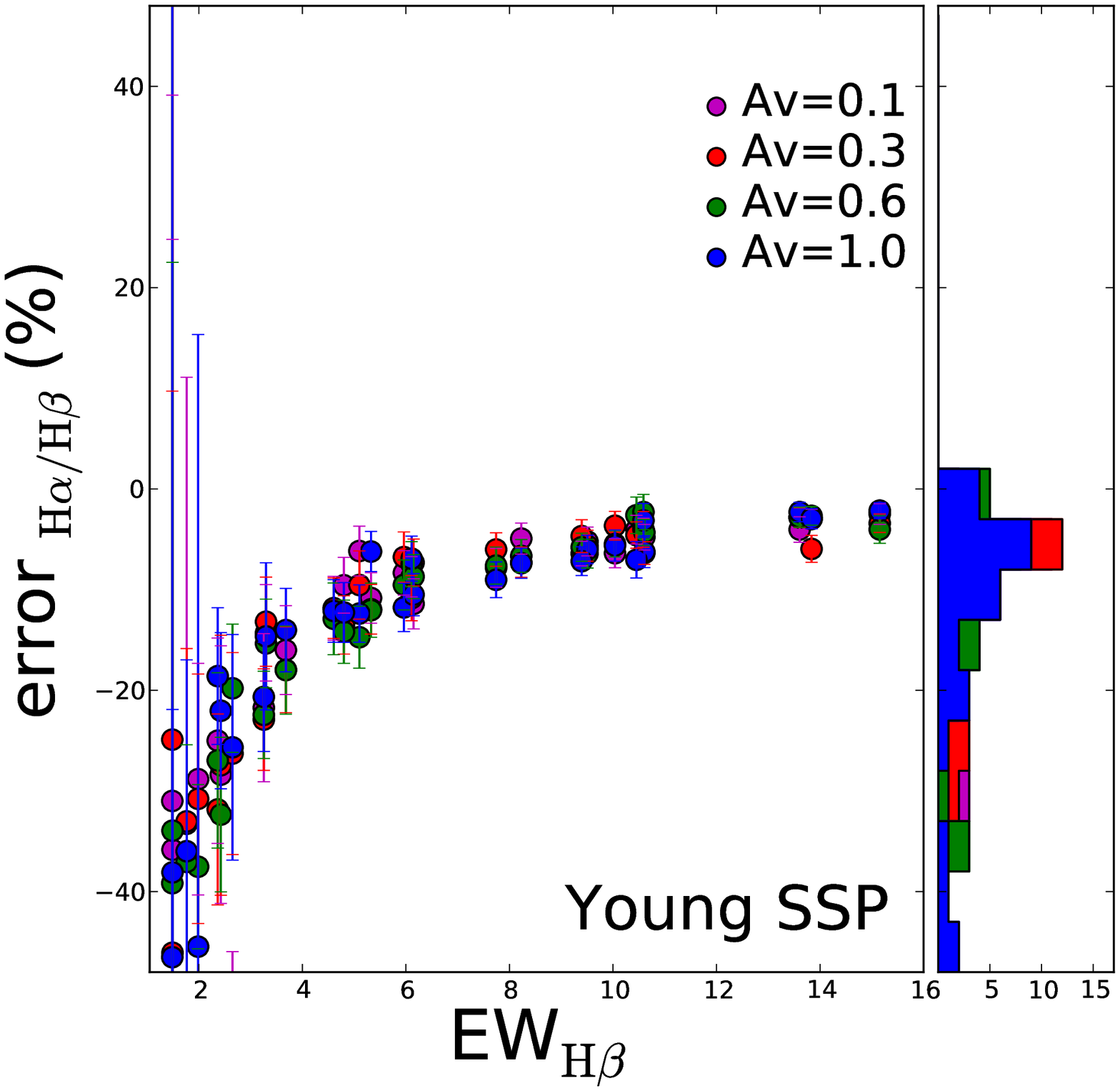}}
\caption{Study of the recovering of the properties of the ionised gas for the
for observations with the medium resolution grism (V600).
Estimated relative random errors in the measurements of the H$\alpha$/H$\beta$ ratio
as a function of the equivalent width of H$\beta$ emission lines (EW$_{\rm
H\beta}$) on mock galaxies with an old (left panels) and a young (right panels)
underlying stellar population and different values for the dust attenuation:
magenta for $A_V=0.1$, red for $A_V=0.3$, green for $A_V=0.6$ and blue for
$A_V=1.0$. Top panels are for dust attenuation fixed in the program, and bottom
panels for dust attenuation fitted in the program.  Each block of the histogram
represents an error of 5\%.}  
\label{Av_plot} 
\end{figure}

The Integral Field Spectroscopy allows the study of the resolved physical
properties of the ionised gas in galaxies, such as (i) local nebular reddening
estimates based on the Balmer decrement; (ii) the oxygen abundance
distributions based on a suite of strong line diagnostics incorporating
reddening-corrected line ratios (e.g., H$\alpha$, H$\beta$, [O\,{\sc ii}],
[O\,{\sc iii}], [N\,{\sc ii}], and [S\,{\sc ii}]); and (iii) measurements of
ionisation structure in H\,{\sc ii} regions and diffuse ionised gas using the
well-known and most updated forbidden-line diagnostics in the oxygen and
nitrogen lines. For all these studies, it is necessary measurements of the
fluxes of the emission lines in the individual spectra. To test the accuracy in
measured fluxes for the ionised gas, we followed the procedure explained above
to create a set of mock galaxies at redshift $z=0.02$ with a velocity
dispersion $\sigma = 150~{\rm km~s^{-1}}$ and S/N = 50 (per pix) in the
continuum. We measured the main emission lines (see \S~\ref{sec:method}) in 29
different regions in NGC628, and we created 29 spectra with gaussians of these
values. In this way, the emission line ratios of the mock galaxies are
realistic and known in advance. Then we have applied the decoupling method and
measured the emission line fluxes.  Finally, we have compared these results
with the input data from NGC628. In Fig.~\ref{Av_plot} we plot the relative
random error of the ratio of H$\alpha$/H$\beta$ as a function of the equivalent width
for the H$\beta$ emission line. We obtain relative errors of the ratio of
H$\alpha$/H$\beta < 10$\% when EW$_{\rm H\beta} > 5~\AA$ for old stellar
populations, while we obtain errors of $< 20$~\% always that EW$_{\rm H\beta} >
6~\AA$ for young stellar populations. There is a tendency to over-estimate the
emission of H$\beta$ for old stellar populations (left panels), while we find
the opposite behaviour for young stellar populations (right panels). This is
expected since the Balmer absorption lines are deeper for a young stellar
population and an accurate decoupling is more difficult if the emission lines
are not very intense. In addition, this effect is clearer when the dust
attenuation is also fitted in the decoupling method (lower panel in
Fig.~\ref{Av_plot}). 

\subsubsection{Stellar Populations}\label{sub_sec:pop}

\begin{figure}
\centering
\includegraphics[width=0.49\columnwidth]{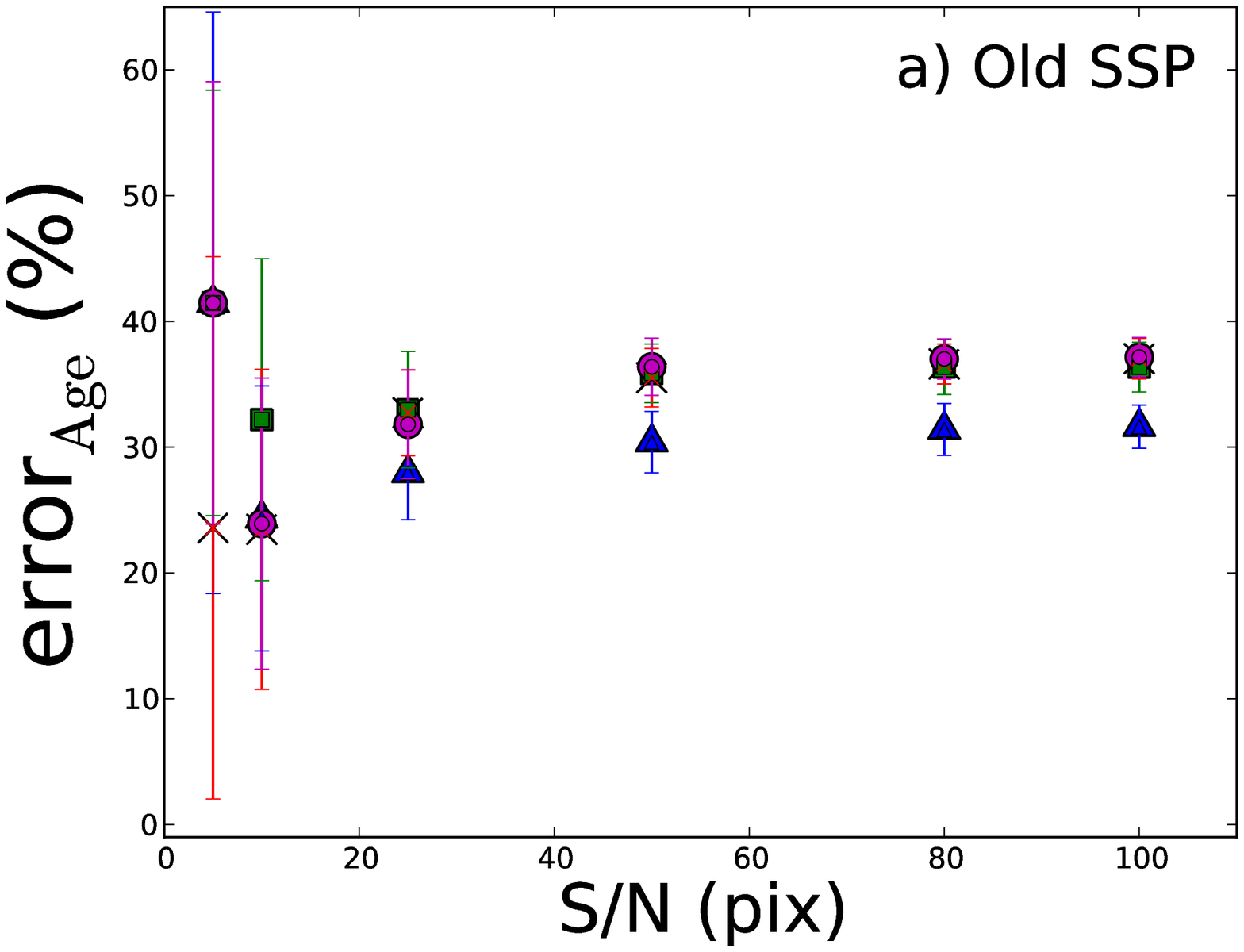}
\includegraphics[width=0.49\columnwidth]{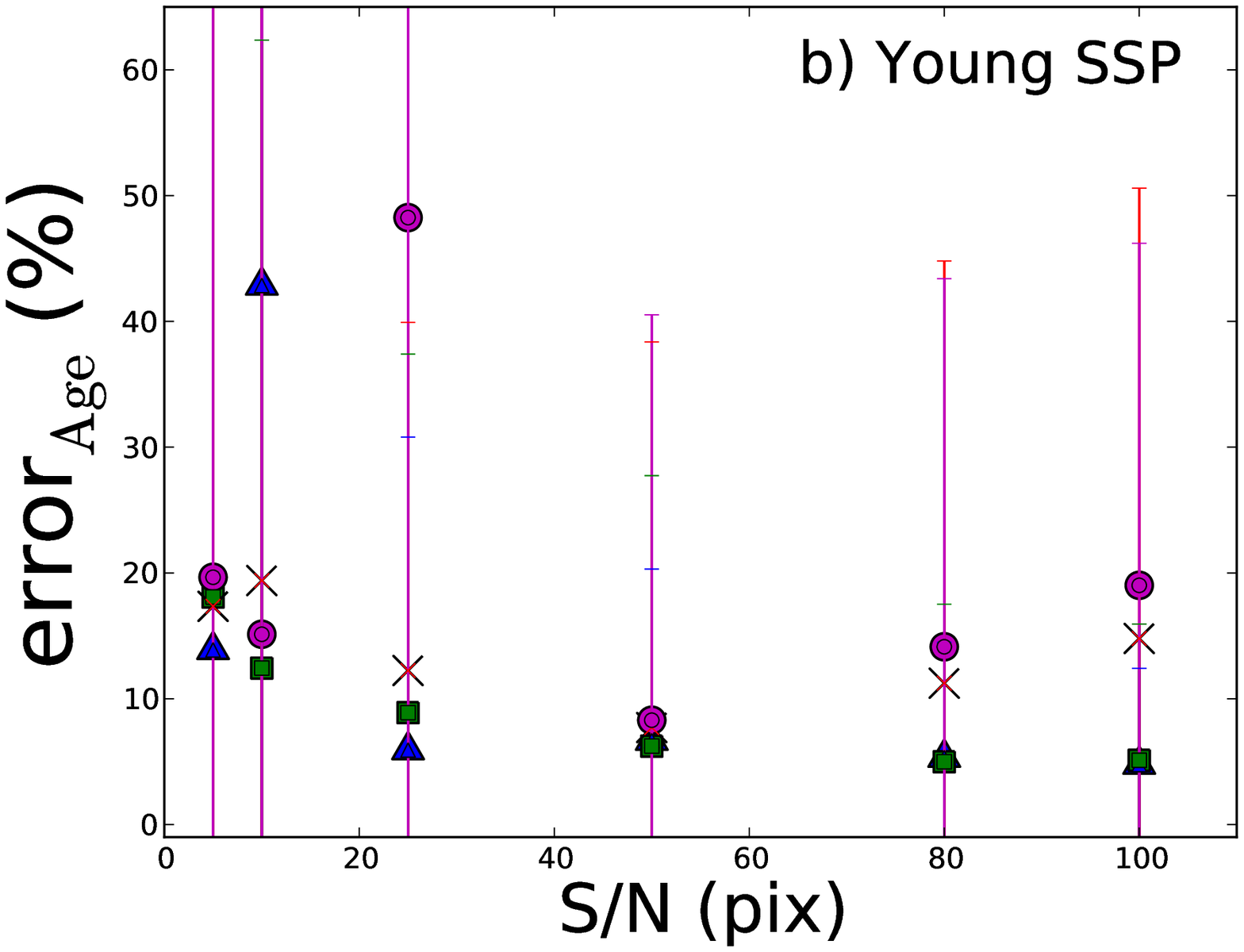}
\includegraphics[width=0.49\columnwidth]{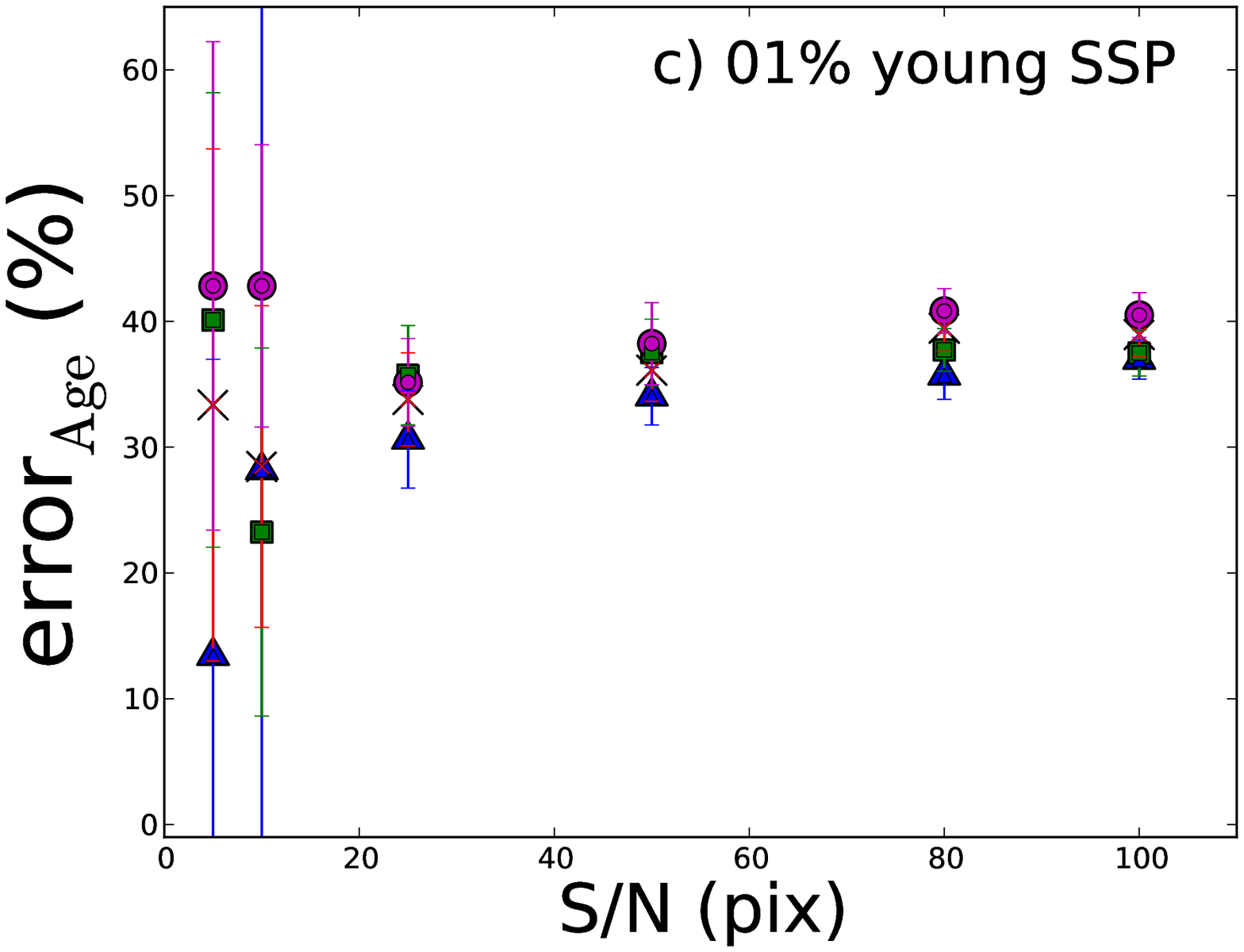}
\includegraphics[width=0.49\columnwidth]{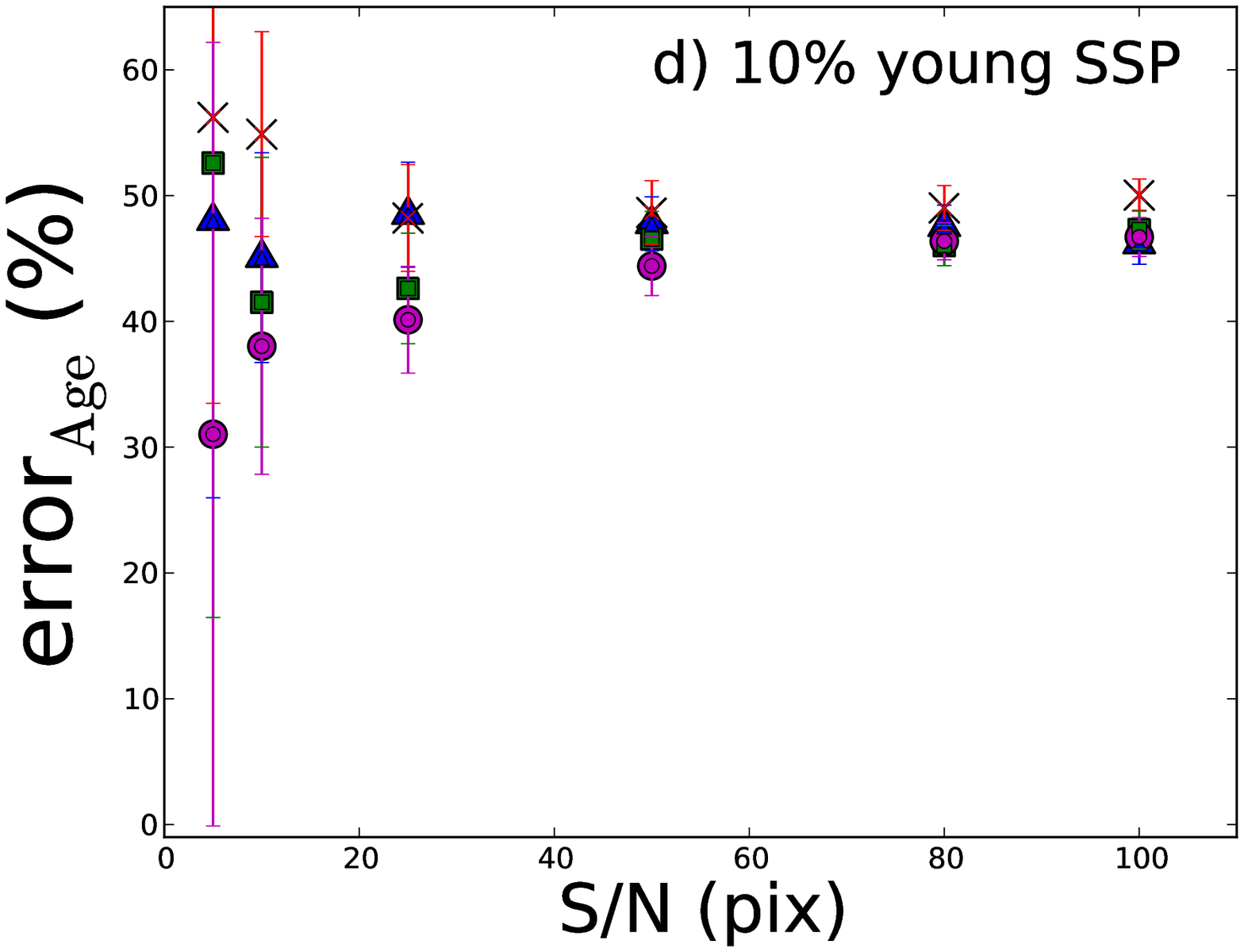}
\caption{Study of the recovering of the luminosity-weighted age as a
function of the S/N in the continuum for the V600 grid. In this case, the dust
attenuation has been considered known and fixed when applying the decoupling
method. Different colours are used for different values of H$\beta$ emission
line.}
\label{SSP_plot_fijo}

\includegraphics[width=0.48\columnwidth]{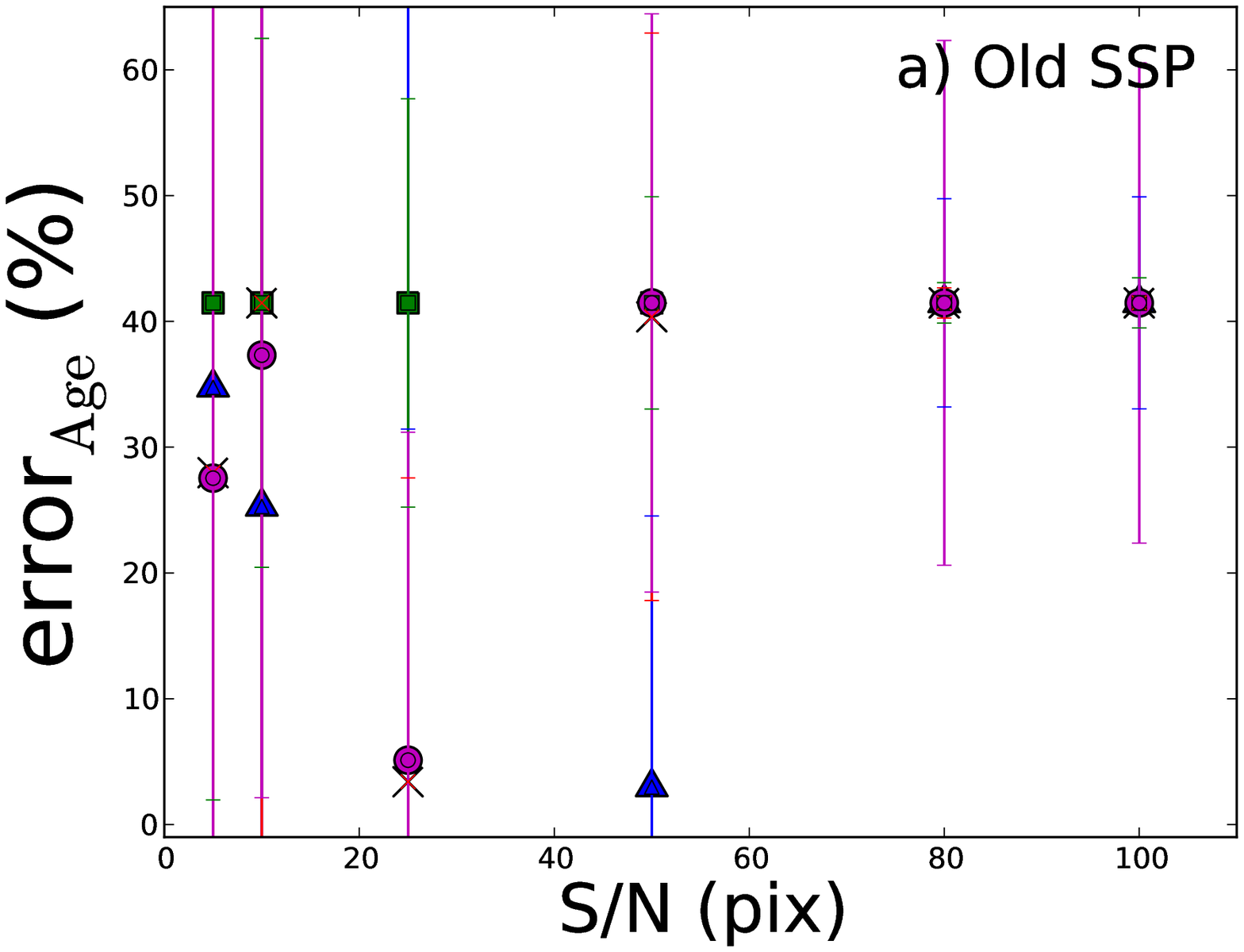}
\includegraphics[width=0.48\columnwidth]{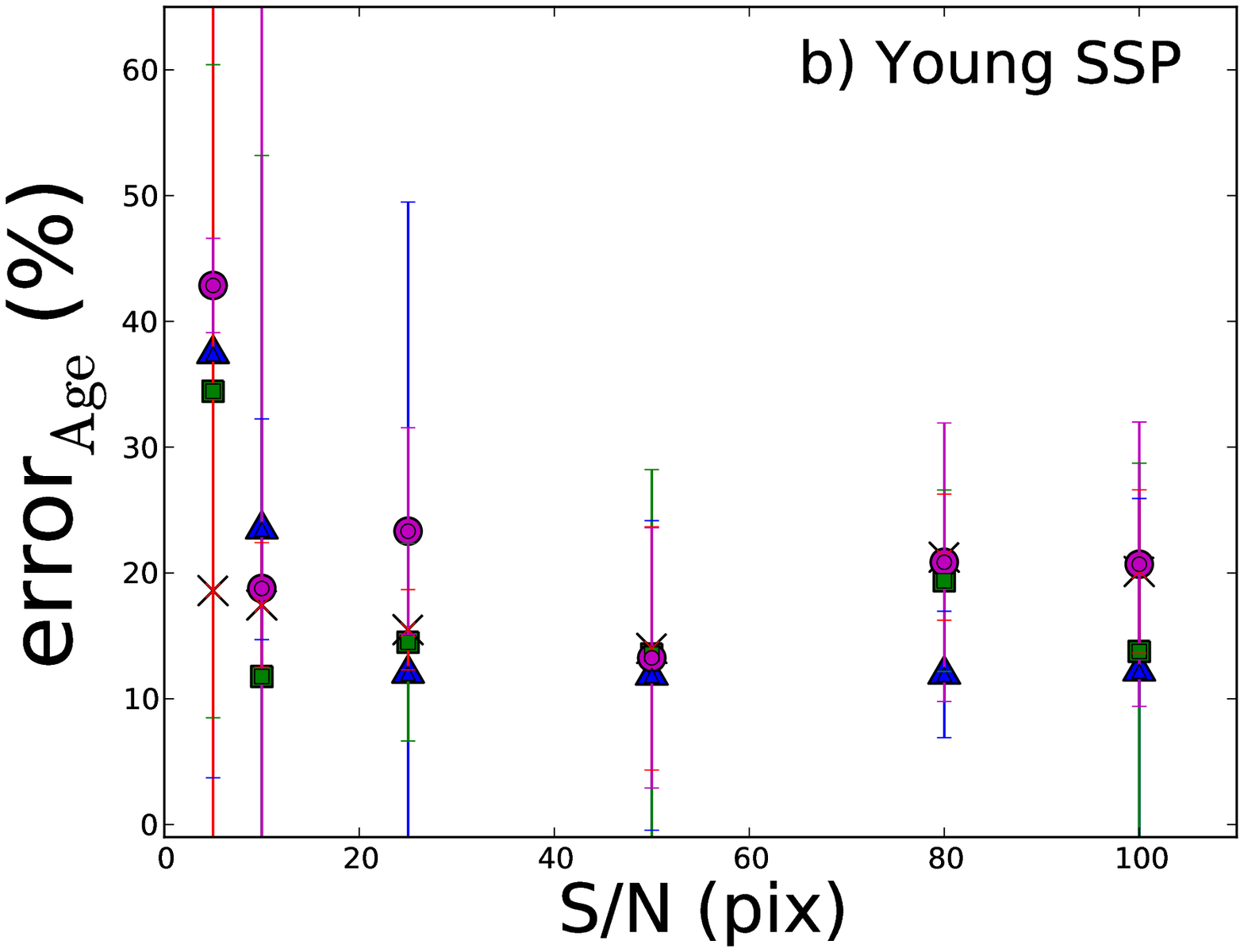}
\includegraphics[width=0.48\columnwidth]{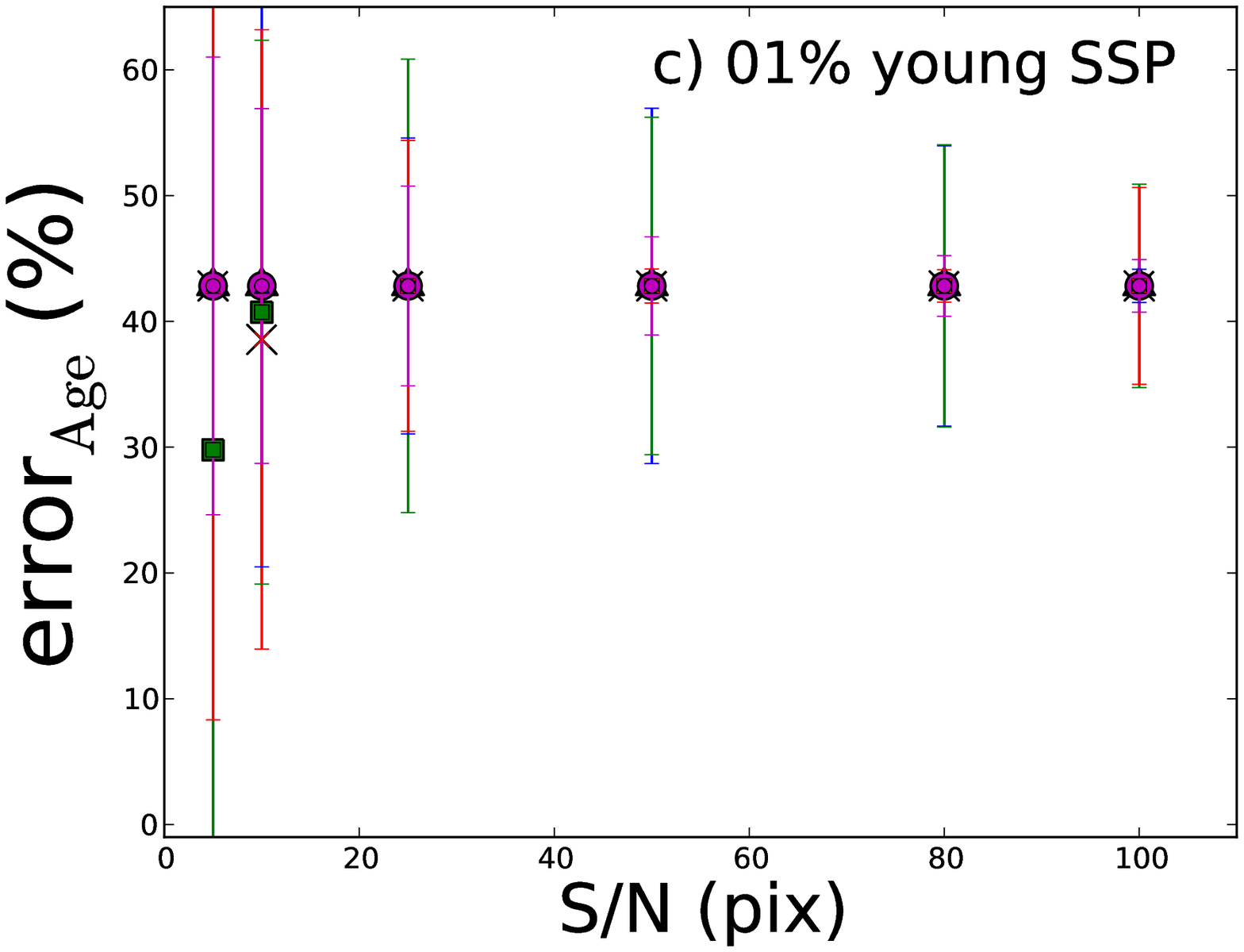}
\includegraphics[width=0.48\columnwidth]{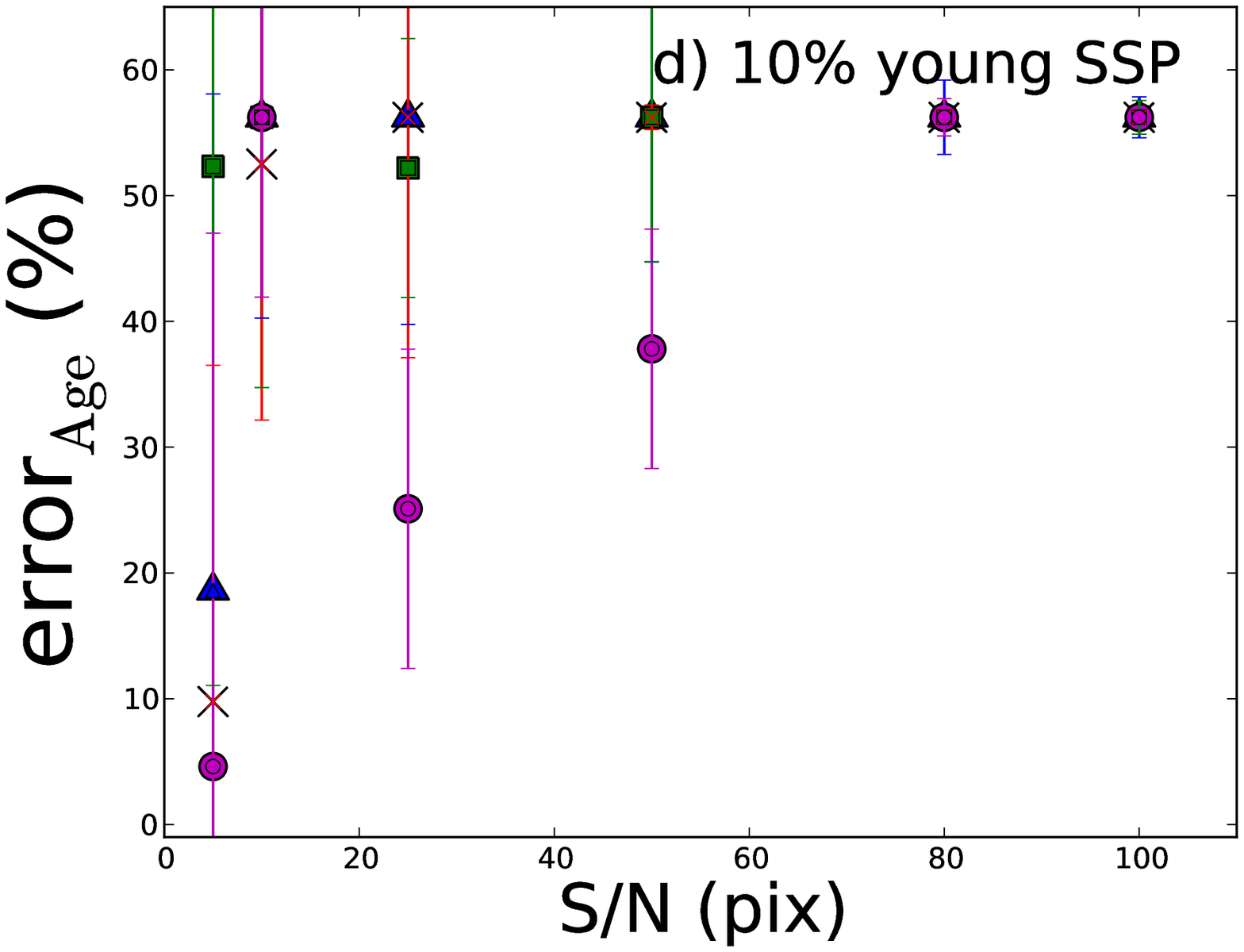}
\caption{Same figure than Fig.~\ref{SSP_plot_fijo} when the dust
attenuation is also computed with the decoupling method. In this case,
the estimated random errors (and their standard deviations) are significantly higher
than in the previous example.}
\label{SSP_plot_libre}
\end{figure}

As explained before, the decoupling method also provides with the best
combination of stellar populations to reproduce the underlying stellar
population of galaxies. In this section we study the recovering of
the luminosity-weighted age and metallicity from the data obtained in this
work. With this purpose, we applied the decoupling method on spectra
of mock galaxies with different underlying stellar populations and gas content
(see \S~\ref{mock-galaxies}), at redshift $z=0.02$ with a velocity dispersion
$\sigma = 150~{\rm km~s^{-1}}$. We considered different S/N (= 5, 10, 25, 50,
80, 100 per pixel) in the continuum of the spectra. In addition, we created a
grid of spectra from each mock galaxies for error analysis, following the
procedure explained in previous sections. Then, we applied the decoupling
method in two cases: (i) the dust attenuation is a well known and fixed value
(results for the luminosity-weighted age in Fig.~\ref{SSP_plot_fijo}), and (ii)
the dust attenuation is a free parameter in the fitting procedure procedure
(results in Fig.~\ref{SSP_plot_libre}). In both cases, we represent the
relative error in the luminosity-weighted age for the four cases of underlying
stellar population. The error bars show the standard deviation for these values
computed from 100 simulations, as in previous sections. 

As it can be seen, the relative random errors in the luminosity-weighted age for
mainly old stellar populations (panels {\it a}, {\it c} and {\it d} in
Figs.~\ref{SSP_plot_fijo} and \ref{SSP_plot_libre}) are $\sim 30-55\%$ while
the estimated errors and their standard deviation are higher, in average, when
considering younger stellar populations (panel {\it b}). In addition, the
relative errors are significantly higher in the case of fitting simultaneously
the dust attenuation (Fig.~\ref{SSP_plot_libre}). It is clear from this
analysis that the spectral range (similar for data obtained with V300 and V600
grids) is not wide enough to break the age-metallicity-dust degeneracy.  
Additional information in other spectral bands, as ultraviolet and infrared, 
are desirable in order to obtain a reliable value for an independent estimate of
the dust attenuation. In that case, dust attenuation would be fixed to this 
value in the fitting process and a more accurate determination of the rest of 
parameters will be obtained.

In addition, the relative random errors in the luminosity-weighted 
metallicity
are extremely high, above the 100\% in all the cases. This indicates that
the grid of templates, based in six stellar population models, are too
simple for the estimate of metallicity. As noted by \citet{MacArthur2009}
and showed by \citet{Sanchez2011}, the analysis of the stellar
populations is limited by the templates used. Tentative tests with a grid
of 20 and 25 SSPs decrease the error in the luminosity-weighted
metallicity to $\sim 35$~\%, and therefore a study of the stellar
populations is possible with this data by assuming the long computational
time necessary. It is important to highlight, in any case, that the
recovering of the kinematics and the properties of the ionised gas are
well performed with the simple grid considered in this paper, and a
further analysis of the stellar population parameters is 
feasible with an appropriate grid of templates.

\begin{figure}
\centering
\includegraphics[width=0.49\columnwidth]{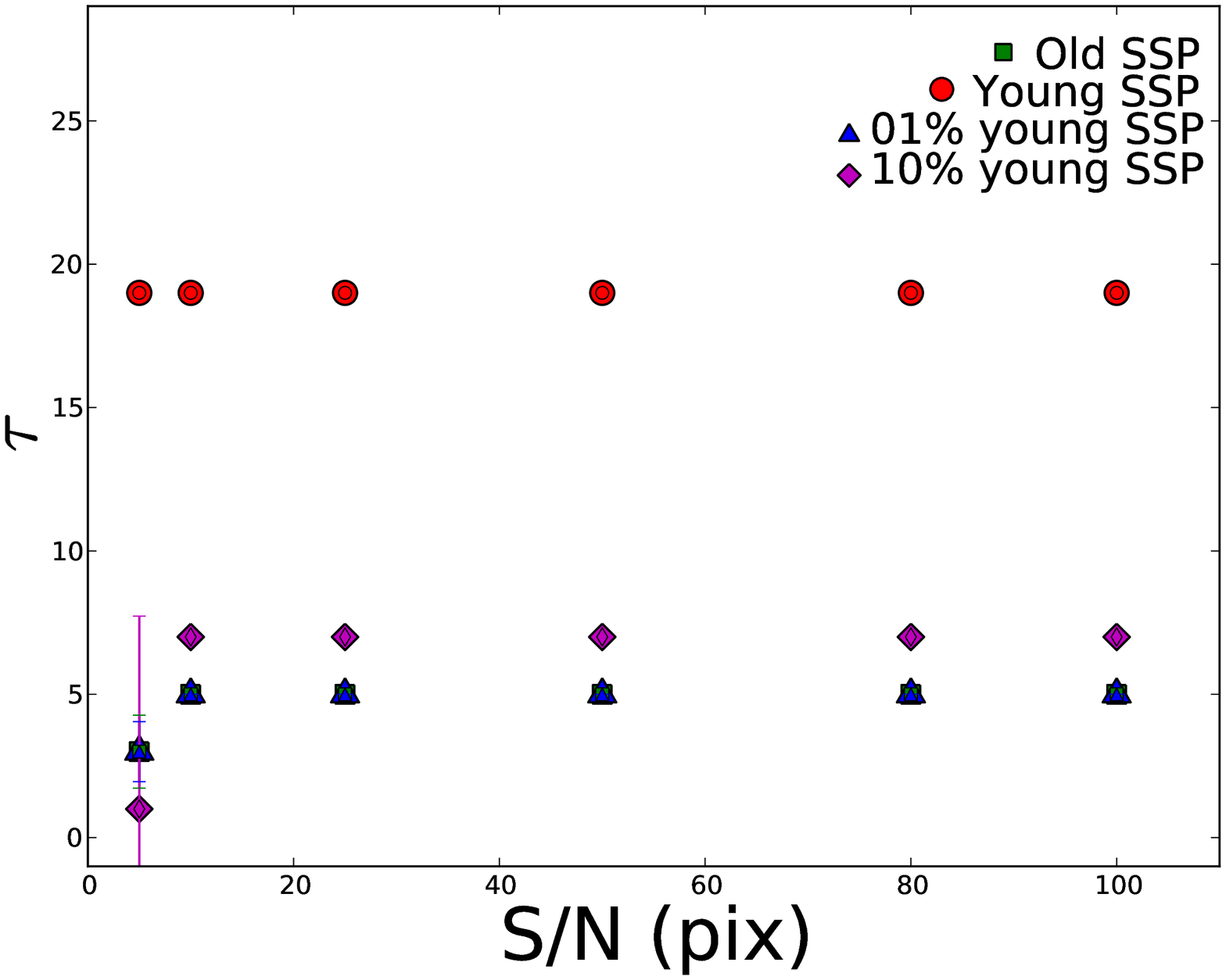}
\includegraphics[width=0.49\columnwidth]{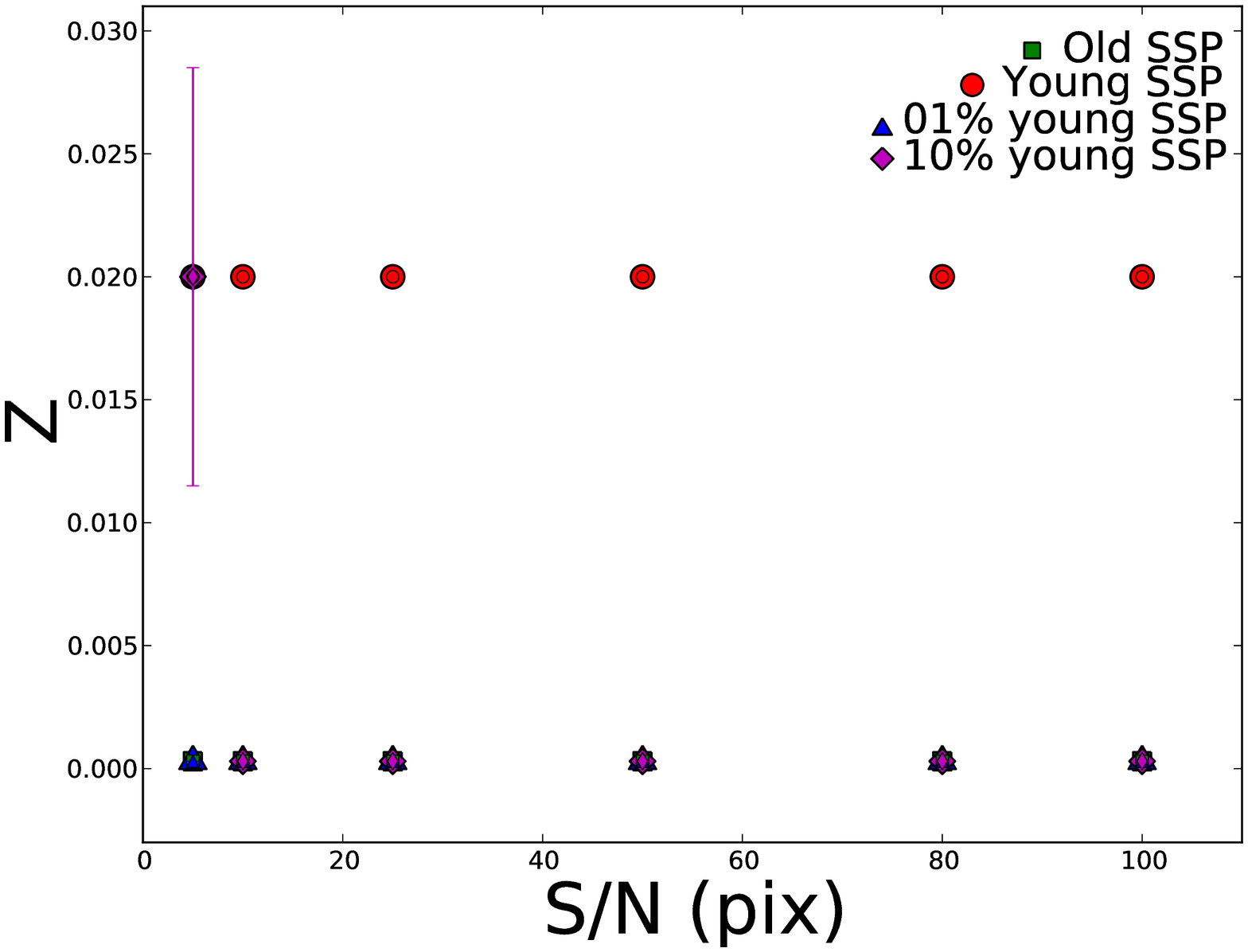}
\includegraphics[width=0.49\columnwidth]{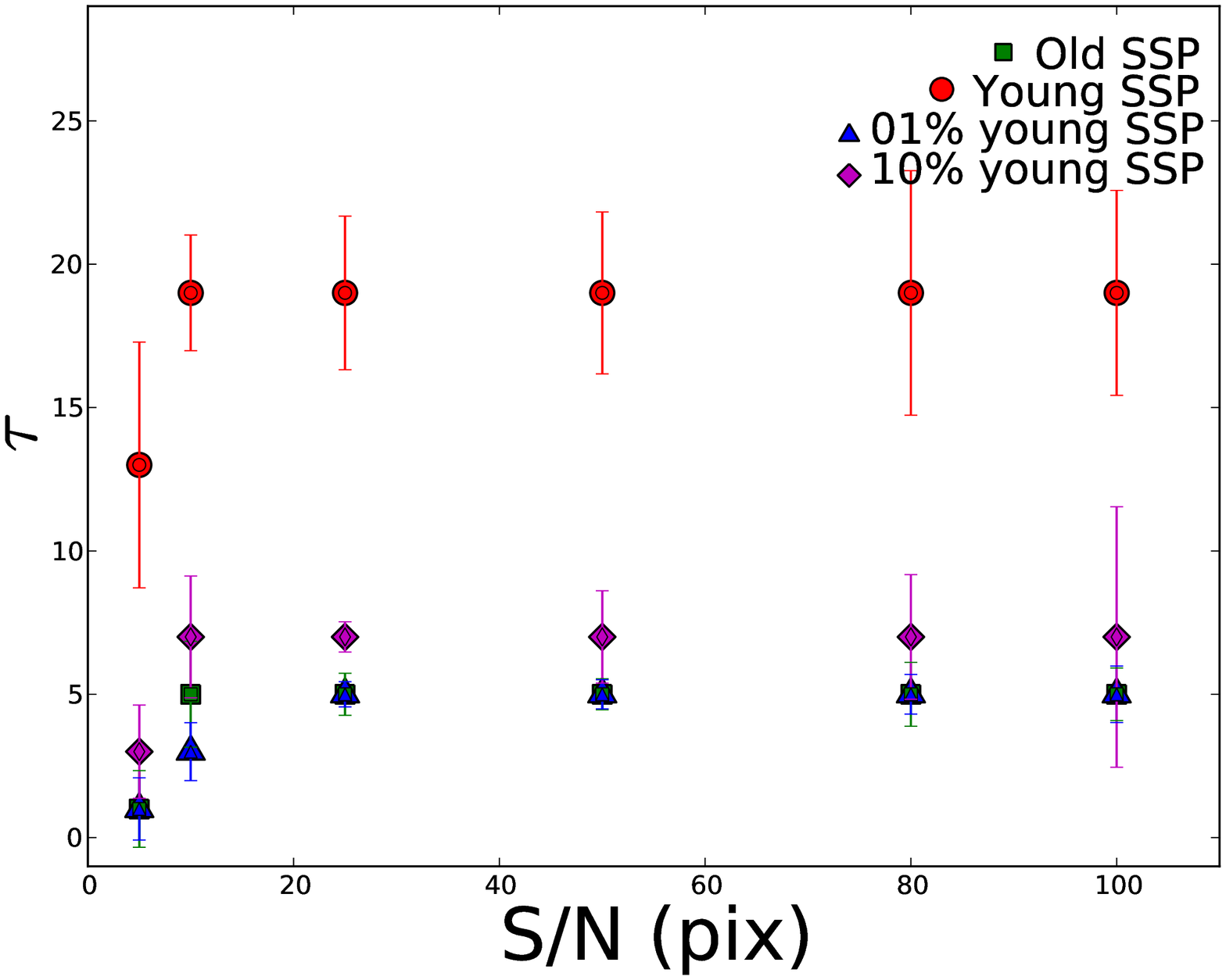}
\includegraphics[width=0.49\columnwidth]{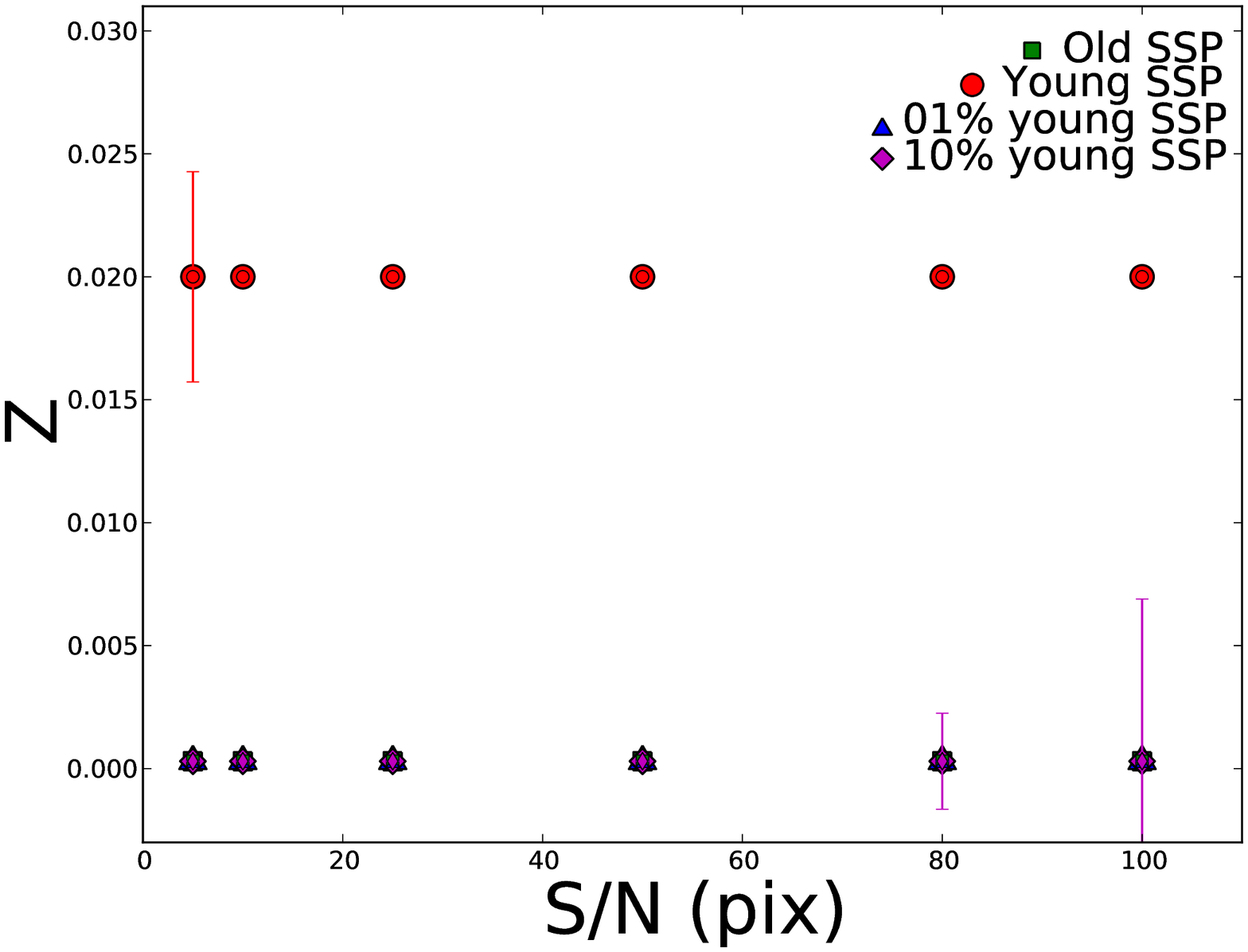}
\caption{Parameters of the most similar SFH spectra ($\tau$ -left panels- and
metallicity -right panels) obtained with the decoupling method, as a function
of the S/N in the continuum for the V600 grid. In this case, the dust
attenuation has been considered known and fixed in the top panels. Results when
the dust attenuation has been fitted are presented in the bottom panels.
Symbols and colours are used for different underlying stellar populations of
the mock galaxies, as it is indicated in the label.}
\label{SSP_plot_SFH} 
\end{figure}

Due to the uncertainties obtained in the luminosity-weighted age and
metallicity when a grid of SSPs is considered in the decoupling method, we
decided to compare our mock galaxies with spectra obtained from a given SFH
(see \S~\ref{sec:SFH}). As part of the fitting procedure explained in
\S~\ref{sec:method}, there is an additional option to compute the most similar
spectrum to the input one from a collection of spectra. This option is
independent of the global fitting to decouple the ionised gas and the stellar
population, and we can feed the procedure with a different set of template
spectra for this experiment. We used this option to explore which SFH
spectra (\S~\ref{sec:SFH}) is more appropriate to describe our mock galaxies.
In Fig.~\ref{SSP_plot_SFH} we represent the $\tau$ parameter (left panels) and
metallicity (right panels) for these mock galaxies, when the dust attenuation
is fixed (upper panels) of fitted (lower panels). The results in both cases are
similar, obtaining low values of $\tau$ ($4 < \tau < 6$) for predominantly old
stellar populations and a high value ($\tau \sim 20$) for younger stellar
populations. In both cases, the metallicity is well determined, as expected
since we consider two metallicity values for this experiment (see
\S~\ref{sec:SFH}). 


\subsection{Comparison with the SDSS data} \label{sec:comparison_SDSS}

\begin{figure}
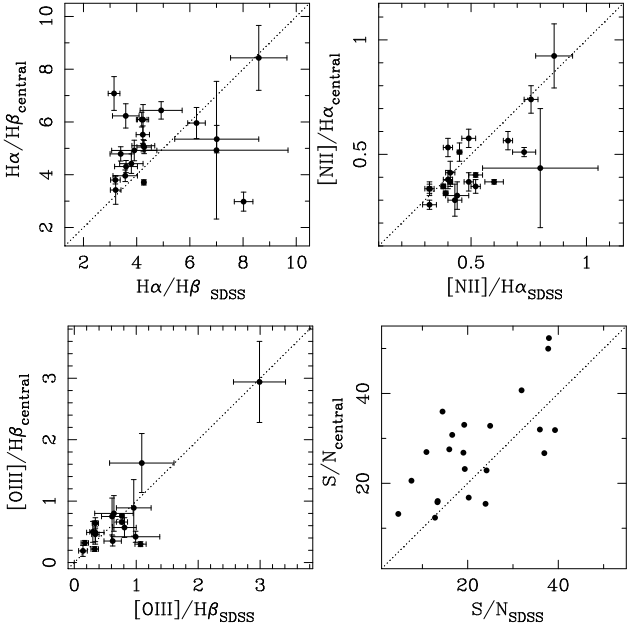

\centering
\includegraphics[angle=-90,width=0.45\columnwidth]{Ha_Hb.ps}
\includegraphics[angle=-90,width=0.45\columnwidth]{NII_Ha.ps}
\vspace{0.3cm}

\includegraphics[angle=-90,width=0.45\columnwidth]{OIII_Hb.ps}
\includegraphics[angle=-90,width=0.45\columnwidth]{SN.ps}
\caption{Comparison of the H$\alpha$/H$\beta$, [N\,{\sc ii}]$\lambda
6583$/H$\alpha$ and [O\,{\sc iii}]$\lambda 5007$/H$\beta$ ratios, derived from
central spectra for the galaxies in this sample and the SDSS spectral, using the
multi-SSP fitting procedure, plus the comparison between the S/N in both
spectra.}
\label{plot_comparison_SDSS2}
\end{figure}

There are 28 galaxies observed in the sample with available SDSS spectroscopy
(marked in Table~\ref{data_sample}). In order to compare the quality of both
data sets, we extracted an integrated spectrum inside a circular aperture of
$5\arcsec$ from the datacubes, centred in the intensity peak of the galaxy at
the $V$-band. The extracted aperture is selected to match the aperture of the
SDSS fibers, taking into account possible errors in the centroid accuracy and
seeing effects in the SDSS observations, and the dithering scheme adopted in
this work. Then we applied the decoupling method to the spectra of both
samples, deriving their physical properties. A full comparison is shown in
Fig.~\ref{plot_comparison_SDSS2}. Four of the galaxies from SDSS present
S/N~$<10.0$ (per pixel, computed over the full spectra), while just one of the
galaxies in our sample are below this level. As expected, there is a good
agreement between the values from both samples, especially for the stronger,
clearly de-blended emission lines, as [O\,{\sc iii}]$\lambda 5007$/H$\beta$
(determination coefficient for a least squares fits to the data $r^2 = 0.79$).
For those galaxies with different measurements, it is possible that the SDSS
fiber would not be well centred in the galaxy, and we are just comparing
different regions of the galaxy. In addition, the S/N of the galaxies observed
in this work is higher in average than the presented for the SDSS data.

\begin{figure*}
\centering
\includegraphics[width=0.46\columnwidth,angle=270]{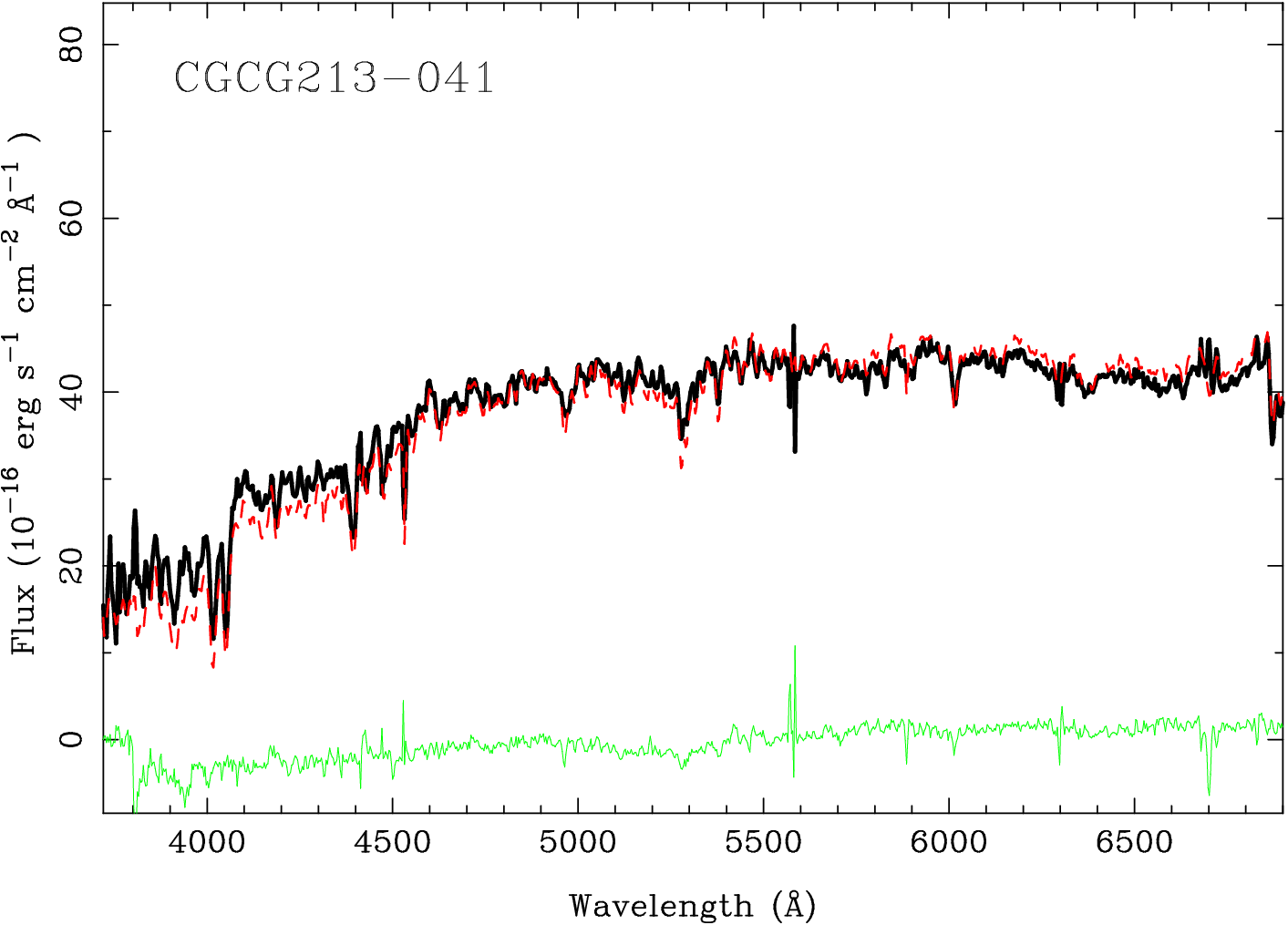}
\includegraphics[width=0.46\columnwidth,angle=270]{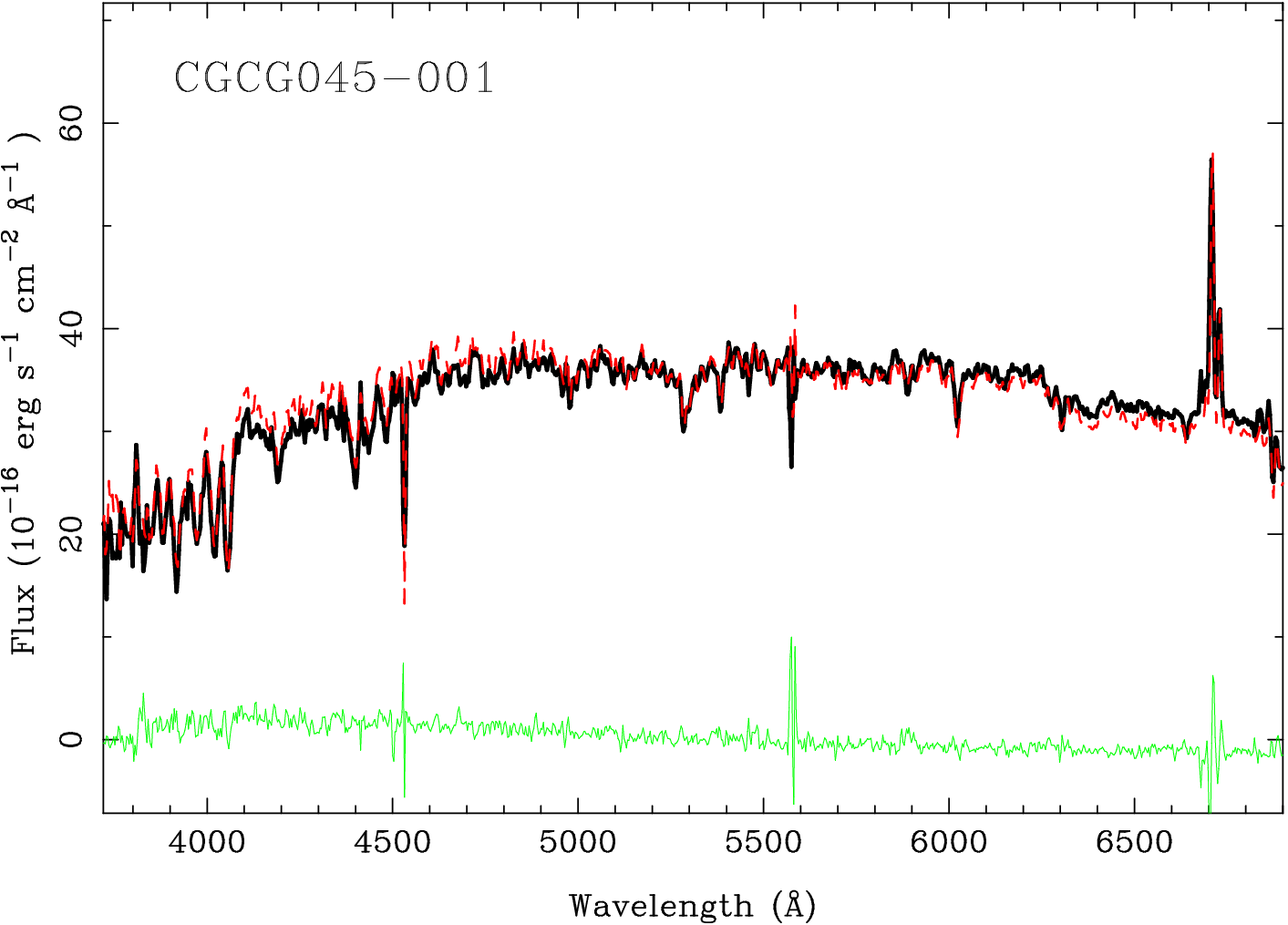}
\includegraphics[width=0.46\columnwidth,angle=270]{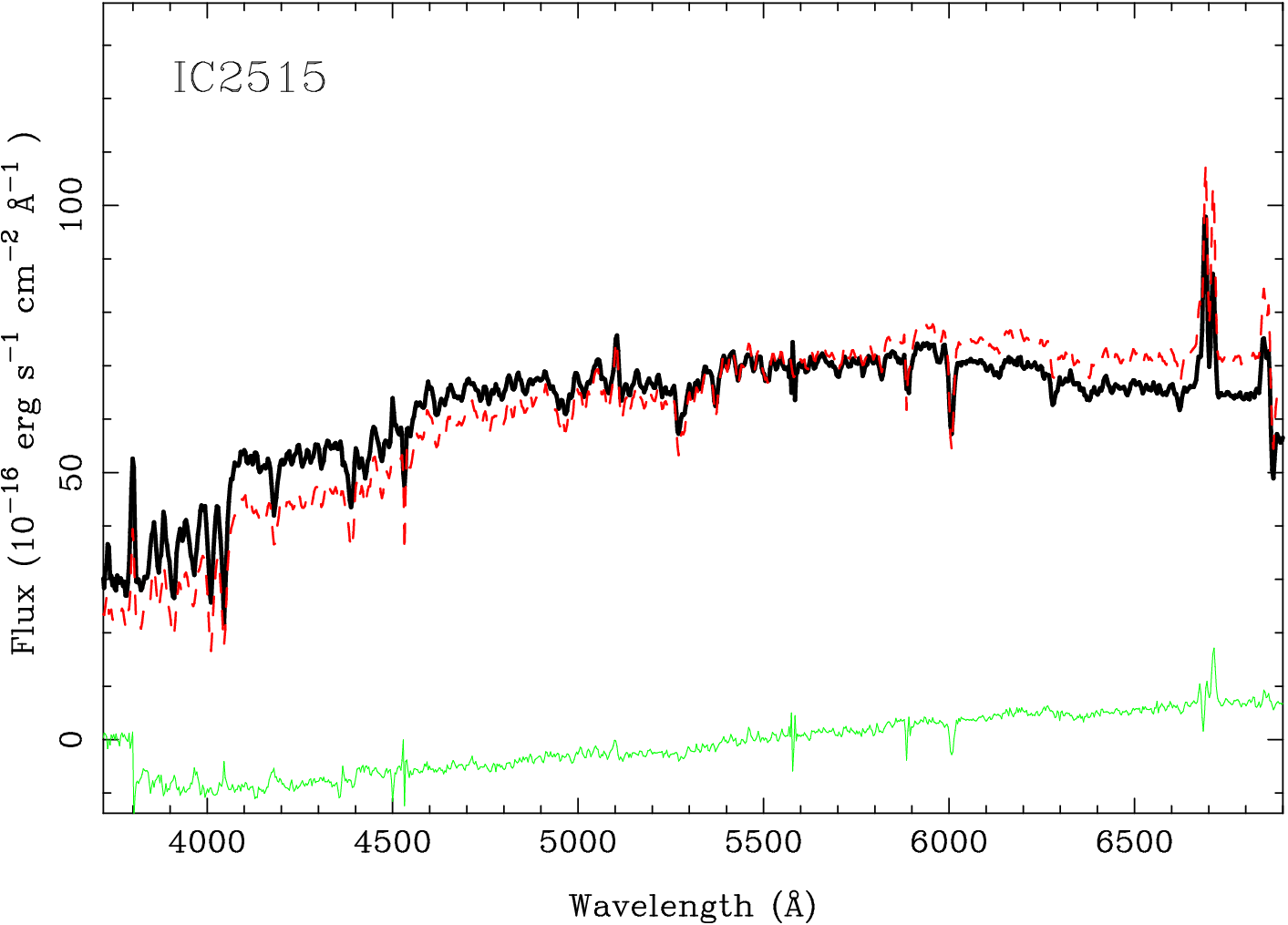}
\includegraphics[width=0.46\columnwidth,angle=270]{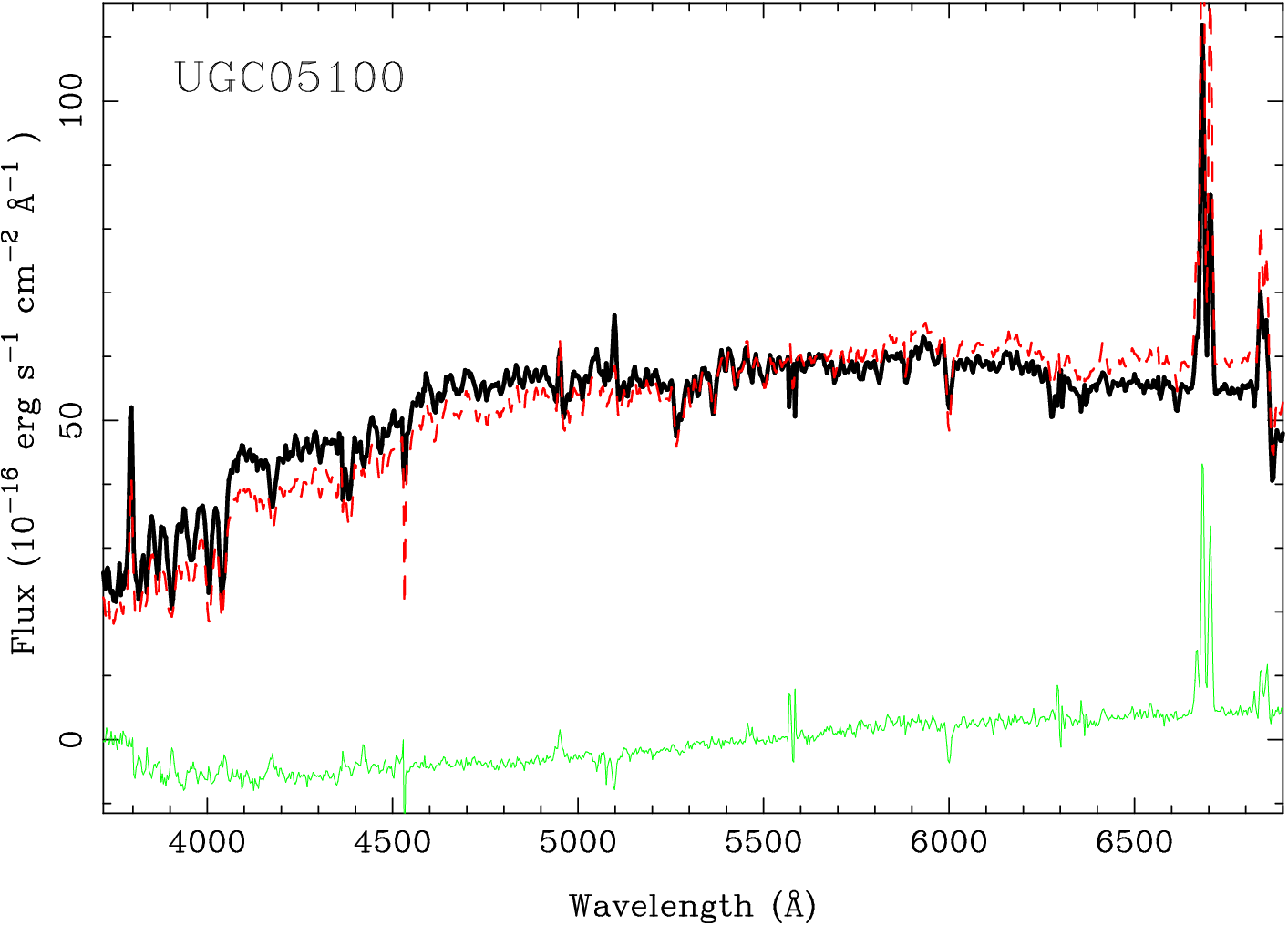}
\includegraphics[width=0.46\columnwidth,angle=270]{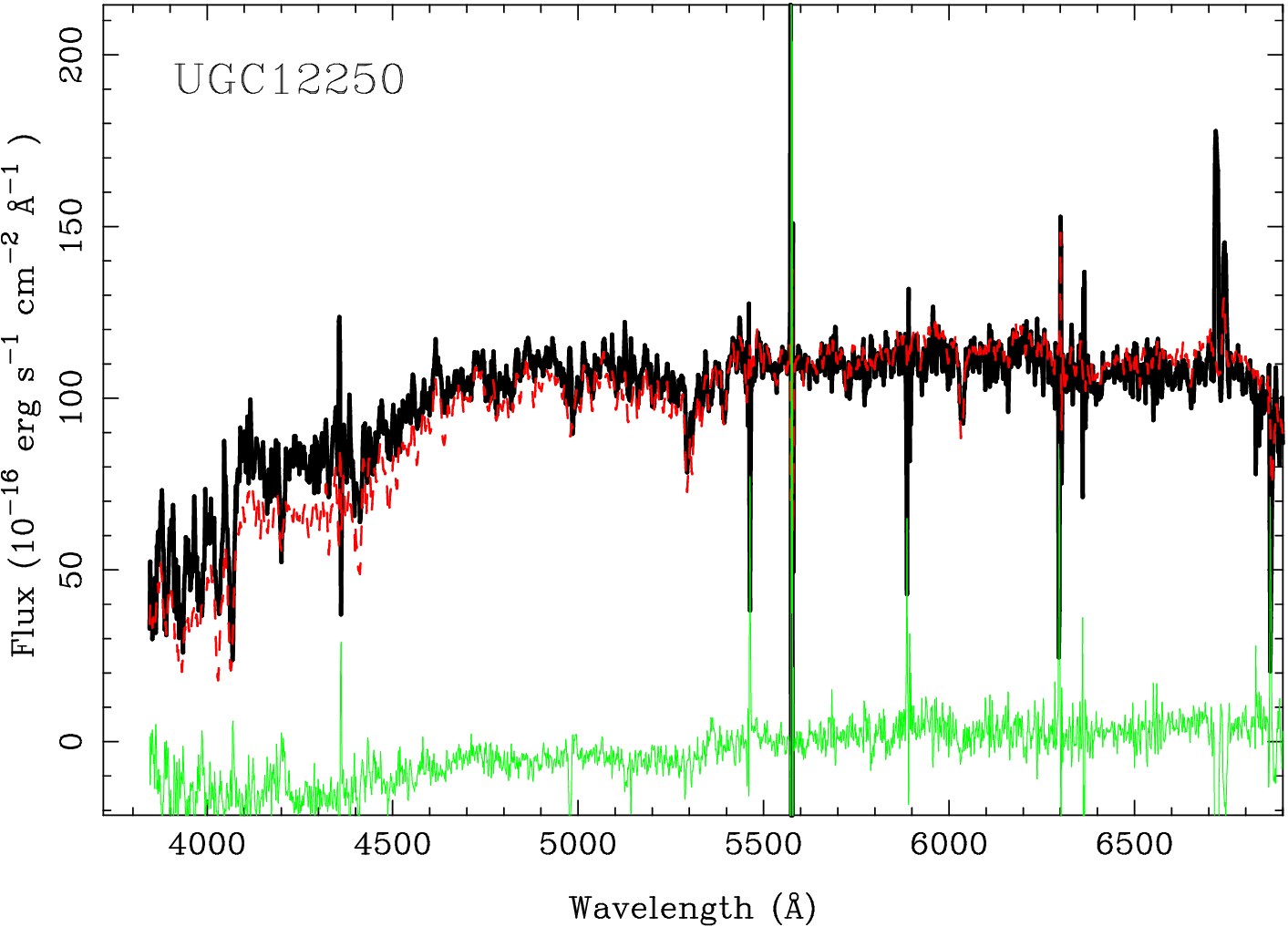}
\includegraphics[width=0.46\columnwidth,angle=270]{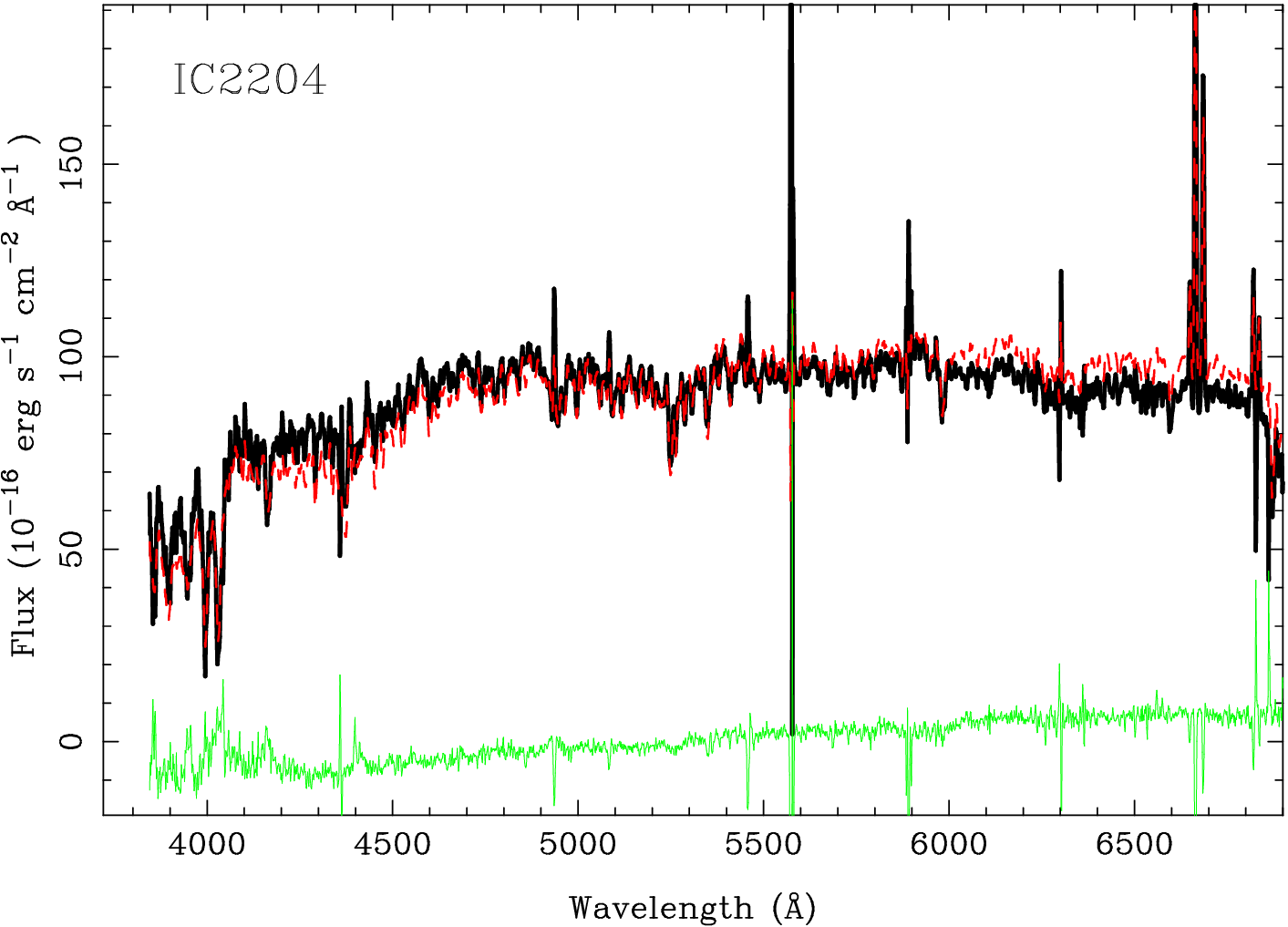}
\includegraphics[width=0.46\columnwidth,angle=270]{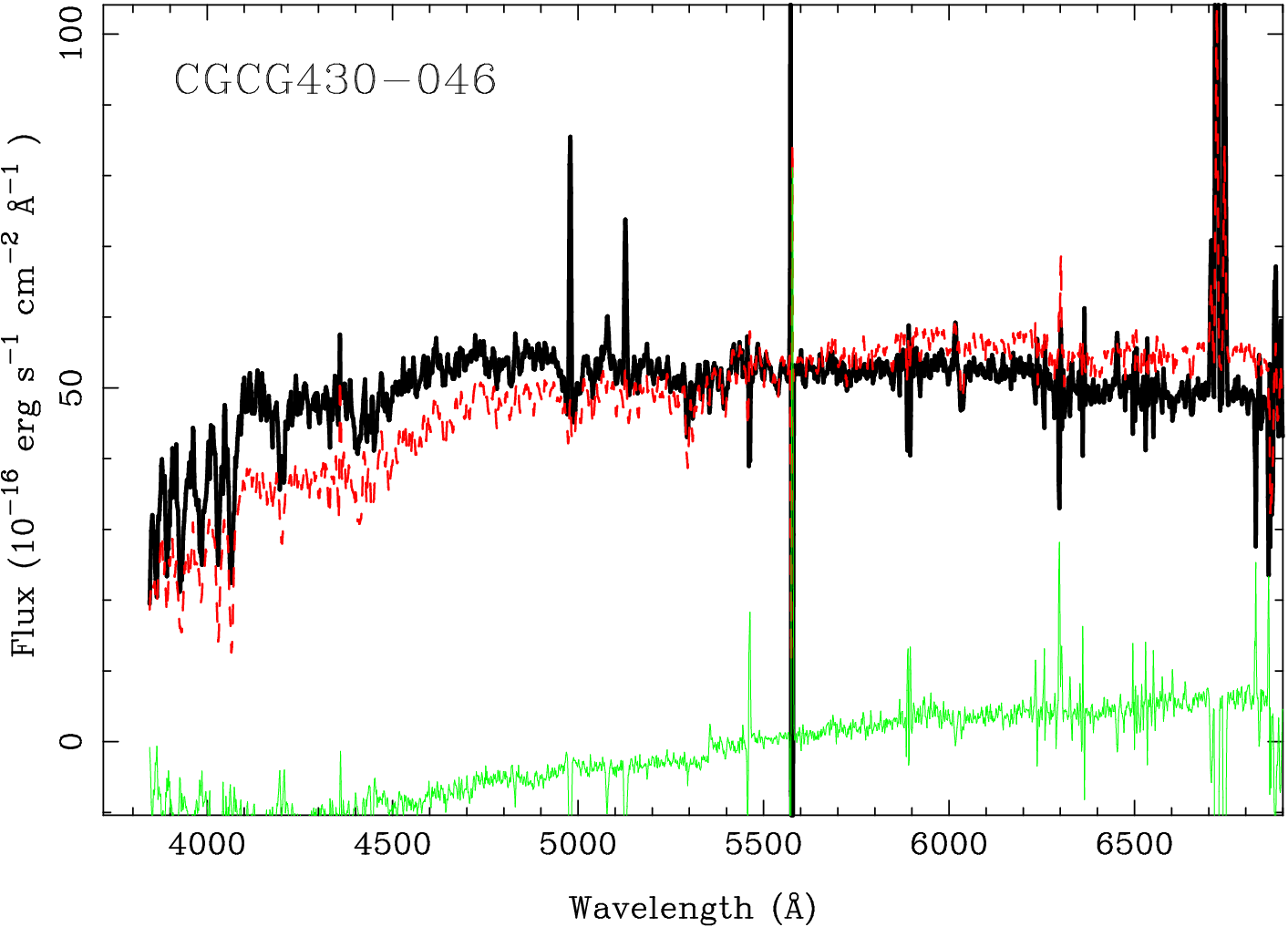}
\includegraphics[width=0.46\columnwidth,angle=270]{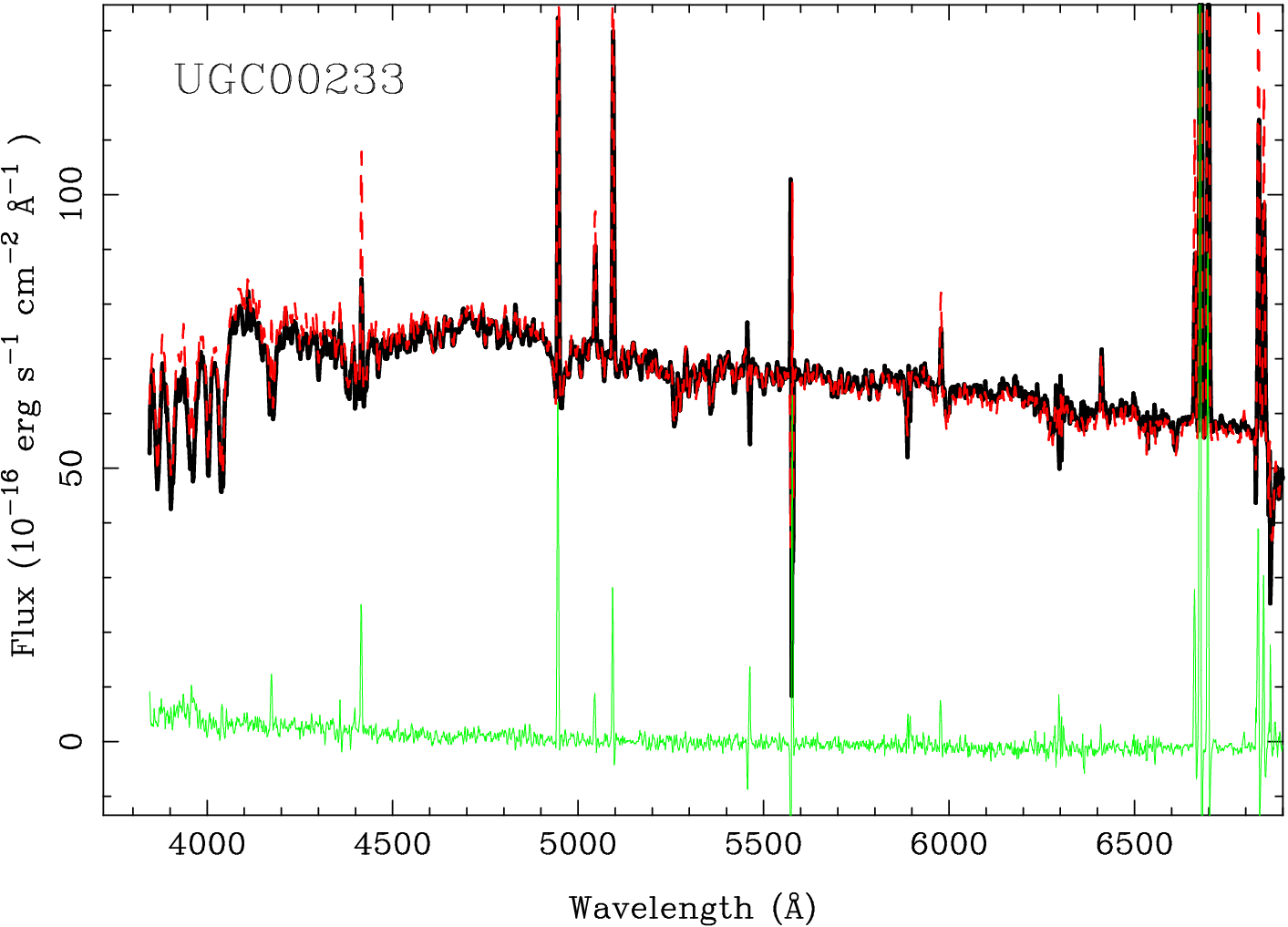}
\includegraphics[width=0.46\columnwidth,angle=270]{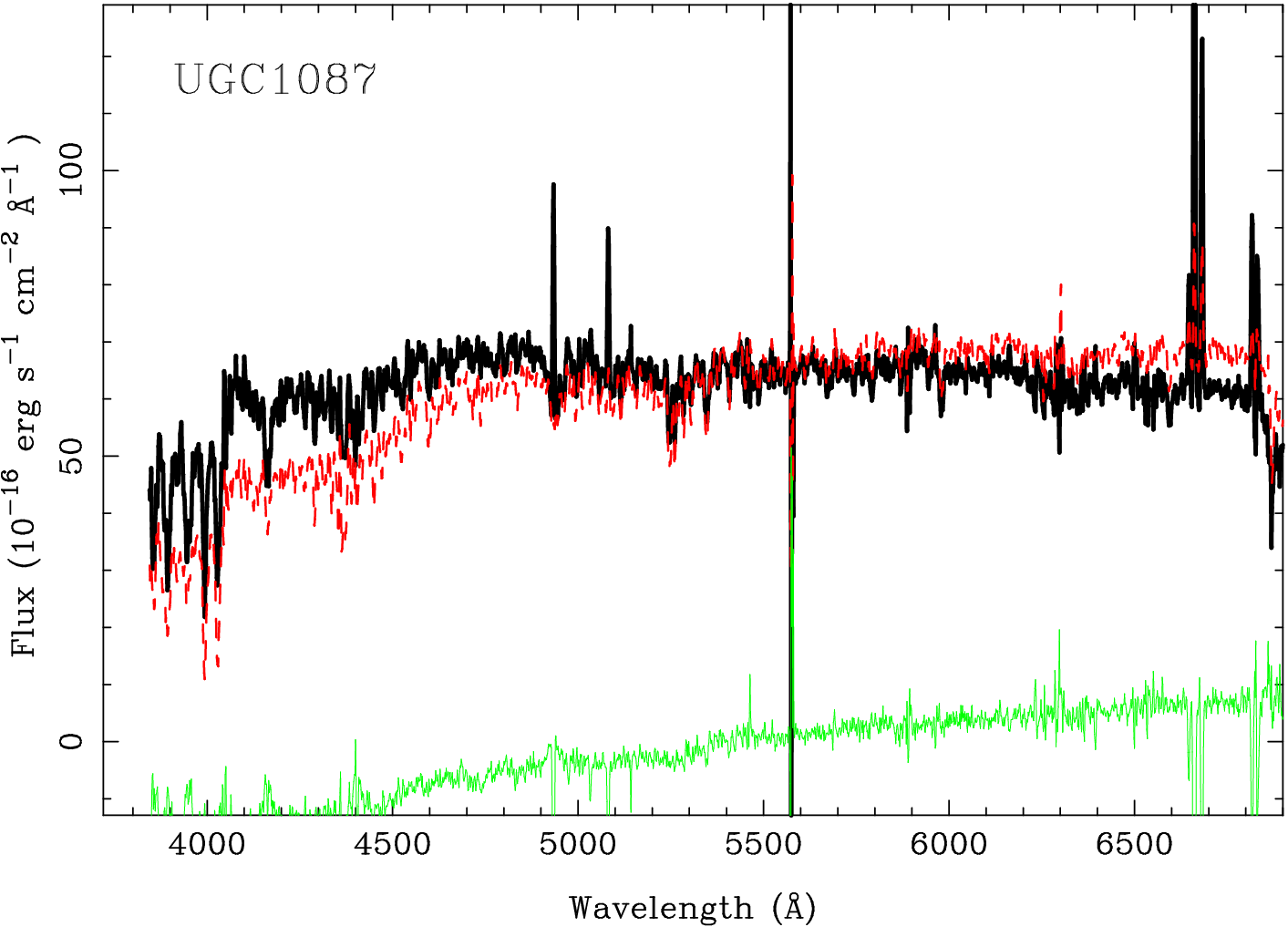}
\caption{Comparison of the central (red) and the integrated (black) spectra for
the subset of galaxies presented in Fig. \ref{example_galaxies}, normalized to
the flux of the central spectrum at $5000-6000\AA\AA$. Differences 
between both spectra are plotted in green.}.
\label{central_vs_integrated}
\end{figure*}

\subsection{Aperture effects in spectroscopy}
\label{sec:apertures}

One of the advantages of Integral Field Spectroscopy is the possibility of
integrating different combinations of the observed spectra for particular
studies. For example, \citet{PINGS} analysed aperture selected spectra of
individual H\,{\footnotesize II} regions within the galaxies of the PINGS
survey, and \citet{Sanchez2011} \citep[and][]{Viironen2011} explored the
aperture effects in the galaxies of this survey, using the IFU as a large
aperture spectrograph. 

\begin{figure}
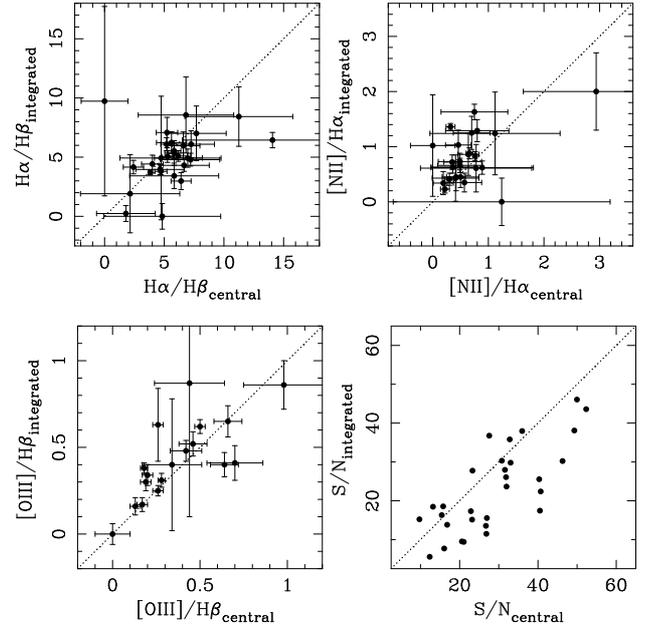

\centering
\includegraphics[angle=-90,width=0.45\columnwidth]{Ha_Hb_pilot.ps}
\includegraphics[angle=-90,width=0.45\columnwidth]{NII_Ha_pilot.ps}
\vspace{0.3cm}

\includegraphics[angle=-90,width=0.45\columnwidth]{OIII_Hb_pilot.ps}
\includegraphics[angle=-90,width=0.45\columnwidth]{SN_pilot.ps}
\caption{Same as in Fig.~\ref{plot_comparison_SDSS2} for the comparison of
the values derived from central and integrated spectra for the galaxies in this
sample.} \label{plot_comparison_pilot}
\end{figure}

In this section, we analyse two aperture spectra derived for each galaxy in the
sample: a central spectrum, obtained as explained in the previous section, and
a $30\arcsec$ diameter aperture spectrum, which comprises most of the observed
regions of each galaxy above the $5\sigma$ detection limit. This spectrum is a
good representation of the integrated properties of each galaxy. In 
Fig.~\ref{central_vs_integrated} we present both aperture spectra for the 
subset of galaxies in Fig.~\ref{example_galaxies}. As it can be seen, the 
integrated spectra present stronger emission lines than the central ones, as 
expected since they are covering also the external parts and different 
structures of galaxies with usually more amount of ionized gas. There are also 
differences in the stellar continuum of both apertures for some of the 
galaxies, with redder colours in the central spectra than in the integrated 
one, which point to older stellar populations in the central part of the 
galaxies. For a further comparison, we 
applied the fitting technique presented above to both spectra, deriving the 
main properties for the galaxies in the sample. We represent the values derived from both spectra in
Fig.~\ref{plot_comparison_pilot} for the main emission line ratios. Again, it can be seen
differences between both measurements, as expected since they are integrating
different regions in the galaxies. In particular, the central aperture is
mainly dominated by the bulge of galaxies, where little ionised gas is usually
found. On the other hand, the integrated aperture covers the most of the galaxy
size and includes arms, bars, and other structures in the galaxies with
different gas and stellar content.

\begin{figure}[h]
\centering
\includegraphics[angle=-90,width=8cm]{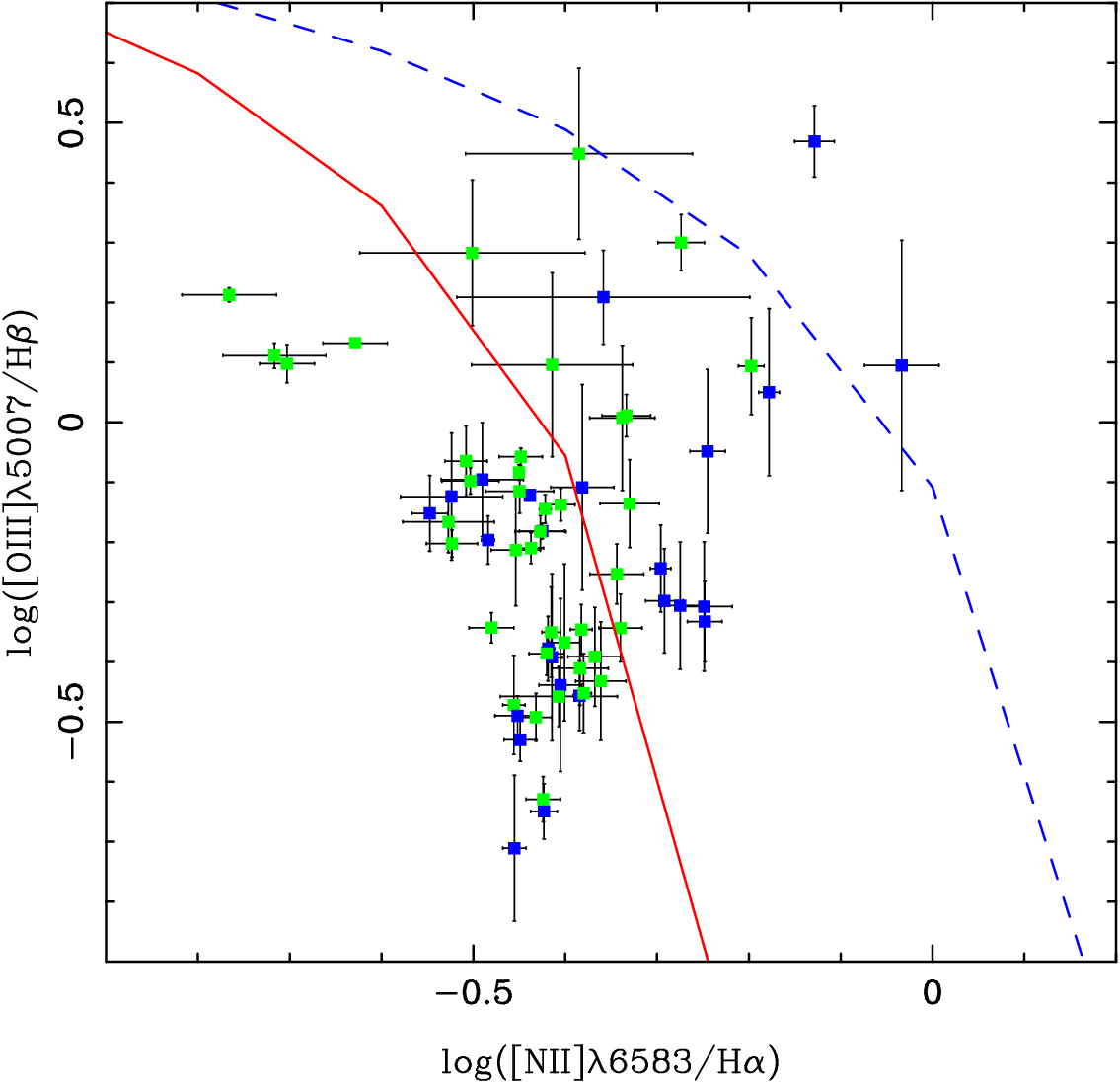} \caption{The BPT diagnostic
diagram for the central (blue filled squares) and integrated (green filled
circles) spectra for the galaxies in the sample. The blue dashed and red solid
lines separate the zones of H\,{\footnotesize II}-like galaxies (below the
lines) and AGNs (above the lines) as defined by \citet{Kauffmann2003} and
\citet{Kewley2001}, respectively.}
\label{comparison_diagnostic}
\end{figure}

As an example of the impact of the differences obtained when analysing both
spectra, in Fig.~\ref{comparison_diagnostic} we present the classic diagnostic
diagram for emission lines galaxies \citep[e.g.,][]{Veilleux1987}, using the
[O\,{\sc iii}]$\lambda 5007$/H$\beta$ and [N\,{\sc ii}]$\lambda 6583$/H$\alpha$
ratios from central (blue filled squares) and integrated (green filled circles)
spectra. The \citet{Kauffmann2003} (blue dashed line) and \citet{Kewley2001}
(red solid line) demarcation curves usually invoked to distinguish between star
forming and AGN galaxies, respectively, are also plotted. For the integrated
spectra, 34 of 46 galaxies (i.e., 73.91~\% of the sample) present accurate
emission line ratios (with an error of $<30$~\%). With these data it is found
that 30 galaxies are classified as pure star forming galaxies (88.23~\% of the
galaxies with accurate emission line ratios), 4 galaxies are in intermediate
region (11.76~\%) and there are no AGN candidates. 2 galaxies in the sample
show little gas ($F_{\rm H\alpha}<0.1~10^{-16}$~cgs Units). Based on the
properties of the central spectra, just 25 (of 46) galaxies present accurate
emission lines ratios (i.e., 54.35~\% of the sample), 14 of them are classified
as star forming galaxies (56.00\%), 2 galaxies (8.00\%) are AGN candidates: IC
2515, classifies as Sy2 in the BAT
catalogue\footnote{\url{http://www.pa.iasf.cnr.it/cgi-bin/bat/main/fonti.cgi}}
\citep{BAT-catalogue}, and CGCG 428-059, galaxy already detected in radio and
X-rays \citep{CGCG428-059}, what indicates that probably hosts an AGN. There
are 9 galaxies in the intermediate region (36.00\%). In particular, 2MASX
J13062093+5318232, classified as AGN in NED database (see
Table~\ref{data_sample}, appears in the intermediate region, but the error in
the [N\,{\sc ii}]$\lambda 6583$/H$\alpha$ ratio is  $>30$~\% and it will be
compatible with its classification as AGN. Finally, in this case 18 galaxies
show little gas in their central spectra.

\begin{figure*}
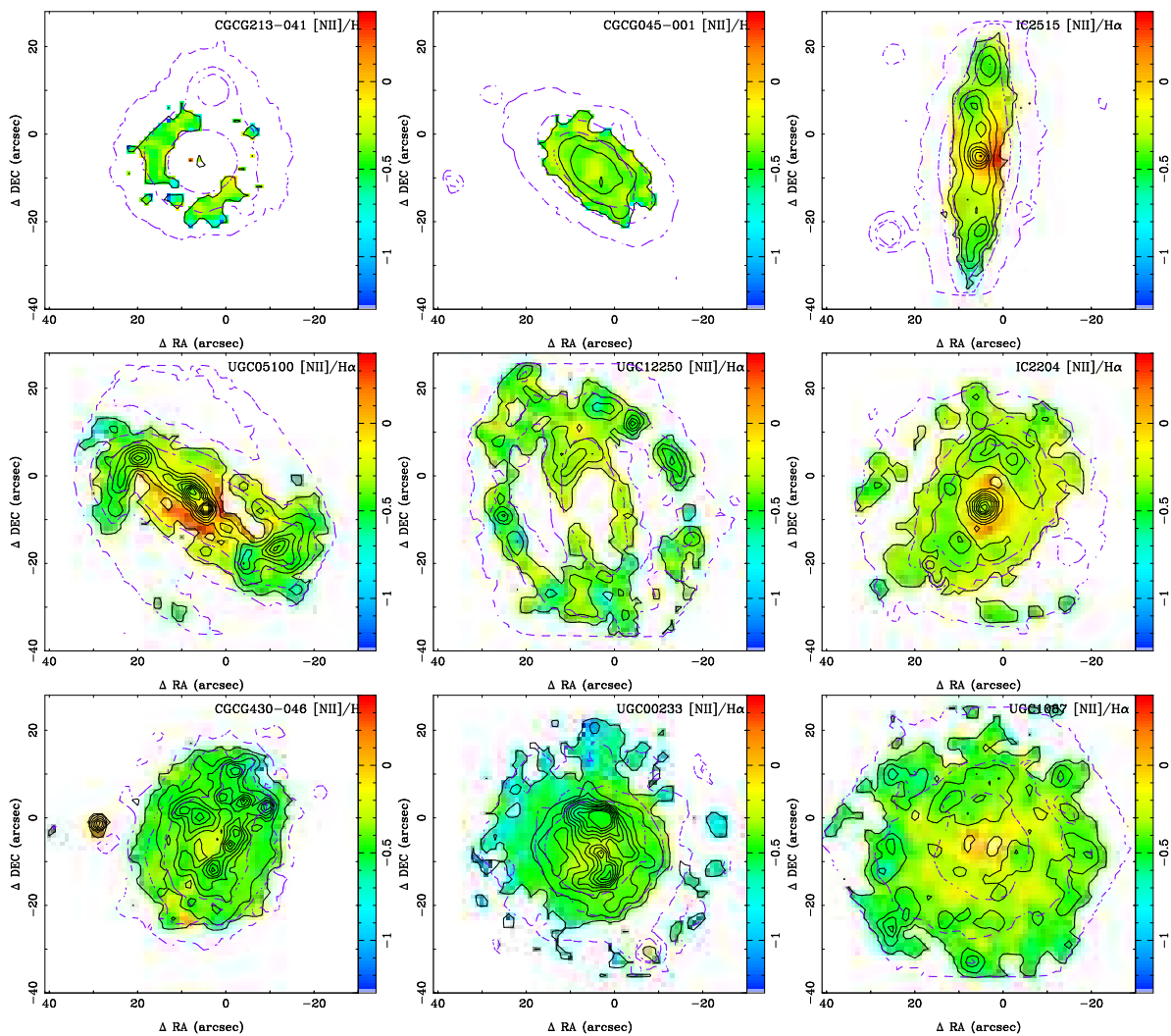

\centering
\includegraphics[width=0.51\columnwidth,angle=270]{rat_NII_Ha_CGCG213-041.ps}
\includegraphics[width=0.51\columnwidth,angle=270]{rat_NII_Ha_CGCG045-001.ps}
\includegraphics[width=0.51\columnwidth,angle=270]{rat_NII_Ha_IC2515.ps}
\includegraphics[width=0.51\columnwidth,angle=270]{rat_NII_Ha_UGC05100.ps}
\includegraphics[width=0.51\columnwidth,angle=270]{rat_NII_Ha_UGC12250.ps}
\includegraphics[width=0.51\columnwidth,angle=270]{rat_NII_Ha_IC2204.ps}
\includegraphics[width=0.51\columnwidth,angle=270]{rat_NII_Ha_CGCG430-046.ps}
\includegraphics[width=0.51\columnwidth,angle=270]{rat_NII_Ha_UGC00233.ps}
\includegraphics[width=0.51\columnwidth,angle=270]{rat_NII_Ha_UGC1087.ps}
\caption{Colour map of the [N\,{\sc ii}]$\lambda 6583$/H$\alpha$ line ratio
derived from the fitting procedures described in the text, for a sub-set of the
observed galaxies. The black solid contours shows the H$\alpha$ observed
intensity, starting at 0.3~\Funits, and with a step of 1~\Funits~for each
contour. The blue dashed contours shows the continuum intensity at the
$R_C-$~band, starting at 0.03~\Funits, with a step of 0.1~\Funits~for each
contour.} 

\label{rat_NII_Ha_maps}
\end{figure*}

It is clear from this comparison that important differences are found when when
analysing and comparing spectra extracted with different apertures over a
galaxy. The two dimensional data from IFS surveys can be used to analyse these
aperture effects \citep[e.g.,][]{Viironen2011}, which could be critical for the
comparison and interpretation of galaxies at different redshifts. A large
number of observed objects using this technique will allow to obtain empirical
calibrations to compare in an appropriate way the values obtained in the Local
Universe with galaxies at higher redshifts.


\subsection{Spatially resolved properties} \label{sec:resolved_properties}

The main advantage of using IFS for the study of galaxies is the ability to
derive spatially resolved spectroscopic properties \citep[e.g.,][]{SAURON_2,
PINGS, VENGA,Castillo-Morales2010}. Using IFS, we obtain a spectrum for each
position of the galaxy, and therefore the decoupling method explained before
can be applied for each of the individual spectra within the obtained datacubes
with enough signal-to-noise. In this way, we obtain two-dimensional
distributions (or maps) of the parameters derived by this analysis. Although it
is out of the scope of this article to analyse in detail each of them, we
present here a few examples to show the quality of the derived maps, focused in
this case on the properties of the ionised gas. 

\begin{figure*}
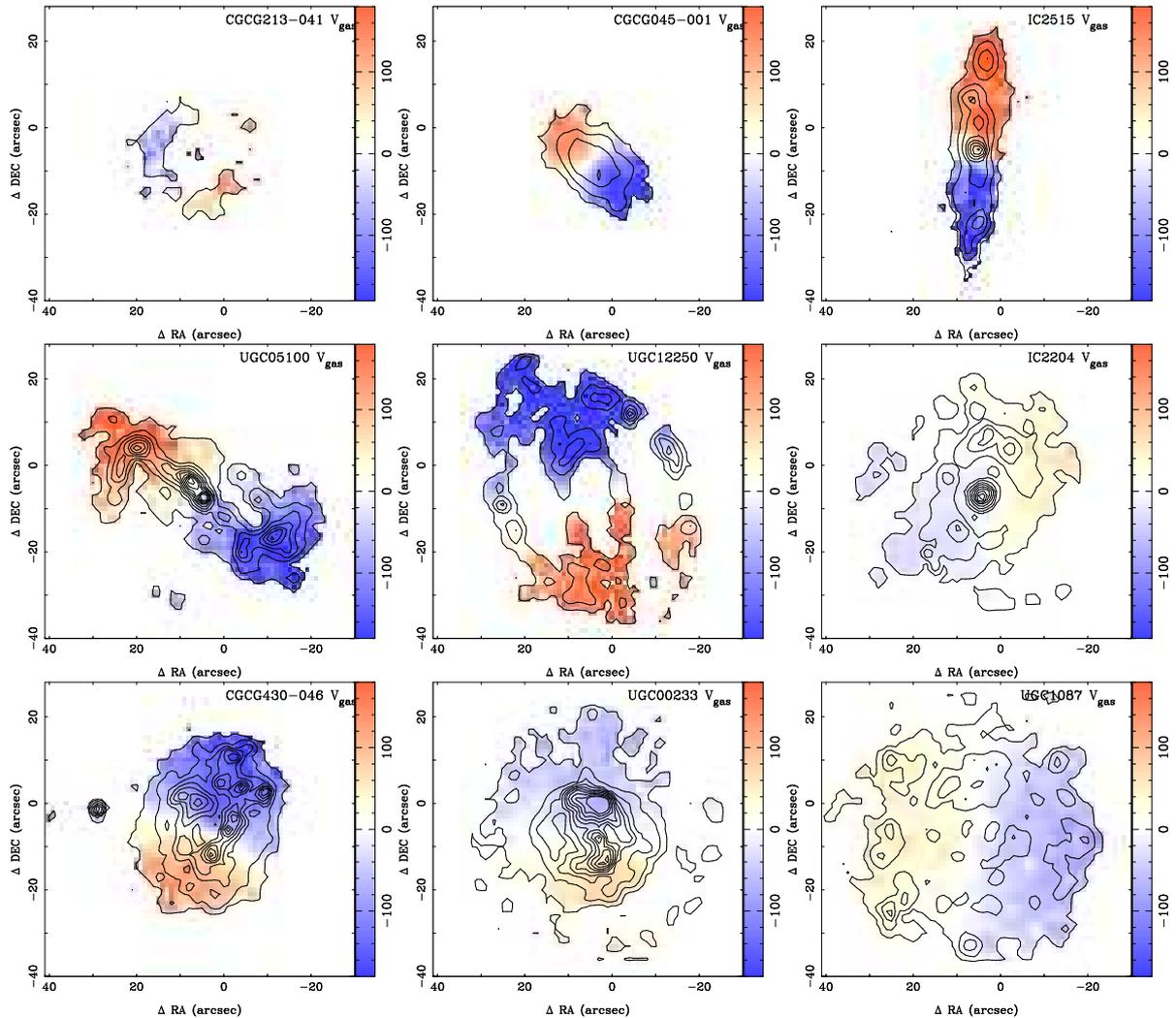

\centering
\includegraphics[width=0.51\columnwidth,angle=270]{Ha_vel_CGCG213-041.ps}
\includegraphics[width=0.51\columnwidth,angle=270]{Ha_vel_CGCG045-001.ps}
\includegraphics[width=0.51\columnwidth,angle=270]{Ha_vel_IC2515.ps}
\includegraphics[width=0.51\columnwidth,angle=270]{Ha_vel_UGC05100.ps}
\includegraphics[width=0.51\columnwidth,angle=270]{Ha_vel_UGC12250.ps}
\includegraphics[width=0.51\columnwidth,angle=270]{Ha_vel_IC2204.ps}
\includegraphics[width=0.51\columnwidth,angle=270]{Ha_vel_CGCG430-046.ps}
\includegraphics[width=0.51\columnwidth,angle=270]{Ha_vel_UGC00233.ps}
\includegraphics[width=0.51\columnwidth,angle=270]{Ha_vel_UGC1087.ps}
\caption{For the same sub-set of galaxies than in previous figures, colour map
of the velocity obtained from the H$\alpha$ emission, once corrected by the
corresponding systemic velocity. The black solid contours shows the H$\alpha$
observed intensity, starting at 0.3~\Funits, and with a step of 1~\Funits~for
each contour, as in Fig.~\ref{rat_NII_Ha_maps}.}
\label{Ha_vel_maps}
\end{figure*}

Figure \ref{rat_NII_Ha_maps} shows the two dimensional distribution of the
[N\,{\sc ii}]$\lambda 6583$/H$\alpha$ line ratio derived from the fitting
procedures described in the text for the sub-set of galaxies in
Fig.~\ref{example_galaxies}. The blue dashed contours shows the continuum
intensity at the $R_C-$~band. The black solid contours shows the H$\alpha$
observed intensity, starting at 0.3~\Funits. Notice that the data presented in
this work enable us to reach sensitivity limits of $\sim$~\Funits, fainter than
previous imaging surveys in the Local Universe
\citep[e.g.][]{Perez-Gonzalez2003, Helmboldt2004, Meurer2006, Kennicutt2008,
Karachentsev2010AJ}, what will allow a more accurate determination of the SFR
in the Local Universe. In addition, we present in Fig.~\ref{Ha_vel_maps} the
velocity map of the ionised gas (from the H$\alpha$ emission), derived with the
previously described procedure. In most of cases it is seen a clear rotational
pattern, mostly due to the bias towards late-type galaxies in the presented
sub-set.

\begin{figure*}
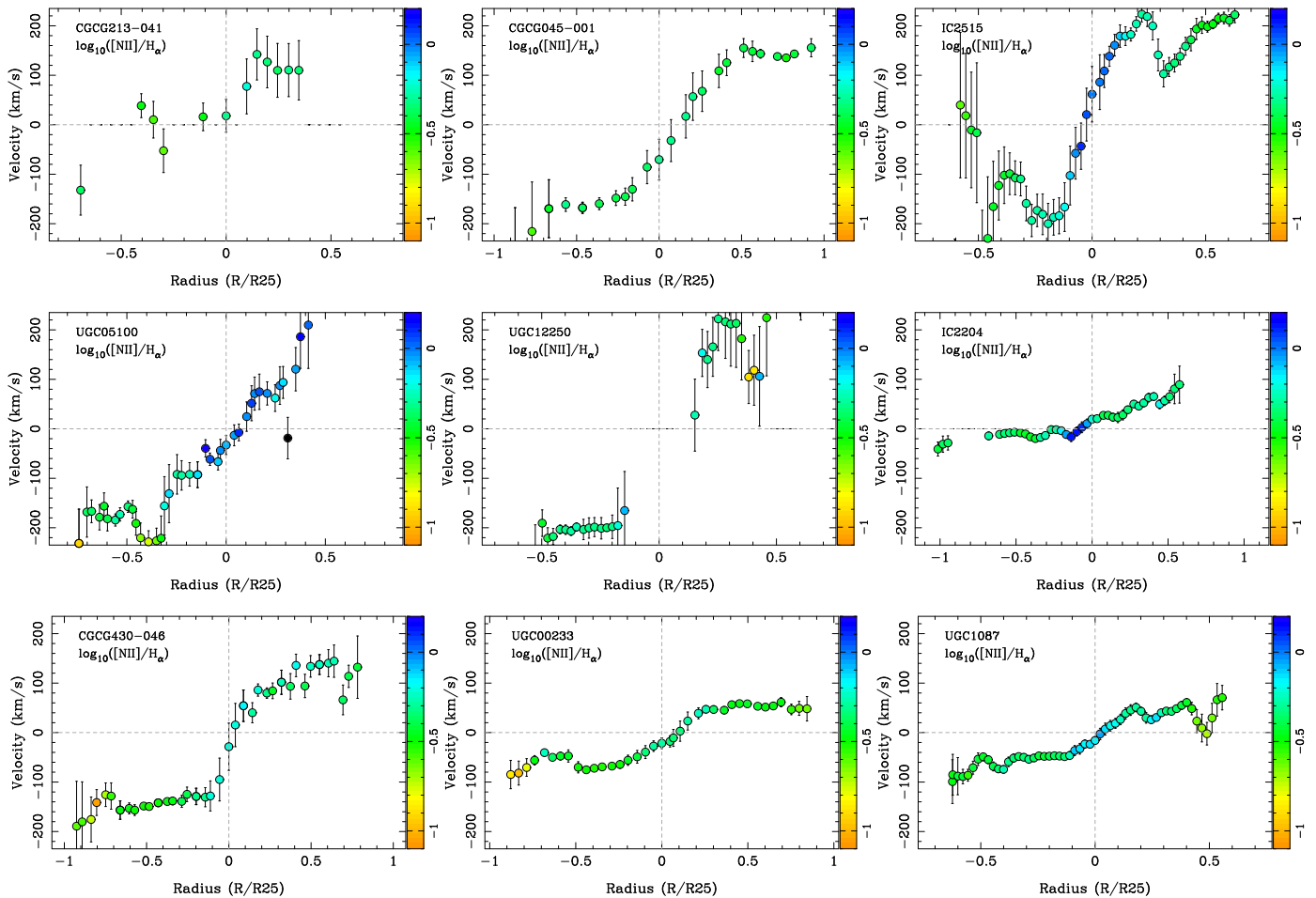

\centering
\includegraphics[width=0.44\columnwidth,angle=270]{vel_rat_CGCG213-041.ps}
\includegraphics[width=0.44\columnwidth,angle=270]{vel_rat_CGCG045-001.ps}
\includegraphics[width=0.44\columnwidth,angle=270]{vel_rat_IC2515.ps}
\vspace{0.3cm}

\includegraphics[width=0.44\columnwidth,angle=270]{vel_rat_UGC05100.ps}
\includegraphics[width=0.44\columnwidth,angle=270]{vel_rat_UGC12250.ps}
\includegraphics[width=0.44\columnwidth,angle=270]{vel_rat_IC2204.ps}
\vspace{0.3cm}

\includegraphics[width=0.44\columnwidth,angle=270]{vel_rat_CGCG430-046.ps}
\includegraphics[width=0.44\columnwidth,angle=270]{vel_rat_UGC00233.ps}
\includegraphics[width=0.44\columnwidth,angle=270]{vel_rat_UGC1087.ps}
\caption{Rotational curves derived from the maps shown in the previous figure
along the axis of maximum rotation. The colour code indicates the
[N\,{\sc ii}]$\lambda 6583$/H$\alpha$ ratio, in logarithmic scale. The
distances are normalised by the $R_{25}$ radius.}
\label{rotation_maps}
\end{figure*}

Finally, Fig.~\ref{rotation_maps} shows the rotational curves along the 
kinematic major axis, derived from the velocity maps of H$\alpha$
(Fig.~\ref{Ha_vel_maps}). The colour code indicates the [N\,{\sc ii}]$\lambda
6583$/H$\alpha$ line ratio presented in Fig.~\ref{rat_NII_Ha_maps}. Values for
$R_{25}$ are taken from the RC3 catalogue (see Table~\ref{data_sample}). From
the sub-sample shown in Fig.~\ref{rotation_maps}, we can see that it is
possible a two dimensional study up to $\sim 1.0~R_{25}$ with the data
presented in this work, although in most of cases we do not reach so large
extension. This easy experiment shows the importance of the sample selection to
accommodate the galaxies in the FOV of the PPAK instrument. Closer galaxies
with high angular sizes will reach $R_{25}$ out of the FOV, while the S/N of
small (and faint) galaxies will be not enough for most of two dimensional
studies. This work shows, therefore, the importance of the sample selection not
only in volume and limit magnitude, but also in the apparent size of galaxies
as previous works showed \citep[e.g.,][]{Bershady2010a}.

These figures illustrate how suitable is the considered experimental setup for
the study of these spectroscopic properties within the full optical extension
of the considered galaxies (characterised by the continuum emission). A
detailed analysis of the resolved properties of these galaxies will be
presented in forthcoming papers.

\section{Summary and Conclusions}

We present in this paper a IFS study of nearby galaxies, exploiting the
capabilities of the wide-field PPAK IFU of the PMAS spectrograph mounted on the
3.5~m telescope. In particular, we explore the requirements to perform a
detailed study of the most relevant spatially resolved spectroscopic properties
of galaxies in the Local Universe along all their optical extension, focused
on: (i) The main properties of the ionised gas; (ii) The luminosity-weighted age
and metallicity of the stellar population; (iii) The main kinematic properties
of both the ionised gas and stellar population.

Two sample selections have been explored, based on (1) a pure redshift
selection and (2) a diameter selected sample, and two different instrumental
setups has been adopted, covering in both cases a similar FOV of
$\sim$~1~arcmin$^2$ and spectral range ($\sim 3700-7050~\AA\AA$), but with
different spectral resolutions: R~$\sim 500$ and R~$\sim 900$. We show that it
is possible to adopt a fix instrumental setup and observing strategy to observe
a large sample of nearby galaxies in a reasonable amount of time (i.e., a
couple of hours per target, including overheads), covering most of the optical
extension of these galaxies and wide spectroscopic optical range with PPAK.

We present a detailed description of the data reduction, and we demonstrate
that the large amount of data produced in this kind of observations can be
handled if a dedicated automatic reduction pipeline is implemented. In addition,
an automatic analysis procedure is used to handle the large number of spectra
in the datacubes derived for each individual galaxy, and to obtain the their
physical properties.

We perform a set of simulations to explore the accuracy of the physical
properties derived from the data. These simulations indicate that the adopted
observational strategy, and in particular, the adopted resolutions and
integration times are suitable to study both the integrated and spatially
resolved spectroscopic properties of these galaxies. The simulations show that
the spectral range of the data is not wide enough to determine simultaneously
the dust attenuation and the stellar populations, and additional information
from other spectral bands, as UV and infrared, is needed to break the
age-metallicity-dust degeneration. The recovering of the properties of the
ionised gas and the kinematics is well performed with the simple grid of
templates used here, although the determination of the luminosity-weighted age
and, especially, metallicity depends on the grid of templates used, as
expected. 

From the comparison of the galaxy data of this work with previously published
imaging and spectroscopic data from the SDSS survey, we show that it is
feasible (1) to obtain spectrophotometric calibrated IFS spectra, and (2) the
derived spectroscopic properties are recovered equally well than with single
fiber spectra. A showcase analysis of the two dimensional properties for a
sub-set of galaxies of this sample indicates that it is feasible to derive the
spectroscopic properties for different apertures, and to obtain accurate 2D
distributions of these properties within the optical extension of these
galaxies.

Finally, we want to highlight that Integral Field Spectroscopy constitutes a
powerful approach for the understanding of the global evolution of nearby
galaxies from the analysis of their resolved physical properties. This study
was part of the preparatory works for a larger IFS survey, the CALIFA
survey\footnote{http://www.caha.es/CALIFA/} ($\sim 600$ galaxies at $0.005 < z
< 0.030$), which will be an important resource for detailed studies of galaxies
in the Local Universe.


\begin{acknowledgements}

We thank the referee, Matt Bershady, for his useful comments.
We acknowledge the {\it Viabilidad, Dise\~no, Acceso y Mejora} funding program,
ICTS-2009-10, and the {\it Plan Nacional de Investigaci\'{o}n y Desarrollo}
funding program AYA2010-22111-C03-03 and AYA-2010-10904-E, of the Spanish
Ministerio de Ciencia e Innovaci\'{o}n (MICINN), for the support given to this
project. EMQ acknowledges financial support from the research projects
AYA2007-67752-C03-03 and AYA2010-21322-C03-02. EMQ, RM and AGP thank support
from Consolider-GTC and AstroMadrid S2009/ESP-1496 -- from the Comunidad de
Madrid --.  RM and AGP acknowledge financial suppor from the project
AYA2005-09413-C02-02. JVM and JIP are funded by the grants AYA2007-67965-C03-02
-- from the Spanish MICINN -- and TIC114 -- from the Junta de Andaluc{\'{\i}}a
--.  

We also thank the director of CEFCA, Dr. M. Moles, for his sincere support to
this project.

This paper makes use of the Sloan Digital Sky Survey data. Funding for the SDSS
and SDSS-II has been provided by the Alfred P. Sloan Foundation, the
Participating Institutions, the National Science Foundation, the U.S.
Department of Energy, the National Aeronautics and Space Administration, the
Japanese Monbukagakusho, the Max Planck Society, and the Higher Education
Funding Council for England. The SDSS Web Site is \url{http://www.sdss.org/}.

The SDSS is managed by the Astrophysical Research Consortium for the
Participating Institutions. The Participating Institutions are the American
Museum of Natural History, Astrophysical Institute Potsdam, University of
Basel, University of Cambridge, Case Western Reserve University, University of
Chicago, Drexel University, Fermilab, the Institute for Advanced Study, the
Japan Participation Group, Johns Hopkins University, the Joint Institute for
Nuclear Astrophysics, the Kavli Institute for Particle Astrophysics and
Cosmology, the Korean Scientist Group, the Chinese Academy of Sciences
(LAMOST), Los Alamos National Laboratory, the Max-Planck-Institute for
Astronomy (MPIA), the Max-Planck-Institute for Astrophysics (MPA), New Mexico
State University, Ohio State University, University of Pittsburgh, University
of Portsmouth, Princeton University, the United States Naval Observatory, and
the University of Washington.

\end{acknowledgements}


\bibliography{E_Marmol_Queralto2011}
\bibliographystyle{aa}

\end{document}